\documentclass[twocolumn,longbibliography,floats,floatfix,superscriptaddress,aps,rmp]{revtex4-1}  

\usepackage{amsfonts}
\usepackage{amssymb}
\usepackage{amsmath}
\usepackage{calc}
\usepackage{graphicx}
\usepackage{bm}
\usepackage{color,xcolor}

\usepackage{epstopdf}
\usepackage{soul}

\def\Black{}
\def\black{\Black}

\def\bk{\color{black}} 
\def\bu{\color{blue}} 

\def\be{ \begin{equation}}
\def\ee{ \end{equation}}
\def\bea{ \begin{eqnarray}}
\def\eea{ \end{eqnarray}}
\def\bse{ \begin{subequations}}
\def\ese{ \end{subequations}}
\def\beas{\begin{eqnarray*}}
\def\eeas{\end{eqnarray*}}
\def\bt{\begin{tabular}}
\def\btc{\begin{tabular}{c}}
\def\et{\end{tabular}}
\def\bM{\left[\begin{array}{cccccc}}
\def\eM{\end{array}\right]}

\def\ket#1{\vert #1 \rangle}

\def\Dk{\Phi_0} 
\def\Dkt{\Phi_0\t} 

\def\Adt#1{\Phi_{#1}\t} 

\def\ddt{\frac{d}{dt}}
\def\e{\,\text{e}}
\def\eig{\varepsilon} 
\def\etal{\emph{et al. }}

\def\fromto{\leftrightarrow}
\def\f{\text{f}} 
\def\i{\text{i}} 
\def\ii{{\rm i}}
\def\mix{\vartheta} 
\def\mixa{\chi} 
\def\mixb{\phi}

\def\s#1{#1} 

 \def\Stokes{$S$ }    \def\pump{$P$ } 
\def\Stokesx{$S$}    \def\pumpx{$P$} 
\def\control{\mbox{\em C }} 

\def\t{(t)}

\def\calA{\mathcal{A}} 
\def\Amin{\calA_{\rm \min}} 
\def\Amin{\mathcal{A}_{\rm min}} 
\def\B{\mathbf{B}} 
\def\C{\mathbf{C}}
\def\D{\mathbf{D}} 
\def\DP{\Delta_{P}} \def\DS{\Delta_{S}}  

\def\deff{\delta_{\rm eff}} \def\Deff{\Delta_{\rm eff}} 

\def\E{\mathbf{E}} 
\def\calE{\mathcal{E}} 

\def\H{\mathsf{H}} 

\def\calN{\mathcal{N}}

\def\Op{\Omega_{P}}  \def\Os{\Omega_{S}} 
\def\Orms{\Omega_{\rm rms}} 

 \def\Omax{\Omega_{\rm max}}  
\def\case{\tfrac} 

\newcounter{ckb}   

\def\mycite#1{\cite{#1}}
\def\mytcite#1{\textcite{#1}}
\def\location#1{{\bu Reprinted from #1 \bk}} 
\def\locationa#1{{\bu Adapted from #1 \bk}} 
\def\location#1{{Reprinted from #1}} 
\def\locationa#1{{Adapted from #1}} 

\def\NMR{Lev79,Fre80,Lev86,Fre97}
\def\polarization{Wes49,Des49,Pan55a,Pan55b,Pan55c,Har64,McI68}

\def\composite{Tor13}
\def\LZSM{Lan32a,Zen32,Maj32,Stu32}

\def\CAP-theory{Tor11}
\def\CAP-experiment{Sch13}

\def\locfigs{./RMPfigsN/}
\graphicspath{\locfigs}

\def\mycite#1{{\bu\cite{#1}}}
\def\mytcite#1{{\bu\textcite{#1}}}

\def\L{$\Lambda$}

\def\Also#1{} 

\def\btc{} \def\et{} 

\def\bt{\begin{tabular}}
\def\et{\end{tabular}}

\def\coloronline{(Color online) }
\def\area{\mathcal{A}}

\def\AT{\Phi_{\rm AP}}

\def\wg{WG}
\def\wgs{WGs}

\def\bws{}

\begin{document} 



\author{Nikolay V. Vitanov}
\affiliation{\mbox{Faculty of Physics, St Kliment Ohridski University of Sofia, James Bourchier 5 blvd, 1164 Sofia, Bulgaria}}
\author{Andon A. Rangelov}
\affiliation{\mbox{Faculty of Physics, St Kliment Ohridski University of Sofia, James Bourchier 5 blvd, 1164 Sofia, Bulgaria}}
\author{Bruce W. Shore}
\affiliation{\mbox{BWShore Quantum Consulting, 618 Escondido Circle, Livermore, CA 94550, USA}}
\author{Klaas Bergmann}
\affiliation{\mbox{Fachbereich Physik und Forschungszentrum OPTIMAS, Technische Universit\"at Kaiserslautern, Germany}}
\title{Stimulated Raman adiabatic passage in physics, chemistry and beyond}
\date{\today }

\begin{abstract}
The technique of stimulated Raman adiabatic passage (STIRAP), which allows efficient and selective population transfer between quantum states without suffering loss due to spontaneous emission, was introduced in 1990 (Gaubatz \emph{et al.}, J. Chem. Phys.  \textbf{92}, 5363, 1990).
Since then STIRAP has emerged as an enabling methodology with widespread successful applications in many fields of physics, chemistry and beyond.
This article reviews the many applications of STIRAP emphasizing the developments since 2000, the time when the last major review on the topic was written (Vitanov \emph{et al.}, Adv. At. Mol. Opt. Phys. \textbf{46}, 55, 2001).
A brief introduction into the theory of STIRAP and the early applications for population transfer within three-level systems is followed by the discussion of several extensions to multilevel systems, including multistate chains and tripod systems.
The main emphasis is on the wide range of applications in atomic and molecular physics (including atom optics, cavity quantum electrodynamics, formation of ultracold molecules, etc.), quantum information (including single- and two-qubit gates, entangled-state preparation, etc.), solid-state physics (including processes in doped crystals, nitrogen-vacancy centers, superconducting circuits, semiconductor quantum dots and wells), and even some applications in classical physics (including waveguide optics, polarization optics, frequency conversion, etc.).
Promising new prospects for STIRAP are also presented (including processes in optomechanics, precision experiments, detection of parity violation in molecules, spectroscopy of core-nonpenetrating Rydberg states, population transfer with X-ray pulses, etc.).
\end{abstract}

\maketitle
\tableofcontents


\section{Introduction \label{sec-intro}}

Stimulated Raman adiabatic passage (STIRAP) originated as a technique for efficiently transferring
population {between} two discrete quantum states by coupling them with two radiations fields via an intermediate state, which is usually a   radiatively decaying   state.
A great variety of techniques exist for producing such transfer, each with its own advantages and disadvantages \mycite{Sho90,Sho13}.
Population transfer by STIRAP is notable because
\begin{itemize}
\item[(i)] it is  immune against loss through spontaneous emission from the intermediate state, despite the fact that radiative coupling may last much longer than the radiative lifetime;
\item[(ii)] it is  robust against small variations of  experimental conditions, such as laser intensity, pulse timing and pulse shape.

\end{itemize}
\vspace{-.5em}
Because of these   features  STIRAP, initially developed for and applied to the excitation of molecular vibrations,
  has subsequently found widespread use, within the last 25 years, not only in atomic and molecular physics and chemistry,
 but also to a variety of other fields of science and engineering.

\begin{figure}[tb]
\includegraphics[width=0.60\columnwidth]{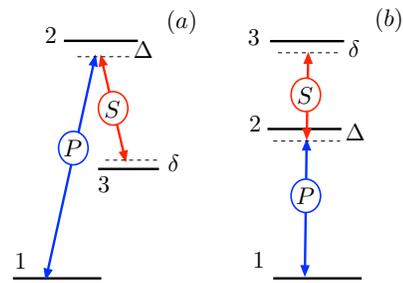}
 \caption{\coloronline
$(a)$ The \L~linkage pattern
showing  \pumpx-field and \Stokesx-field linkages,  the single-photon detuning $\Delta \equiv \Delta_{P}$ and the two-photon detuning $\delta \equiv \Delta_{P}-\Delta_{S}$.
$(b)$ The ladder linkage pattern, showing
the two-photon detuning $\delta \equiv \Delta_{P}+\Delta_{S}$.
For STIRAP  it is necessary that $\delta = 0$.
 The relative ordering of energies $E_2$ and $E_3$ does not matter when using the customary rotating-wave approximation (RWA).
}
\label{fig-links2dN-r}
\end{figure}

The concept of STIRAP was first fully presented,  with  experimental data and the basic underlying theory  by \mytcite{Gau90}.
That work followed the earlier presentation of some preliminary data by \mytcite{Gau88}
and the discussion of an essential aspect,  the condition for adiabatic evolution, by  \mytcite{Kuk89}. 
The acronym STIRAP was coined because the process was first studied in the  \L~linkage, see Fig. \ref{fig-links2dN-r}, which is reminiscent of a stimulated Raman process.
Nowadays, the acronym is used for any transfer process that exhibits the features defining STIRAP.

An early comparison with other methods for population transfer was presented by \mytcite{He90}.
A summary of some features of STIRAP was given by \mytcite{Ber95}; its counterintuitive aspects were discussed in detail by \mytcite{Sho95}.
A presentation with a tutorial approach can be found in \mytcite{Ber98}, and a review by \mytcite{Vit01b} addressed primarily the chemistry community.
The progress during the decade after the presentation of the concept has been thoroughly reviewed by \mytcite{Vit01a}.
Detailed discussion of the theory appear in \mytcite{Sho08} and Chapter 14 of \mytcite{Sho11}.
Selected aspects of STIRAP have been reviewed by \mytcite{Rice00,Sha03,Kra07}.
The  individual concepts (most notably adiabatic following,  population trapping, torque equations and Autler-Townes splitting) that combine to create and interpret the STIRAP procedure have earlier origins, as discussed by \mytcite{Sho13}.
Brief discussion of what motivated STIRAP and how it was found are given in Sec. IA of \mytcite{Vit01a} as well as in Sec. I of \mytcite{Ber15}.

The next section  summarizes the basic features of  STIRAP.
Section  \ref{sec-three-state}  describes the features of STIRAP that are relevant for the three-state quantum systems, which were of interest for the first researchers.
They remain relevant for contemporary applications.
Sections \ref{Sec:multi}  and \ref{sec-AMO} describe extensions of the basic concepts to include multistate systems,
with discussion of both theoretical aspects
and   experimental results in atomic and molecular physics.
The remaining sections  \ref{Sec:QInfo}-\ref{sec-classical} emphasize mainly developments from the years after  the review of \mytcite{Vit01a}.
Section \ref{Sec:QInfo} discusses application of STIRAP to quantum information processing, Sec.~\ref{sec-solid} looks at STIRAP processes in solid-state environments,
and Sec.~\ref{sec-classical}  discusses STIRAP-inspired processes  in classical systems.
Plans for very promising new applications of STIRAP are discussed in Sec.~\ref{sec-perspectives}.
Finally, an Appendix  shows a list of STIRAP-related acronyms.

\section{STIRAP basics \label{sec-basics}}

\subsection{Three-state linkages \label{sec-basic-eqns}}

In its simplest version \mycite{Gau90}  STIRAP allows, in principle,  the complete transfer of population  along a three-state chain
$1 - 2 - 3$, from an initially-populated quantum state 1 to a target quantum state 3,
induced by two coherent-radiation fields that couple the intermediate state 2 to states 1 and 3, labeled the \pump (pump) or \Stokes (Stokes) lasers, respectively.
Figure \ref{fig-links2dN-r} shows the linkage pattern of radiative interactions that are relevant to STIRAP.

In coherent  atomic excitation, the internal dynamics of the atom is described by the time-dependent  Schr\"odinger equation,
 \be \label{SchrodingerEq}
\ii \hbar \ddt \Psi\t = \H\t \Psi\t,
\ee
where   $\H\t$ is the Hamiltonian   matrix for the system and its interaction with  the pulsed  fields.
  For three discrete states the statevector $\Psi\t$ is a three-component column-vector with probability amplitudes $C_n\t$ as elements,
$\Psi\t \equiv \C\t = [C_1\t, C_2\t, C_3\t]^T$.
Treatment of the dynamics of STIRAP is traditionally done within the rotating-wave approximation (RWA) \mycite{Rab54,Sho90}, for which the RWA Hamiltonian matrix is typically written as
 \footnote{ With the RWA comes the use of rotating coordinate vectors   $\psi_n$ in the underlying Hilbert space \mycite{Sho13}.}
\be
\H\t = \hbar \left[\begin{array}{ccc}
0  & \frac12\Omega_{P}\t & 0 \\
\frac12\Omega_{P}\t & \Delta & \frac12\Omega_{S}\t \\
0 & \frac12\Omega_{S}\t & \delta
\end{array}\right] .  \label{def-H}
\ee

For the traditional application of STIRAP to atomic and molecular excitation the Rabi frequencies $\Op\t$ and $\Os\t$  are
evaluated from the interaction energy $-{\bf d}\cdot \E\t$ proportional to the projections of  dipole-transition moments
${\bf d}_{nm}$ for the $n \fromto m$ transition onto the electric field at the center of mass, $\E\t$.
    In RWA the carrier frequencies $\omega_{P}$ and $\omega_{S}$ are factored from the
\pump and \Stokes electric fields $\E_{P}\t$ and $\E_{S}\t$ respectively, leaving slowly varying
amplitudes $\calE_{P}\t$ and $\calE_{S}\t$. The two Rabi frequencies are then evaluated as
\be
  \Omega_{P}\t = - {d}_{12}  \calE_{P}\t /\hbar ,\qquad
  \Omega_{S}\t = - {d}_{23} \calE_{S}\t /\hbar,
\ee
where  ${d}_{12}$ and ${d}_{23}$ are components of the dipole-transition moments along their respective electric-field vectors.

In both of the linkage patterns of Fig. \ref{fig-links2dN-r} the excited states can undergo spontaneous emission to lower-lying states.
Those emission processes that lead to levels outside the three-state system lead to an undesirable probability loss, usually described by adding an imaginary term to the appropriate diagonal element of the Hamiltonian.
Spontaneous emission processes back to states 1 or 3 are incoherent and thus are also undesirable.

\begin{figure}[tb]
\includegraphics[width=0.78\columnwidth]{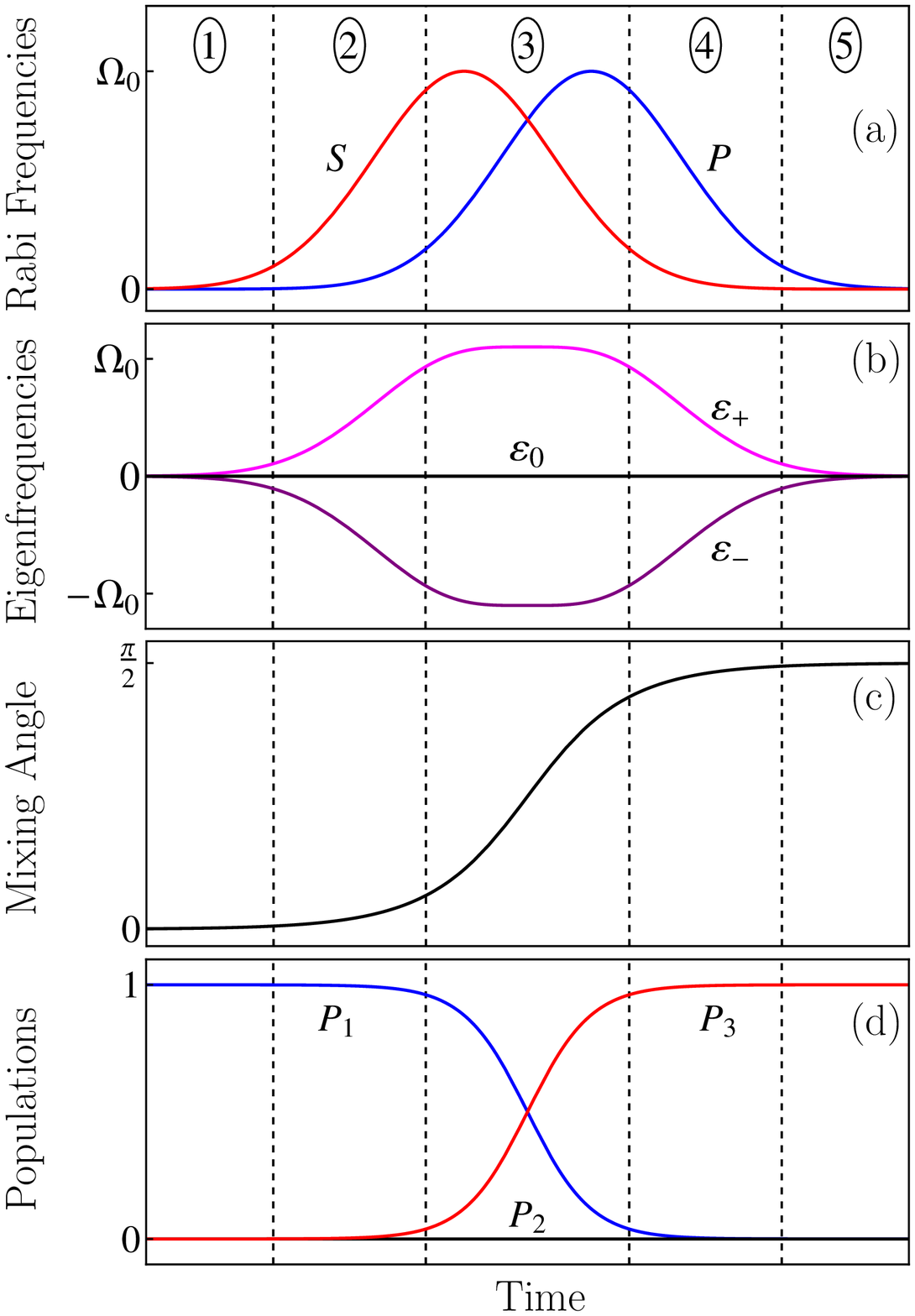}
\caption{\coloronline
Example of STIRAP induced by Gaussian pulses of equal peak value $\Omega_0$ and single-photon resonance, $\DP = \DS = 0$:
(a) time dependences of the \pump   and \Stokes   Rabi-frequencies $\Op\t$ and $\Os\t$;
(b) adiabatic eigenfrequencies $\eig_-\t$, $\eig_0\t$, $\eig_+\t$;
(c) mixing angle $\mix\t$;
(d) populations $P_n\t$ ($n=1,2,3$).
Dotted vertical lines separate the five phases of STIRAP discussed in  this section.
\locationa{Fig. 8 of \mytcite{Vit01a}.}
}
\label{fig-stirap-time}
\end{figure}

The individual state energies $E_n$ appear in these equations only indirectly, as constituents of detunings
 of carrier frequencies $\omega_{P}$ and $\omega_{S}$ from
transition frequencies,
\be
\hbar \Delta_{P}  =  E_2 - E_1 - \hbar \omega_{P }, \qquad \hbar \Delta_{S} = E_2 - E_3 -\hbar \omega_{S}.
\label{def-detune}
\ee
The two-photon detuning $\delta$ appearing in
the RWA Hamiltonian \eqref{def-H}
 is either the sum (for the ladder linkage) or the difference (for the  \L~linkage) of the laser detunings
 $\Delta_P$ and $\Delta_S$.
 STIRAP requires $\delta = 0$.
For the \L~linkage, which we take as the standard, this means that the two single-photon detunings are equal,
 $\Delta = \Delta_{P} = \Delta_{S}$.

\subsection{Eigenenergies and eigenstates}

On two-photon resonance ($\delta=0$), one of the eigenvalues of the Hamiltonian of Eq.~\eqref{def-H} vanishes, $\varepsilon_0 = 0$.
The corresponding eigenstate (or adiabatic state) of the Hamiltonian reads \mycite{Gau90}%
 \footnote{The addition of a constant value to the diagonals of the Hamiltonian, as was done  by \mytcite{Gau90}, will shift all the adiabatic energies accordingly \mycite{Sho13};   only with the convention used in Eq.~\eqref{def-H}, reckoning all excitation energies from state \s1,  does the dark state have zero as its eigenvalue.}
\be\label{dark state}
\Dkt =  \cos \mix\t \psi_1 - \sin\mix\t \psi_3,
\ee
where $\psi_k$ ($k=1,2,3$) are the wavefunctions of the unperturbed states of the \L-system, and the mixing angle $\mix\t$ is given by
\be
\tan \mix\t = \Op\t / \Os\t. \label{def-mix}
\ee
This is the so-called ``coherent population trapping'' (CPT)  state \mycite{Lam52,Gra78,Dal82a,Dal82b} or ``dark'' state \mycite{Ari76,Ari96,Gau90},
 which includes no component of state 2.
It is therefore immune against loss of population from the three-state system through spontaneous emission from state 2.
In order to have, prior to the transfer process, $\Dkt$ coincide with $\psi_1$ (the state that carries the population), $\mix = 0$ is needed, which demands $\Op(t)/\Os(t) \to 0$.
The transfer process is completed when $\Dkt$ coincides with $\psi_3$, requiring $\mix = \pi$, which demands $\Os(t)/\Op(t) \to 0$.
Therefore, the so-called counterintuitive ordering of  $\Op\t$ and $\Os\t$ is needed, with the system exposed to $\Os$, coupling the initially unpopulated states, prior to $\Os$.
A suitable overlap is, however, necessary to guarantee adiabatic evolution, see Sec.~\ref{sec-adiabatic}, i.e. smooth flow of the population from state 1  to  state 3, without putting transient population into state 2.

The other two adiabatic states are
\bse
\label{STIRAP-adiabatic-states}
\begin{align}
\Adt+ &= \psi_1\sin \mix\t\sin \varphi\t +\psi_2 \cos \varphi\t \nonumber\\
 &\quad +\psi_3\cos \mix\t \sin \varphi\t,  \label{Phi_plus }
  \\
\Adt- &= \psi_1\sin \mix\t \cos \varphi \t -\psi_2 \sin \varphi\t \nonumber\\
 &\quad  +\psi_3\cos \mix \t\cos \varphi\t,  \label{Phi_minus}
\end{align}
\ese
where the second mixing angle $\varphi \t$ is defined by
\bse
\begin{align}\label{STIRAP-angles}
\tan 2\varphi \t  &= \frac{\Orms \t}{\Delta }, \\
\Orms\t &= \sqrt{\Op\t^2 + \Os\t^ 2}.
\end{align}
\ese
The adiabatic energies corresponding to these eigenstates --
 the eigenvalues of the Hamiltonian \eqref{def-H} -- are $\hbar\varepsilon_+\t$ and $\hbar\varepsilon_-\t$, where (for $\delta = 0$)
 \be\label{STIRAP-eigenvalues}
\varepsilon_{\pm}\t = \tfrac12 \left[ \Delta \pm \sqrt{\Delta^{2}+\Orms\t^{2}}\right] .
\ee
The three eigenvalues $\varepsilon_0(t)$ and $\varepsilon_\pm(t)$ are shown,  for $\Delta=0$, in
 Fig. ~\ref{fig-stirap-time}(b).
 When $\Orms = \Op = \Os = 0$,   i.e. at very early and very late times, the three eigenvalues are degenerate (and zero).
 When either of the Rabi frequencies is nonzero, the degeneracy of  the eigenvalues $\eig_{\pm}\t$ is lifted [Autler-Townes splitting, see \mytcite{Aut55,Coh96}] but  the eigenvalue $\eig_0\t$ stays zero.
At very early time $\eig_0\t$ is related to state 1, at very late times it is related to state 3.

Having a constant eigenvalue for the Hamiltonian \eqref{def-H}
  requires maintaining the two-photon resonance  condition $\delta = 0$   throughout the transfer process.
Any deviation from this condition will inevitably populate state 2 \mycite{Few97}, with ensuing losses.

\subsection{STIRAP process step by step}

For STIRAP, a time interval is needed where initially $|\Os\t| > 0$ while $\Op\t =  0$ (or $|\Os\t| \gg |\Op\t| $), meaning $\mix = 0$,
and at the end $|\Op\t| > 0$ while $ \Os\t  = 0$ (or $|\Os\t| \ll |\Op\t|$), meaning $|\mix|  = \pi/2$.
At some intermediate time,
 the two Rabi frequencies will have equal magnitudes, $|\Op| = |\Os|$.
 Moreover, the variation of the Rabi frequencies must be smooth, to assure adiabatic evolution (Sec. \ref{sec-adiabatic}).
 The single-photon detuning  $\Delta = \Delta_{P} = \Delta_{S}$
    remains constant during the interaction.
In most cases STIRAP works best for $\Delta = 0$.

A main benefit of STIRAP, and its most surprising feature, is the elimination of spontaneous decay from state \s2 during the transfer process, despite the fact that the laser fields are tuned to resonance (or near resonance) with the transitions to state \s2,
and that the duration of the radiative interaction may well exceed the radiative lifetime by several orders of magnitude.
The lossless transfer occurs because, by design, the statevector $\Psi\t$ is aligned at all times with the dark state $\Dkt$.

Figure \ref{fig-stirap-time} shows characteristics of a representative STIRAP process.
The time dependence of the Rabi frequencies can be imposed either by suitably delayed laser pulses interacting with particles that do not change their position significantly during the pulse duration, or for particles in a beam, by spatially suitably displaced cw fields of the \pump and \Stokes lasers.
There is a smooth population transfer from state \s1 to state \s3, with negligible population in state \s2 at any time.
This figure is the basis of our introductory discussion of the basic features of STIRAP.

The mechanism of STIRAP can be understood \mycite{Vit01a,Sho11} by dividing the interaction into five stages,   deliniated by dashed vertical lines in
    Fig.   \ref{fig-stirap-time}    and distinguished by   the ratio   of the \pump  and \Stokes  fields.

\textbf{Stage 1:} \emph{\Stokesx-induced   Autler-Townes phase}.
Only the \Stokes   pulse is present 
  linking states 2 and 3,  causing Autler-Townes splitting   \mycite{Aut55} of   the related adiabatic energy levels \eqref{STIRAP-eigenvalues}.
 The population in state $\s1$   is unchanged.
 The state vector   coincides  with the dark eigenvector, $\Psi\t \equiv \Phi_0\t$,   and is equal to $\psi_1$.

\textbf{Stage 2:} \emph{\Stokesx-induced CPT phase.} 
The \Stokes   pulse is strong, while the \pump  pulse has just arrived and   is much weaker than the \Stokes pulse.
The state vector $\Psi\t \equiv \Phi_0\t$ deviates only slightly from the basis vector  $\psi_1$.
The \pump  field
does not induce transitions to state 2 because this process is suppressed by the same mechanism that leads to
  electromagnetically-induced transparency (EIT) (Sec.~\ref{sec:EIT}): \black 
   destructive interference causes    cancellation of the transition rate from the ground state to the two Autler-Townes states   produced from states \s2 and \s3 by the strong \Stokes field.

\textbf{Stage 3:} \emph{Adiabatic passage phase}.
Both fields are strong, with \Stokes decreasing and \pump  increasing.
Consequently, the mixing angle increases from $0$ toward $\pi/2$, and the state vector $\Psi\t \equiv \Phi_0\t$ departs from $\psi_1$ toward $-\psi _3$, while remaining in a linear combination of $\psi_1$ and $\psi _3$, thereby leaving state $\s2$ unpopulated.

\textbf{Stage 4:} \emph{\pumpx-induced CPT phase}.
The state vector $\Psi\t \equiv \Phi_0\t$ is almost aligned with $-\psi _3$.
The population is now almost completely deposited in state $\s3$.
The weak \Stokes  pulse does not induce transitions to state 2  because  the strong \pump field  couples states  1 and 2.
The related Autler-Townes splitting protects the population in state 3, just like the \emph{S} laser protects the population of state 1 in stage 2.


\textbf{Stage 5:} \emph{\pumpx-induced Autler-Townes phase}.
The \Stokes  pulse is gone and the \pumpx-induced Autler-Townes splitting   gradually reduces to zero.
The state vector $\Psi\t \equiv \Phi_0\t$ is equal to $-\psi _3$.
STIRAP is completed.

\subsection{Typical signatures of STIRAP \label{sec-signatures}}

\begin{figure}[tb]
\includegraphics[width=0.76\columnwidth]{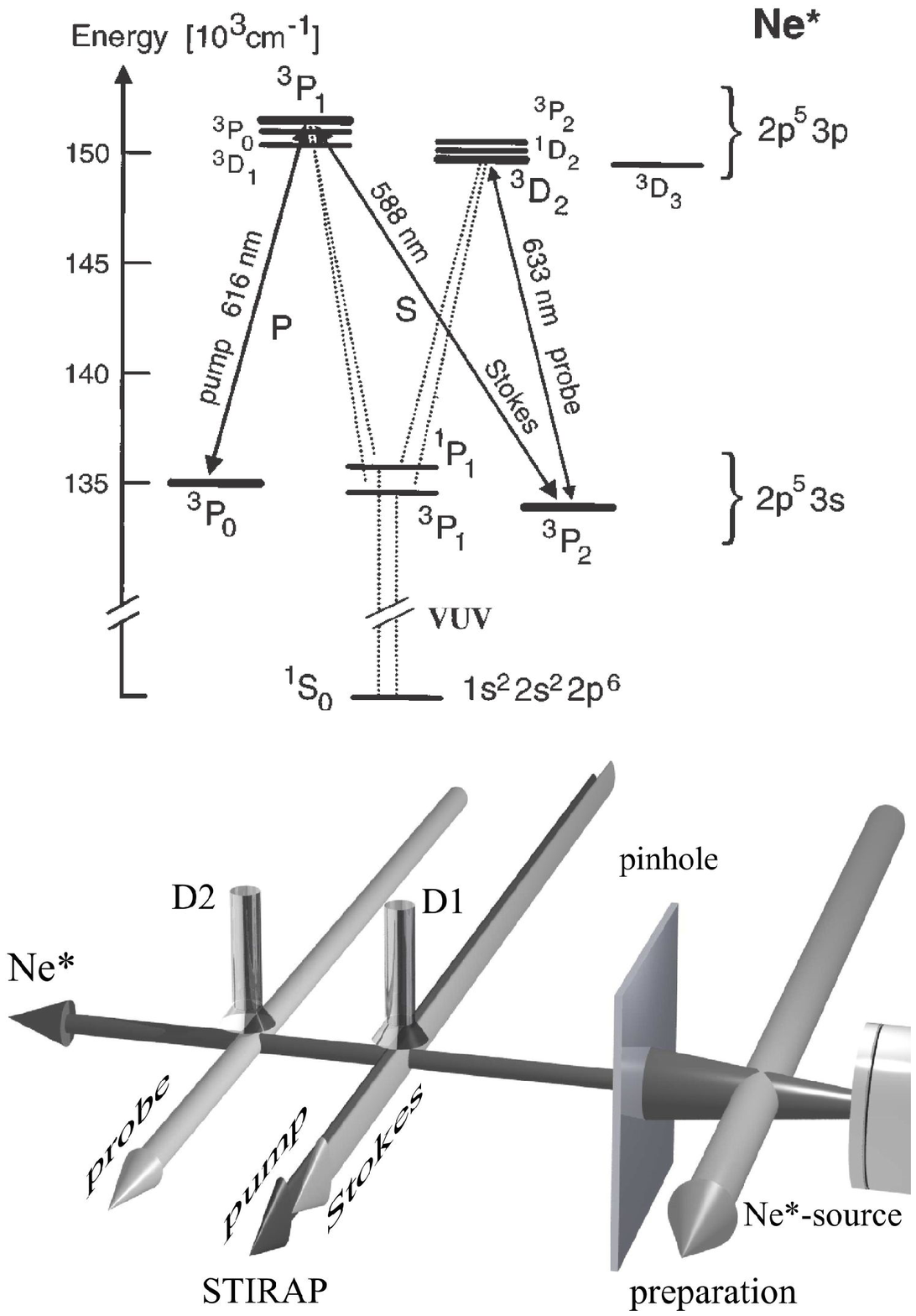}
\caption{
\emph{Top}:  Energy levels of  Ne$^*$ used in the experiment by
\mytcite{The98}.
\emph{Bottom}: Schematic of the experimental setup of crossed  molecular and laser beams,
showing the discharge source, a collimating aperture, the \pump and \Stokes  laser-beams, and channeltron photon-detectors $D1$ and $D2$.
\locationa{Figs. 6  and 7 of \mytcite{Ber98}.}
\Also{Figs. 1 and 4  of \mytcite{Mar96} and Fig. 13 of \mytcite{Vit01a}.}
}
\label{Fig: Ne levels}
\end{figure}

Regardless of the medium in which it is implemented, STIRAP has   characteristic signatures that distinguish it from other coherent population transfer techniques.
Here we present   two  examples of such signatures, as seen in experiments of a beam of metastable Ne$^*$ atoms crossing two spatially displaced but overlapped laser beams (\pump  and \Stokesx) at right angles
\mycite{The98}.
Figure \ref{Fig: Ne levels} shows the relevant energy levels   of Ne$^*$ (top) and a schematic view of the experimental setup (bottom).
STIRAP transfers  population  from   the initially populated state ${}^3P_0$ (state 1) to a Zeeman sublevel  of the target level ${}^3P_2$ (state 3) via a
sublevel of   ${}^3P_1$ (state 2).
The populations are measured by detecting light-induced fluorescence from states 2 and 3.


\begin{figure}[tb]
 \includegraphics[width=0.68\columnwidth]{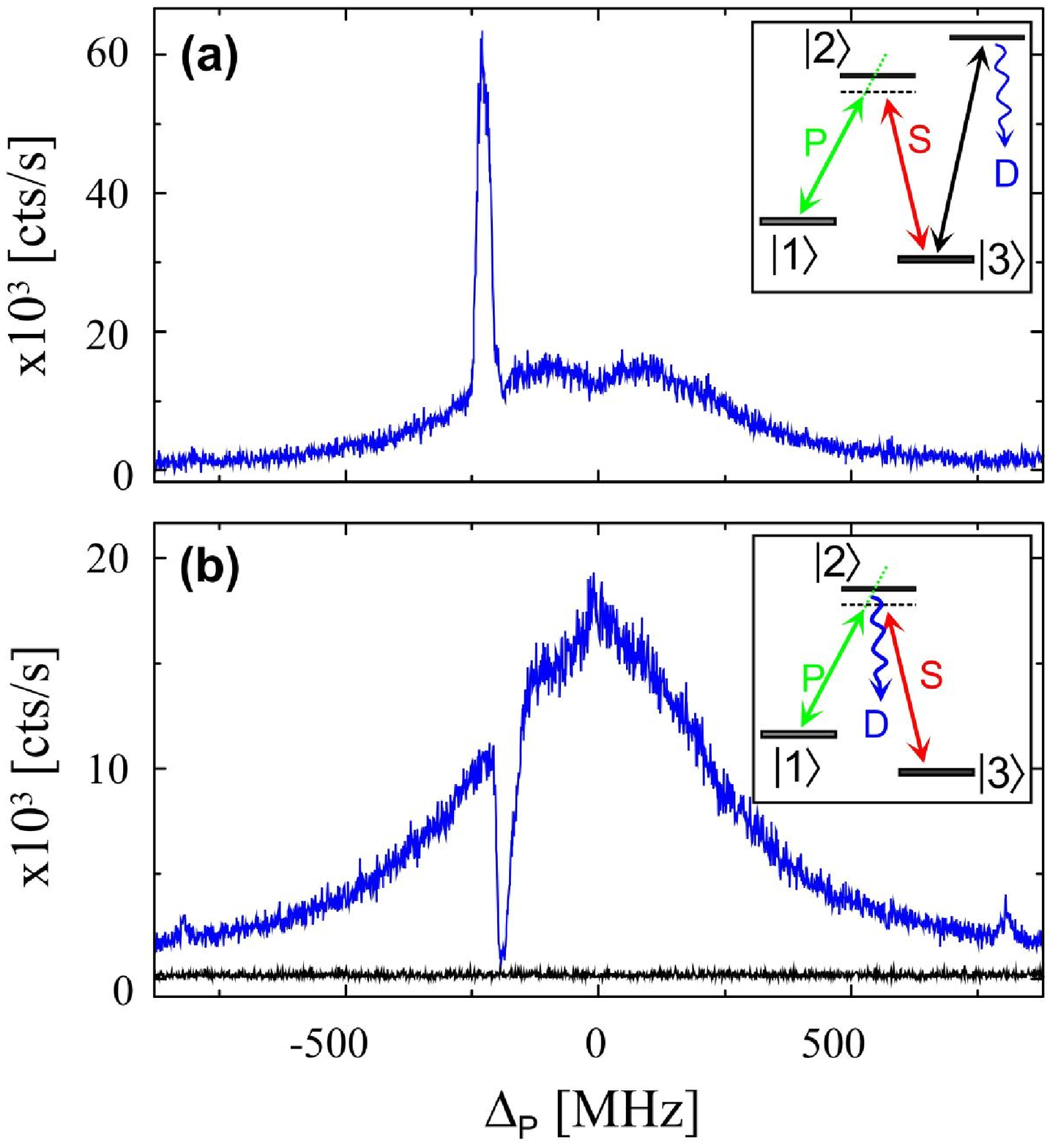}
\caption{\coloronline
Final populations   vs \pumpx-field detuning $\Delta_{P}$ in the Ne$^*$ experiment of \mytcite{Mar96}.
(a) Target state  population $P_3$.
(b) Intermediate excited-state  population $P_2$.
The peak in $P_3$,  frame $(a)$, and the dip in $P_2$, frame $(b)$,  are typical (and mandatory) signatures of STIRAP.
\locationa{Fig. 8 of \mytcite{Mar96}.}
\Also{Fig. 18 of \mytcite{Ber98} and Fig. 16 of \mytcite{Vit01a}.}
}
\label{Fig: Ne detuning}
\end{figure}

Figure \ref{Fig: Ne detuning} shows typical STIRAP signatures in the final populations $P_2$ and $P_3$ plotted vs the \pump detuning $\Delta_{P}$ for fixed \Stokes detuning $\Delta_{S}$ ($\approx 200$ MHz).
The broad feature centered around $\Delta_P= 0$ is descriptive of single-photon excitation of state \s2,  followed by spontaneous emission into state \s3.
Upon this background there is a narrow feature of each curve, a peak in $(a)$ and a dip in $(b)$, centered around {$\Delta_{P} = \Delta_S$}, the condition for two-photon resonance.
When this resonance condition is fulfilled there is a strong increase of population transfer to state \s3, and a consequent suppression of population $P_2$ in the intermediate state.
Both the peak in $P_3$ and the dip in $P_2$ are necessary signatures that must be present to validate a claim of STIRAP.


\begin{figure}[tb]
\includegraphics[width=0.80\columnwidth]{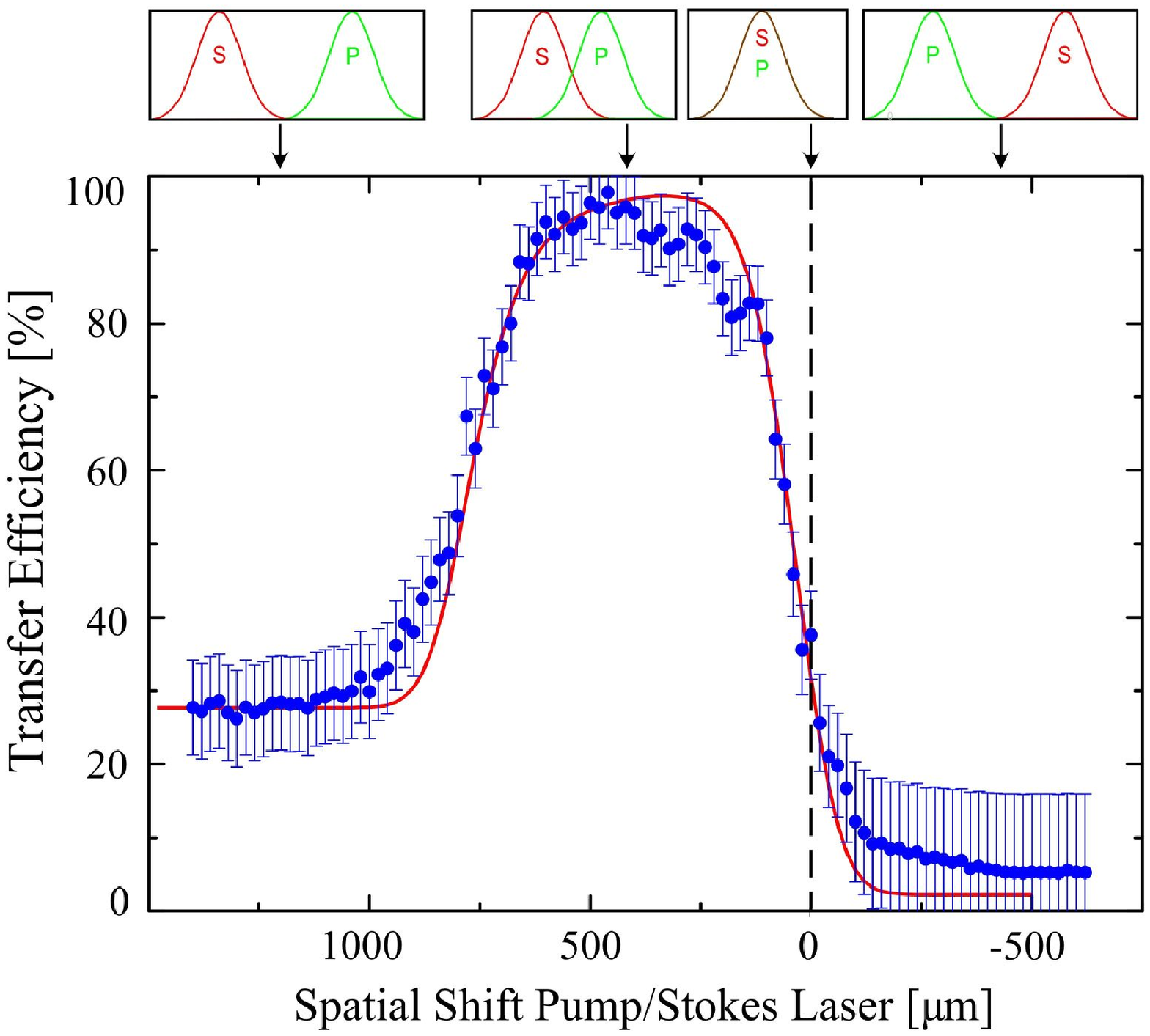}
\caption{\coloronline
Population of the target state 3 vs the pulse delay in   Ne$^*$ experiments. 
The large population efficiency {on the left side of the vertical line} is a signature of  STIRAP.
\locationa{Fig. 9 of \mytcite{Ber98}.}
\Also{Fig. 17 of  \mycite{Vit01a}  and Fig. 11 of \mytcite{Vit01b}.}
}
\label{Fig: Ne delay}
\end{figure}

Figure \ref{Fig: Ne delay} shows another typical signature of STIRAP: the population of   state \s3 plotted vs.~the pulse delay, here shown as spatial displacement of the \Stokes and \pump laser beams.
In this experiment, the pulse durations   were comparable to the lifetime of the middle state 2.
Therefore, if   state \s2 is populated during the process then its population decays to other   states and is irreversibly lost.
Data to the left of the vertical dashed line corresponds to the STIRAP pulse sequence.
High population transfer efficiency results for a range of spatial shifts.
Data to the right of the dashed line corresponds to the intuitive pulse sequence, when state \s2 is populated.
Radiative decays from this state take population outside the three-states system, and very little population reaches    state 3.


\subsection{Adiabatic evolution \label{sec-adiabatic}}

Maintaining the alignment of the statevector with the dark state (i.e. adiabatic following)
  is  {a} defining feature of the STIRAP process.
This alignment requires suitably slow (adiabatic) variation of the mixing angle.
 The conditions for adiabatic evolution have been described in detail  \mycite{Kuk89,Gau90,Kuh92,Ber98,Vit01a,Sho13,Ber15}.
Adiabatic evolution is required to prevent (nonadiabatic) coupling between the adiabatic states.
To this end,  the rate of change of the mixing angle must be small compared with   the difference of the adiabatic eigenvalues \mycite{Mes62}.
Hence the timing of the \pump and \Stokes pulses must be designed such that the splitting of the eigenvalues is maximal
 when the rate of change of the mixing angle is largest, Fig. \ref{fig-stirap-time}(c).
Below we list the conditions for adiabatic evolution with emphasis on three different aspects.

 \subsubsection{Local adiabatic conditions\label{sec-local adiabat}}

 The condition for adiabatic evolution  during STIRAP was derived by \mytcite{Kuk89} and reads
\be
\Orms\t \gg  | \dot{\mix}\t | =
\frac{|\Os\t \dot{\Omega}_{P}\t -\Op\t \dot{\Omega}_{S}\t|}{\Op\t^2 + \Os\t^2}.
\label{eq-AdiabCondition-local}
\ee
This condition quantifies the smoothness required for the pulses:
 the relationship must hold at any time during the transfer process (hence ``local'' condition).
When the adiabatic condition is fulfilled, the completeness of STIRAP is insensitive to small variations of the laser intensity, the duration and the delay of the pulses, {as well as to variations in the transition dipole moments}.

 \subsubsection{Global adiabatic conditions\label{sec-global adiabat}}

A very useful ``global'' condition is derived by  integrating  Eq.~\eqref{eq-AdiabCondition-local} over the interaction duration.
 The integral of the rms Rabi frequency is the rms pulse area,
\be
\calA = \int_{-\infty}^{\infty} \Orms\t  \, dt = \int_{-\infty}^{\infty} \sqrt{\Omega_{P}\t ^2+\Omega_{S}\t ^2}\, dt,
\label{def Arms}
\ee
and because the integral over  $\dot{\mix}\t$ produces the value $\pi/2$,
 the inequality \eqref{eq-AdiabCondition-local} reduces to
\be\label{AdiabCondition-global}
\calA \gg {\pi}/{2}.
\ee
Because the temporal pulse areas are proportional to the peak Rabi frequency $\Omax $ and the pulse duration $T$ (assuming, for simplicity,  these are the same for both pulses), $\calA\propto\Omax  T$,  condition \eqref{AdiabCondition-global} demands that the intensities and the pulse durations must be large enough. 
We can write condition \eqref{AdiabCondition-global} as
\be\label{AdiabCondition-Omega}
\Omax  T > \Amin,
\ee
where $\calA_{\min}$ is some minimum pulse area, dependent on the pulse shape and the required population transfer efficiency.
The majority of STIRAP users have been very happy with 95\% efficiency (but see Sec. \ref{Sec:QInfo}).
Usually, pulse areas of $\Amin \gtrsim 3\pi$ have sufficed  to provide efficient population transfer.

Obviously, the  global condition \eqref{AdiabCondition-global} is simpler to evaluate (and less restrictive)
  than the  local condition \eqref{eq-AdiabCondition-local}.
 For smooth pulses the global condition  \eqref{AdiabCondition-global} usually guarantees also the fulfilment of the local condition. 
The global adiabatic condition
is directly applicable to STIRAP with \emph{continuous-wave} (cw) lasers in the crossed-beam geometry
    because cw lasers have very good coherence properties.

For 
{pulsed radiation with transform-limited bandwidth}
 the adiabatic condition \eqref{AdiabCondition-Omega} is more conveniently written as
\be\label{AdiabConditionIntensity}
\Omax ^2 T >   \Amin^2/{T}.
\ee
The term on the left-hand side is proportional to the pulse energy.
 For a given pulse energy the required peak  {\em intensity} increases with decreasing pulse length $T$.
However, Eq. \eqref{AdiabConditionIntensity} shows that the required {\em energy} per pulse
 increases linearly with the inverse of the pulse duration:
 with $\Amin = 10$ we require $\Omax ^2 T \geq  100/T$.
For $T\lesssim 1$ ps the resulting laser intensity or pulse energy will most likely be sufficiently high to triger alternative {detrimental} couplings in atoms or molecules, such as multiphoton ionization.
Furthermore,  the RWA, essential  for the derivation of these equations, may no longer be valid.
The appendices of \mytcite{Ber95} and  of \mytcite{Ber15} offer  guidelines for how to estimate the required laser intensity for efficient population transfer between   rovibrational levels of a diatomic molecule.

\subsubsection{Consequences of phase fluctuations \label{sec-broadband}}

Ideally, the bandwidths of the radiation fields are transform-limited, i.e., the fields do not suffer from any phase fluctuations.
The relative phase between the two fields does not matter as long as it is constant during the transfer.
In a real experiment, phase fluctuations cannot be entirely eliminated.
A suitable measure of their extent is the ratio of the measured bandwidth
$\Delta\omega$
and the transform-limited bandwidth
$\Delta \omega_{\rm TL }$
which is determined not only by the pulse width but also by the pulse shape. 
A detailed analysis \mycite{Kuh92} of the consequences of $\Delta\omega > \Delta \omega_{\rm TL }$
  leads to the  more restrictive   adiabatic condition
\be\label{AdiabCriterionPulse}
\Omax  T > \Amin \sqrt{1+(\Delta \omega /\Delta \omega _{\rm TL})^{2}}.
\ee
This formula quantifies the expectation that  the detrimental consequences of a bandwidth in excess of the transform limit can be reduced by increasing the Rabi frequencies;
it offers  an estimate for the intensity increase needed to compensate for phase fluctuations.
However, phase fluctuations may be fast and thus the local condition
\eqref{eq-AdiabCondition-local}  may be violated even when the global condition \eqref{AdiabCriterionPulse} is satisfied.


%

\begin{figure}[tb]
\includegraphics[width=0.75\columnwidth]{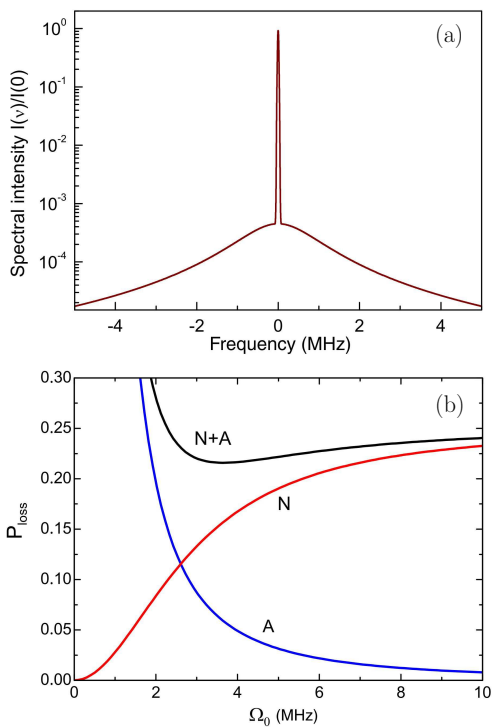}
\caption{\coloronline
$(a)$ Spectral line shape assumed in the analysis of \mytcite{Yat14}, typical of stabilized
diode lasers.
$(b)$ The STIRAP efficiency loss vs the maximum Rabi frequency
{accounting for broadband noise (curve $N$), nonadiabaticity (curve $A$), and both ($N+A$).}
\location{Figs 1 and 2 of \mytcite{Yat14}}.
}
\label{fig-Yat14-1-2}
\end{figure}

{A series of papers by}  \mytcite{Yat98a,Yat02b} and \mytcite{Rom05} {examined the  consequences  of   a stochastic component of the fields upon   STIRAP.}
\mytcite{Yat14} have shown that rapid phase fluctuations, recognizable as excess spectral density in  the wings of the spectral profile are detrimental.
%
{Because} such fluctuations are {usually} uncorrelated {for the \emph{P} and \emph{S} lasers} they result in detrimental deviation from two-photon resonance {and induce nonadiabatic coupling between the dark state and the two bright states}.
%
 Such fluctuations typically accompany the laser stabilization procedures that produce nearly monochromatic light on top of a Lorentz-profile pedestal of much broader bandwidth, which may carry only a few percent of the total power.
 \mytcite{Yat14} found that  the  effects of  these two noise components  differ  qualitatively from   those produced by the fluctuations that have hitherto been considered (for example, phase diffusion).
Figure \ref{fig-Yat14-1-2} shows the laser spectral density assumed in this work and the population transfer efficiency.
Their results {indicate} that  there is an optimum value for the peak Rabi frequency, and that the effect of fluctuations, although small, cannot be eliminated   by increasing the laser intensity.

{These observations underline the fact that efforts are needed to reduce the bandwidth of the radiation fields to very near the transform limit.}
{However, even when all ``technical'' frequency fluctuations   are eliminated there remains a nonreducible bandwidth -- the basic
Schawlow-Townes limit} \mycite{Sha58} {to the laser bandwidth, determined by spontaneous emission.}

\subsubsection{Consequences for degenerate levels  \label{sec:degenerate}}

\begin{figure}[tb]
\includegraphics[width=0.80\columnwidth]{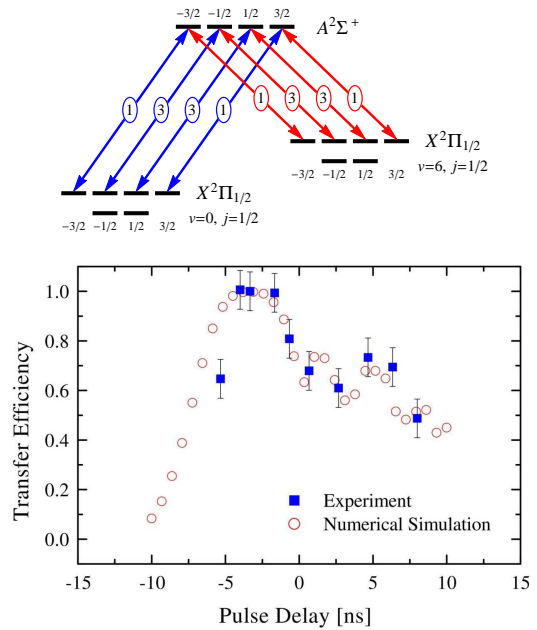}
\caption{
STIRAP in ${}^{14}$N${}^{16}$O molecules.
\emph{Top}: Hyperfine structure of the rovibrational states $X\ ^{2}\Pi _{1/2}\ (v=0,J=\frac12)$, $X\ ^{2}\Pi _{1/2}\ (v=6,J=\frac12)$ and $A\ ^2\Sigma \ (v=0,J=\frac12)$, and linkage patterns for linearly polarized pump and Stokes fields.
The numbers on the arrows indicate the relative coupling strengths of the transitions.
\locationa{Fig. 18 of \mytcite{Ber98}}.
\emph{Bottom}: Population transfer efficiency vs pulse delay.
\locationa{Fig. 2 of  \mytcite{Sch93}.}
}
\label{Fig:Schiemann1993}
\end{figure}

Because the adiabatic condition requires large pulse areas, STIRAP can be accomplished between degenerate states if adiabaticity is assured for the \emph{weakest} radiative transition allowed by the optical selection rules.
Figure \ref{Fig:Schiemann1993}  gives an example for population transfer by nanosecond laser pulses in the electronic ground state of ${}^{14}$N${}^{16}$O molecules \mycite{Sch93,Kuh98}, from the rovibrational state $X\ ^{2}\Pi _{1/2}\ (v=0,J=\frac12)$ to $X\ ^{2}\Pi _{1/2}\ (v=6,J=\frac12)$ via the intermediate state $A\ ^2\Sigma \ (v=0, J=\frac12)$.
Because ${}^{14}$N$^{16}$O has a nuclear spin of $I=1$ each of the three levels is split into two sublevels with $F=\frac12$ and $F=\frac32$.
Therefore, STIRAP operates here in a system of 16 magnetic sublevels. 
For linearly polarized light, and parallel pump and Stokes polarizations, four independent systems are identified:
 two three-level  \L~systems with $F=\frac32, M_{F}=\frac32$ or $M_{F}=-\frac32$, and two five-level systems involving the $|M_F|=\frac12$.
However, because the hyperfine splittings of the initial and final levels are large enough (214 MHz) to be resolved experimentally, the latter reduce to three-level systems which include, depending on the tuning of the \pump and \Stokes lasers, either one of the levels for the $F = \frac32$ or $F = \frac12$ in the initial or final state.
For STIRAP transfer to occur along all parallel paths it is necessary to satisfy the adiabatic condition on the weakest transitions, which is this example are $\ket{F=\frac32,M_F=\pm \frac12} \fromto \ket{F'=\frac32,M_F'=\pm \frac12}$, that are weaker than the other transitions with $M_F =\pm \frac32$ by a factor of 3 due to the corresponding Clebsch-Gordan coefficients.
Because there was sufficient laser power to satisfy the adiabatic condition for STRIAP, highly efficient population transfer has been achieved, despite the complexity of the system, see Fig.~\ref{Fig:Schiemann1993} (bottom).

Similarly, when the adiabatic condition is satisfied for particles in an atomic or molecular beam which cross the spatial wings of the \pump  and \Stokes laser beams, then efficient transfer will happen also for all those particles crossing the laser beams closer to their center.

\subsection{ Optimum pulse delay  \label{Sec:delay}}


The pulse delay $\tau$ between the \Stokes and \pump pulses affects the efficiency of STIRAP through (i) the adiabatic condition, and (ii) the completeness
of the projection of the state vector $\Psi(t)$ onto the dark state $\Dkt$ at the initial and final times $t_{\i}$ and  $t_{\f}$.
The optimum delay is determined by the following arguments.

{\em Coincident pulses.}
In this case, {and for identical pulse shapes,} the mixing angle $\vartheta $ is constant; then the nonadiabatic coupling vanishes ($\dot{\vartheta}=0$) and the evolution is perfectly adiabatic.
However, the state vector $\Psi (t)$ is not initially aligned with the dark state $\Dkt$, but instead
$\Psi(t_{\i}) = [\Dk(t_{\i})+\Phi_-(t_{\i})]/{\sqrt{2}}$, and a similar relation applies at the end time $t_\f$.
The interference between different evolution paths, $\Dkt$ and $\Phi_-(t)$,  from state 1 to state 3 leads to oscillations in the final population $P_3$ of state 3 \mycite{Vit97}, instead of complete population transfer.

{\em Small delay, very large overlap.}
For small delay, the overlap is large and the mixing angle $\vartheta(t)$ is nearly constant during most of the overlap, $\vartheta(t)\approx \vartheta_0$; hence $\dot{\vartheta}(t)\approx 0$ and adiabaticity is good there.
However, due to the small delay, $\vartheta(t) $ rises too quickly from $0$ to about $\vartheta_0$ before the overlap, and then again from about $\vartheta_0$ to $\pi/2$ after the overlap.
These rapid rises generate large nonadiabatic couplings $\dot{\vartheta}(t)$ at early and late times, which cause nonadiabatic transitions from the dark state to the other two adiabatic states.
These two nonadiabatic zones lead again to interference and oscillations in $P_3$.

{\em Large delay, very small overlap.}
The initial state vector is  $\Psi(t_{\i}) = \Phi_0(t_{\i})$.
Because for most of the time only one pulse is present, the mixing angle $\vartheta(t)$ stays nearly constant for most of the excitation:
 $\vartheta(t) \approx 0$ early and $\vartheta(t) \approx \pi /2$ late.
However, $\vartheta(t)$  rises from $0$ to $\pi /2$ during the very short period when the pulses overlap, thereby generating a large nonadiabatic coupling $\dot{\vartheta}(t)$, which ruins the population transfer.

{\em Optimum delay.}
For maximal adiabaticity, the mixing angle $\vartheta (t)$ must change slowly and smoothly in time, so that the nonadiabatic coupling $\dot{\vartheta}(t)$ remains small. 
The optimal value of $\tau$ depends on the pulse shapes: for Gaussian pulses, the optimum delay is slightly larger than the pulse width, $\tau _{\rm opt}\gtrsim T$ \mycite{Vit97}.
In any case,  the Autler-Townes splitting of the eigenvalues should be maximal when $\dot\vartheta(t)$ is largest.

\subsection{Dependence on velocity \label{sec-velocity}}

The Doppler shift may contribute to the detuning $\Delta$  from single-photon resonance, 
 which  has little effect on STIRAP,  but the detuning $\delta$ from two-photon resonance is critical.
What matters for the Doppler shift is the component $v_k$ of the particle velocity ${\bf v}$   along the laser propagation vector  ${\bf k}$ (where $|{\bf k}| \equiv k = 2\pi/\lambda$).
The effective detuning, i.e. the sum of static  laser detuning $\Delta$ and Doppler shift $k v_k$,  is $ \Deff = \Delta + k v_k$.
For the  \L~linkage the velocity-dependent detuning from the two-photon resonance reads $\deff = \Delta_{{\rm eff},P} -\Delta_{{\rm eff},S}$, hence
\be\label{kb-1}
 |\deff | =  | \Delta_P - \Delta_S +  (k_P - k_S) v_k|.
 \ee
Therefore for co-propagating laser beams ($k_P k_S > 0$) and null two-photon detuning ($ \Delta_P = \Delta_S$),
\be\label{kb-2}
|\deff| = |k_P - k_S| v_k.
\ee
Thus for $k_P = k_S$ (i.e., $\lambda_P = \lambda_S$), $\deff$   does not depend on velocity.
Then STIRAP transfers population for the entire ensemble of particles, independent of their velocity.
When $k_P \neq k_S$, the fraction of the velocity distribution that is addressed by STIRAP depends on the two-photon linewidth (Sec.~\ref{Sec:two-photon-detuning}), which increases with increasing laser intensity. Thus, even for $k_P \neq k_S$ (provided $|k_P - k_S| \ll k_P$) the entire velocity distribution can  be addressed if the laser intensity is sufficiently high.

When the  laser beams are counterpropagating ($k_P k_S < 0$) we have, for $\Delta_P = \Delta_S$, the relationship
\be\label{kb-3}
|\deff| = (|k_P| + |k_S|) v_k,
\ee
and an ensuing enhanced sensitivity of $\deff$ to velocity.
In this case, according to Eq. \eqref{kb-1}, it depends on $|\Delta_P - \Delta_S|$ which velocity group will experience the resonance condition $|\deff| = 0$.
Thus, by appropriate choice of  $\Delta_P$ and $\Delta_S$ the experimenter can restrict the STIRAP transfer to particles within a small range of a given velocity component  $v_k$ \mycite{Rai14}, see also Sec.~\ref{sec-bright beams}.
For STIRAP transfer with a ladder linkage, the roles are interchanged:  the velocity dependence of $\deff$  is reduced for counterpropagating \pump and \Stokes  beams.

\subsection{Limitations to the success  of  STIRAP \label{sec-limitations}}

It is already implied in the previous discussion of the adiabatic conditions that STIRAP will not work when
(i) the actual bandwidths of the radiation fields exceed the transform-limited bandwidths by much, because the excess bandwidth signals the presence of detrimental phase fluctuations, or
(ii) the pulse duration is too short (e.g. in the femtosecond regime) because then the pulse energy required to allow adiabatic evolution will be so high that the dynamics may be dominated   by competing processes, such as multiphoton excitation or ionization.
\bws

In the simple linkage patterns discussed hitherto, only three quantum states have direct involvement in the STIRAP process.
However, atoms, molecules and other quantum systems have many other discrete states as well as states from photoionization and photodissociation energy continua.
If the three-state idealization needed for STIRAP is to be satisfactory, none of these states can be linked by near-resonant transitions to any of the STIRAP states
  \mycite{Una00b}.
  However, these additional states are not without influence: 
  they are responsible for induced dipole moments, proportional to field intensities. 
  Specifically, they alter the diagonal elements of the Hamiltonian, producing intensity-dependent (dynamic) Stark shifts. 
Because the \pump and \Stokes fields vary differently with time the related Stark shifts may cause a detrimental time-varying deviation from two-photon resonance \mycite{Ric00,Ran05,Yat02c,Yat99,Nak94}.
 We now mention a few consequences of the dynamic Stark shift (see also Sec. \ref{sec-STIHRAP}).

(i) When the  \pump  and \Stokes fields are tuned to resonance ($\Delta = 0$) or near resonance
($|\Delta | \ll |\Omax |$)  and the global adiabatic condition is fulfilled we usually have
$|\Omax | \gg |\Delta_{\rm Stark}|$.
The detrimental consequences of Stark shifts are negligible as long as $|\Omax | < |E_n - E_m|$.

(ii) When $|E_n - E_m| \leq |\Omax|$  the dynamics can no longer be treated as that of a three-state system.
All nearby levels need to be included in the evaluation of adiabatic eigenstates, with the consequence that avoided crossings of adiabatic-state energies may occur and the adiabatic passage path from state \s1 to state \s3 may be blocked
 \mycite{Mar95}.
 {This is}   particularly relevant  for molecules with high density of energy levels.
 Model calculations that aim  to test the suitability of STIRAP for population transfer will not yield reliable results  unless they include  all states that may radiatively couple to the initial and final states (including one-photon or multiphoton coupling paths as well as off-resonance interactions); see Sec. \ref{sec-Jakubetz}.

(iii) The off-resonant Stark shifts are severely detrimental when one or both of the \pump and \Stokes couplings 
 occurs via a multiphoton process.
 \bws
It is tempting to consider reaching levels in the first electronic states of a molecule such as H$_2$ by two-photon excitation (because radiation sources with suitable coherence properties are not yet available in the VUV region) for populating, e.g., vibrational level $v \gg 1$ of the electronic ground state.
However, as explained in Sec. \ref{sec-STIHRAP}, for such coupling the two-photon Rabi frequency is, like the Stark shift, proportional to the laser intensity and thus the Stark shift and the Rabi frequency are of the same order of magnitude.
Then STIRAP 
 is very likely to fail
\mycite{Gue98a,Yat98}.


\subsection{Comparison with electromagnetically induced transparency (EIT) \label{sec:EIT}}

A  phenomenon  that  was  independently  discovered  and  developed  at  the  same  time  as  STIRAP  is electromagnetically  induced  transparency  (EIT) \mycite{Har90,Bol91,Har97,Fle05}.
The  physics  of  EIT  and  STIRAP  share  some  common  features  and  exhibit distinct differences.
Both schemes exploit the consequences of interference of optically driven probability amplitudes for transitions between states of a quantum system but they address different areas of optical science.
While EIT is mainly seen as a phenomenon associated with the propagation of radiation fields in high-density  media,  STIRAP  is mainly applied in a low-density environment with the aim to precisely control or modify the population distribution over the quantum states.

In EIT, we consider the three-state system of Fig.~\ref{fig-links2dN-r}, as in STIRAP. 
The strong laser field \emph{S} creates a coherent superposition of states 2 and 3 observable  as an Autler-Townes  doublet  in  a  spectroscopic  measurement.
The  transition amplitude of the transition driven by the (much weaker) \emph{P} laser is the sum of the two transition amplitudes to the Autler-Townes components.
Because the latter are 180$^\circ$ out of phase, and (when the frequencies are tuned to resonance  with  the  respective  bare  state  transition)  have  equal  amplitudes,  the  transition  amplitude of the \emph{P} transition vanishes.

EIT allows the propagation of a radiation field through optically thick media.
When alone, the radiation of  the  \pump  field  
  is  strongly absorbed.
However, when  the \Stokes  field is present 
 the quantum coherence induced by that field renders the otherwise optically thick medium transparent for the \pump laser.
In  a  typical  EIT  experiment  both  fields  are simultaneously  applied and  drive  the  quantum  system.
At  the  same  time that system  acts  back  on  the fields  so  that,  after  propagating  some  distance,  the rapidly  varying  components  of the envelopes of the fields are modified so as to match each other \mycite{Har93}.

While in EIT it is the \Stokes laser that leads to the cancellation of the transition amplitude for the \pump laser, this is true in STIRAP only during the initial stage of the transfer process.
In the final stage of the process, the role of the \Stokes and \pump lasers are interchanged.
In both cases the dark state of Eq.~\eqref{dark state}  plays an important role.
In EIT the ratio of the Rabi frequency $\Omega_P/\Omega_S$  is constant or its variation is small.
Therefore a (nearly) stationary dark state is created.
In STIRAP it is essential that the dark state evolves in time because the ratio $\Omega_P/\Omega_S$  changes during the process from zero to infinity.
In both cases the process is robust against small variations of field intensities.
The robustness relies in part on the observation that the phases of the cooperating laser fields do not matter for EIT or STIRAP to be successful as long as they are constant during the particle-field interaction period.

Stage 2  of  STIRAP  (see  Fig.~\ref{fig-stirap-time}  and  part  C  of  this  section)  resembles  EIT  because  due  to  the presence  of  an  already  strong  \Stokes  laser  the  photons  from  the  \pump  laser  do  not  induce  transitions  to  the intermediate  state  2.
In  stage 4 of  STIRAP,  the   \pump  laser  takes  the  role  of  the \Stokes  laser.
The  EIT  and STIRAP cooperate for ``stopping'' light, i.e. for transferring the properties of a light pulse to a medium for storage and read out.
EIT and STIRAP  also underlie \bk~the physics of ``slow light''
\mycite{Hau99,Fle00a,Vit01a,Fle05,Zim08}, in which the strong field alters the refractive index and hence the group velocity of the weak field.

\section{Further aspects of three-state STIRAP \label{sec-three-state}}

This section scrutinizes the basic properties, requirements and restrictions for STIRAP.

\subsection{One- and two-photon linewidths \label{sec-linewidths}}

\begin{figure}[tb]
\includegraphics[width=0.85\columnwidth]{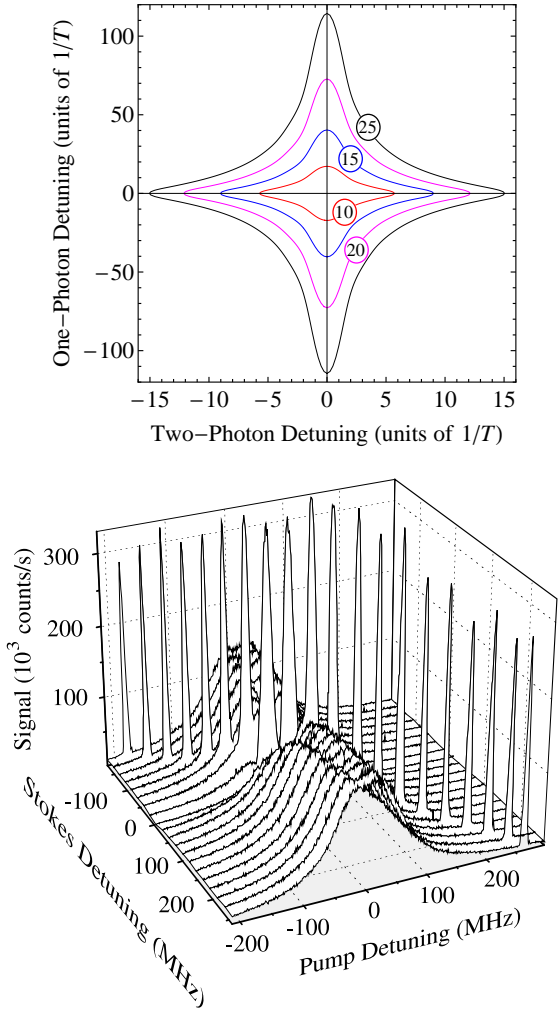}
\caption{\coloronline
\emph{Top}: Numerical simulation of $P_3=0.5$ contour lines for $\Omax  = 10/T,15/T,20/T,25/T$ for Gaussian pulses with delay $\tau=1.2T$.
\emph{Bottom}: STIRAP efficiency vs the \pump  and \Stokes  detunings $\Delta_{P}$ and $\Delta_{S}$ in an experiment with Ne$^\ast$.
\locationa{Fig. 9 of \mytcite{Mar96}.}
}
\label{Fig:DD-num-exp}
\end{figure}

A characterizing feature of STIRAP is the variation of the one- and two-photon linewidths with the detunings $\Delta_{P}$ and $\Delta_{S}$. 
Variation of either  carrier frequency, while keeping the other fixed, will change the two-photon detuning $\delta$, thereby producing the \emph{two-photon profile} $P_3(\delta )$.
Variation of both the \pump  and \Stokes  frequencies, while maintaining the two-photon resonance condition, will produce the \emph{single-photon profile} $P_3(\Delta)$.
The dependences of the transfer efficiency on $\delta$ and $\Delta$ are very different.
STIRAP is very sensitive to the two-photon detuning $\delta$ (cf. Fig.~\ref{Fig:DD-num-exp})
 because the formation of the dark state $\Dkt$ requires two-photon resonance.
On the other hand, {the formation of} the dark state  $\Dkt$ {is not prevented by} the single-photon detuning $\Delta$
 and therefore STIRAP does not depend on $\Delta$ in the adiabatic limit; hence the much broader profiles versus $\Delta$ in Fig.~\ref{Fig:DD-num-exp}.

Figure \ref{Fig:DD-num-exp}(bottom) shows data from experiments on Ne$^*$ atoms by \mytcite{Mar96}.
One sees here broad resonance structures as a function of the single-photon detuning $\Delta_{P}$
on which are superposed  narrow resonances when $\DS=\DP$, i.e. when $\delta = 0$.

\subsubsection{Linewidth for one-photon detuning \label{Sec: single-photon-detuning}}

Careful examination of adiabatic conditions reveals the scaling law of the FWHM $\Delta_{1/2}$ of the single-photon line profile $P_3(\Delta)$:
  $\Delta_{1/2} \propto \Omax ^2$ \mycite{Vit97detune}.
This quadratic dependence is indeed observed in Fig.~\ref{Fig:DD-num-exp}(top).
Because   the peak Rabi frequency  $\Omax $ is proportional to the electric-field amplitude, $\Delta_{1/2}$ is proportional to the peak laser intensity.

\begin{figure}[tb]
\includegraphics[width=0.65\columnwidth]{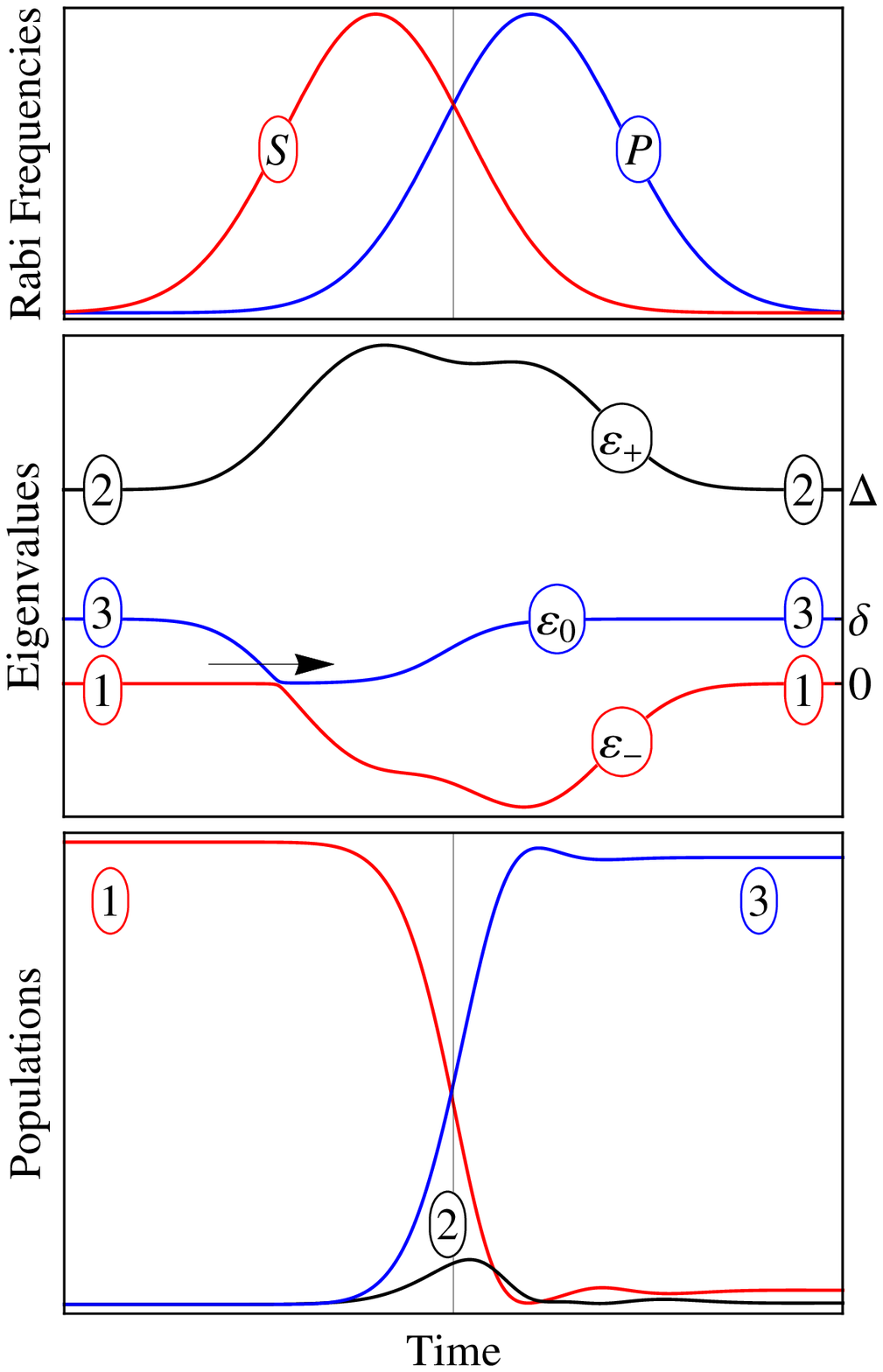}
\caption{\coloronline
Time evolution of the eigenvalues {and the populations} in a three-state system with nonzero two-photon detuning $\delta$.
The path from state $\s1$ to state $\s3$ passes through a diabatic crossing.
\locationa{Fig. 11b of \mytcite{Vit01a}.}
}
\label{Fig:two-photon-detuning}
\end{figure}

\subsubsection{Linewidth for two-photon detuning \label{Sec:two-photon-detuning}}

The effect of nonzero two-photon detuning $\delta$ was studied by
\mytcite{Dan94}
in the absence of population decay, by
\mytcite{Rom97}
in the presence of strong population loss from state \s2, and by
\mytcite{Gri01}
for large single-photon detuning.
For $\delta\neq 0$, the eigenstates and the eigenvalues of the Hamiltonian
 are no longer given by Eqs.~\eqref{dark state}, \eqref{STIRAP-adiabatic-states} and \eqref{STIRAP-eigenvalues}, and there are no null eigenvalue and no dark eigenstate \mycite{Few97}.
\mytcite{Dan94}   found that for $\delta\neq 0$ adiabatic evolution leads to complete population return to the initial state   $\s1$, 
 and the only mechanism by which population can reach state $\s3$ is in a mixed diabatic-adiabatic manner:
  by a nonadiabatic transition between the adiabatic states through an avoided level crossing, which emerges  for small $\delta$.
  Figure~\ref{Fig:two-photon-detuning}   illustrates such a narrow avoided crossing. 
For large $\delta$, the   separation of adiabatic eigenvalues becomes larger, thereby blocking   the diabatic-adiabatic path $\s1\to\s3$.
\mytcite{Dan94} derived an analytic expression for the two-photon linewidth by using the Landau-Zener-St\"uckelberg-Majorana formula \mycite{\LZSM}  to evaluate the nonadiabatic transitions at the crossing.

An alternative approach to estimating the two-photon linewidth makes use of the adiabatic condition
\mycite{Vit01b}
and treats terms, which emerge in the resonant adiabatic basis [Eqs.~\eqref{dark state} and \eqref{STIRAP-adiabatic-states}]
 due to $\delta\neq 0$, as perturbation.
These terms induce additional couplings 
 between the adiabatic states that cause nonadiabatic transitions.
Considerable population transfer $\s1 \to \s3$ can occur if these nonadiabatic couplings are suppressed,
 which leads to the condition $\delta_{1/2} \propto \Omax$ \mycite{Vit01a}.
Hence the two-photon linewidth is proportional to the square root of the peak intensity.

In conclusion, STIRAP is much less sensitive to the one-photon detuning
  than to the two-photon detuning. 
These features are seen in Fig.~\ref{Fig:DD-num-exp}(top) where, as the peak Rabi frequency $\Omax$ increases,
  the high-efficiency region increases linearly versus $\delta$ and quadratically versus $\Delta$.
{Although we only show the $P_3 = 0.5$ contour lines, similar scaling laws are observed for any other value, e.g. $P_3 = 0.9$.}
%

\subsubsection{Asymmetric line shapes \label{Sec: asymmetry}}

The two-photon resonance between states \s1 and \s3 is usually assumed to be a mandatory condition for STIRAP.
This is certainly correct when the peak Rabi frequencies are nearly equal.
However, it has been shown theoretically \mycite{Mol07,Bor10b} and experimentally \mycite{Sor06,Dup15}
that when the \pump  and \Stokes  couplings differ significantly, the population transfer profile becomes asymmetric with respect to the two-photon resonance,
 $\delta = 0$.

\begin{figure}[tb]
\includegraphics[width=0.80\columnwidth]{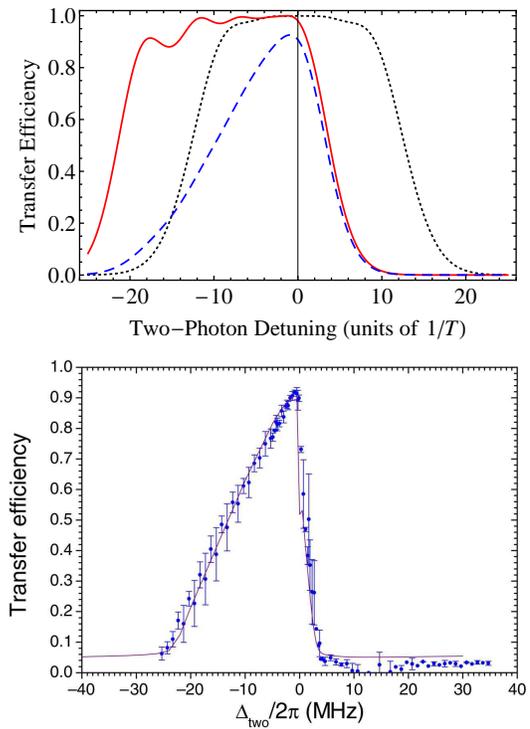}
\caption{\coloronline
\emph{Top}: Numerical simulation of STIRAP efficiency vs the two-photon detuning for equal (dotted) and unequal (by a factor of 3) \emph{P} and \emph{S} peak Rabi frequencies.
In the absence of population loss, the asymmetry just shifts the two-photon profile from resonance (solid).
The irreversible population from state 2 with rate $\Gamma = 5/T$ erodes the off-resonance part and produces a triangular-shaped profile (dashed).
Gaussian pulse shapes are assumed, of $1/e^2$-width $T$.
\emph{Bottom}:
STIRAP efficiency vs  two-photon detuning in the ${}^{40}$Ca$^+$ experiment by
\mytcite{Sor06}.
The magnitudes of the two fields differ by a factor of 2.5.
\location{Fig. 5 of \mytcite{Sor06}.}}
\label{Fig:asymmetric}
\end{figure}

Such situations often emerge in applications of STIRAP where the two interactions have different origins.
A prominent example is vacuum STIRAP (Sec.~\ref{Sec:vacuum-STIRAP}) wherein the \Stokes  laser is replaced by the vacuum field in a cavity.
Another example is when the two fields are derived from different radiation sources, as in many applications of STIRAP, such as a laser and a microwave generator \mycite{Dup15}.
Similar  conclusions hold for equal peak Rabi frequencies but different pulse widths. 
For moderate one-photon detuning ($|\Delta| < \Omax $, where $\Omax $ is the lesser of the two peak Rabi frequencies), the center $\delta_c$ and the width $\delta_w$ of the two-photon profile read \mycite{Bor10}
$\delta_c \approx \tfrac89 \Delta$ and $\delta_w \approx \tfrac43 \Omax$.
%
Figure \ref{Fig:asymmetric}(top) shows asymmetric excitation profiles shifted from two-photon resonance.
Because off-resonance population transfer takes place via an avoided crossing of adiabatic energies (and ensuing inevitable transient population of state \s2),
 in the presence of loss (dashed curve) the off-resonance part of the lossless excitation profile (solid curve) is eroded, thereby producing a triangular excitation profile.
Figure \ref{Fig:asymmetric}(bottom) shows an asymmetric STIRAP profile in an experiment with trapped calcium ions.

\subsection{Bright STIRAP \label{sec-ordering}}

The two opposite pulse sequences (\emph{SP} and \emph{PS}) lead to strikingly different results with regard to the single-photon detuning $\Delta_P$.
The counterintuitive sequence \emph{SP} induces complete population transfer to state $\s3$ via the dark state $\Dkt$ of Eq.~\eqref{dark state} regardless of $\Delta_P$.
On the contrary, the population redistribution caused by the intuitive sequence \emph{PS} strongly depends on $\Delta_P$.

\textbf{Single-photon resonance.}
On single-photon resonance ($\Delta=0$), we have $\varphi=\pi/4$ [see Eq.~\eqref{STIRAP-angles}], and hence both adiabatic states $\Phi_+(t)$ and $\Phi_-(t)$ are initially populated.
The interference between the two different paths from state $\s1$ to state $\s3$ -- via $\Phi_+(t)$ and $\Phi_-(t)$ -- leads to generalized Rabi oscillations in the final population of state $\s3$ \mycite{He90,Sho90,Sho92a,Vit97},
\be
P_{1} = 0,\quad
P_{2} = \sin^2\case12\area,\quad
P_{3} = \cos^2\case12\area,
\ee
with $\area$ given by Eq.~\eqref{def Arms}.

\textbf{Single-photon detuning: Bright STIRAP.}
For nonzero $\Delta$, but with the \pump and \Stokes fields still on two-photon resonance,
 adiabatic evolution may produce complete population transfer $\s1\to\s3$ for the intuitive pulse ordering \mycite{Vit97}.
Indeed, for $\Delta \neq 0$, we have $\varphi_{i} = \varphi_{f} = 0$, and hence the adiabatic state $\Phi_-(t)$ provides an adiabatic connection between states $\s1$ and $\s3$,
$\psi_1\stackrel{-\infty }{\longleftarrow } \Phi_-(t) \stackrel{+\infty }{\longrightarrow }\psi_3$.

This population transfer scenario is named bright STIRAP (b-STIRAP) \mycite{Kle07}.
However, here state 2 receives a significant transient population, $P_2(t) = \sin^2\varphi (t)$ [cf. Eq.~\eqref{Phi_minus}],
 and hence b-STIRAP can produce efficient population transfer only if the lifetime of state $\s2$ is long compared to the pulse duration.

\begin{figure}[tb]
\includegraphics[width=0.95\columnwidth]{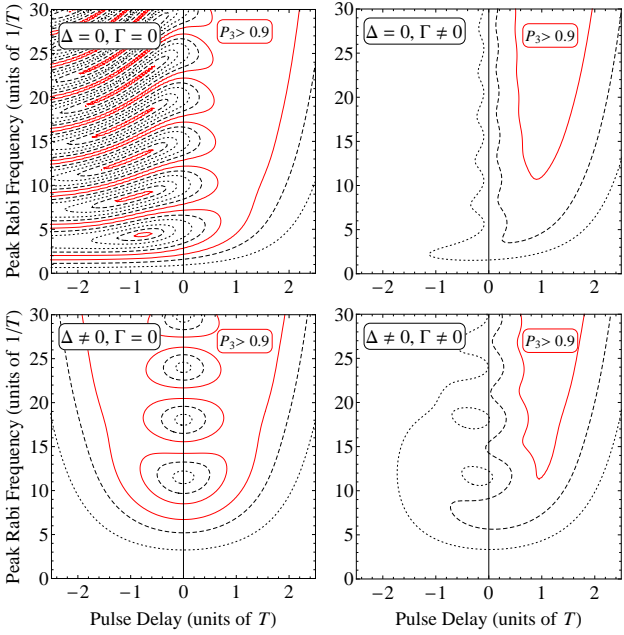}
\caption{\coloronline
Numerically simulated contour plots of population transfer in STIRAP vs the pulse delay $\tau$
(positive for STIRAP) and the peak Rabi frequency $\Omax $.
Solid, dashed and dotted contours show the $P_3=0.9$, 0.5 and 0.1 values, respectively.
We have assumed resonant Gaussian pulses of equal magnitude $\Omax$ and width $T$.
Left frames: no loss, $\Gamma=0$.
Right frames: population loss from state $\s2$ with rate $\Gamma=3/T$.
Top frames: single-photon resonance, $\Delta=0$.
Bottom frames: $\Delta=10/T$.
}
\label{Fig:tau-rabi}
\end{figure}

\def \Deff{\Delta_{\rm{eff}}}  \def\Oeff{\Omega_{\rm{eff}}}

\textbf{Large single-photon detuning.}
When the detuning $\Delta$ is very large ($| \Delta | \gg \Omega_{P},\Omega_{S}$),  then the middle state $\s2$ can be eliminated adiabatically
\mycite{Ore84,Gau90,Sho92b,Vit97detune,Sho11,Sho13}.
An effective two-state model results, with the effective coupling and detuning
\be
\Oeff\t = -\frac{\Omega_{P}\t \Omega_{S}\t}{2\Delta},\quad
\Deff\t = \frac{\Omega_{P}\t^2 -\Omega_{S}\t ^2}{2\Delta}.
\ee
Obviously, for temporally delayed pulses, regardless of the order, the detuning $\Delta_{\rm{eff}}(t)$ crosses resonance at the instant $t_0$ when $\Omega_P(t_0)=\Omega_S(t_0)$.
In the adiabatic limit, this level crossing leads to complete population transfer for both pulse orderings, because the sequence reversal leads to an unimportant change of sign in $\Deff\t$.

These observations are illustrated in Fig.~\ref{Fig:tau-rabi}, which shows the population of state $\s3$ versus the delay and the peak Rabi frequency of the two pulses.
On resonance, a large plateau of high population transfer probability occurs for counterintuitively-ordered pulses ($\tau>0$), while  oscillations
occur for the intuitive pulse ordering  ($\tau<0$) in the lossless case ($\Gamma=0$).
The oscillations disappear when irreversible population loss is present ($\Gamma > 0$).
Off single-photon resonance, robust population transfer occurs for both pulse orderings in the absence of losses.
However, with irreversible population loss only the island where STIRAP occurs has high probability of population transfer.
This is because   the dark state is not affected by the lossy middle state $\s2$, whereas b-STIRAP contains a sizeable component of this state.

{Curiously, in the lossless case the final population of state 1 is the same for either pulse orderings, i.e. it is a symmetric function of the pulse delay} \mycite{Vit99c}.

\subsection{Fractional STIRAP \label{sec-fractional}}

\begin{figure}[tb]
\includegraphics[width=0.60\columnwidth]{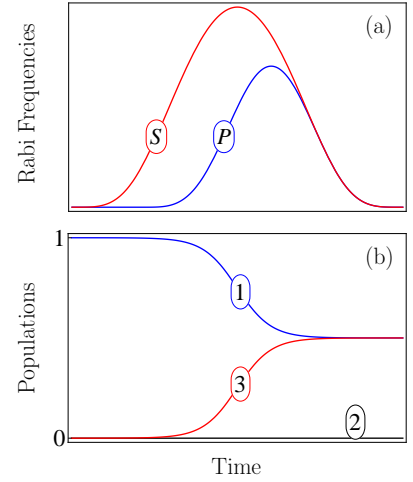}
\caption{\coloronline
Time dependences of the \pump  and \Stokes  Rabi frequencies and the populations in fractional STIRAP.
As in STIRAP, the \Stokes  pulse arrives before the \pump pulse, but here the two pulses vanish simultaneously while maintaining a fixed ratio.
Consequently, instead of complete population transfer $\s1 \to \s3$, a coherent superposition of states $\s1$ and $\s3$ is created.
\locationa{Fig.~2 of \mytcite{Vit99}}.
}
\label{Fig:f-STIRAP}
\end{figure}

As discussed above, the counterintuitive pulse ordering in STIRAP ensures the complete population transfer $\s1 \to \s3$.
It was recognized soon after the discovery of STIRAP that when the ratio of the two Rabi frequencies remains fixed, then so too does the mixing angle, and
 the statevector  will be frozen in a coherent superposition  of states $\s1$ and $\s3$
\mycite{Mar91}.
With  {complex}-valued Rabi frequencies  that satisfy the relations
 \be\label{fractional STIRAP}
0 \stackrel{-\infty\leftarrow t }{\longleftarrow } \frac{\Omega_{P}\t}{\Omega_{S}\t} \stackrel{t\rightarrow +\infty }{\longrightarrow } e^{\ii\alpha} \tan \Theta,
\ee
where $\Theta = \vartheta(+\infty)$,
the superposition reads
\be \label{f-STIRAP}
\Psi = \psi_1\cos \Theta - \psi_3 e^{\ii\alpha} \sin \Theta.
\ee
Thereby, instead of STIRAP, we have \emph{fractional} (or partial) STIRAP, in which only a \emph{controlled} fraction of the population is transferred to state $\s3$.
In particular, if ${\Omega_{P}\t} / {\Omega_{S}\t} \stackrel{t\rightarrow +\infty }{\longrightarrow } 1$, meaning $\Theta=\pi/4$, then an equally-weighted superposition of states $\s1$ and $\s3$ will be created, $\Psi = (\psi_1 - \psi_3)/\sqrt{2}$, a process termed {\em half-STIRAP}.
As in STIRAP, state $\s2$ remains unpopulated in the adiabatic limit.
Instead of suddenly interrupting the evolution of the \pump  and \Stokes  pulses, we can let them vanish simultaneously, in a smooth fashion \mycite{Vit99}, as in Fig.~\ref{Fig:f-STIRAP} (top).

Transfer of population from a single quantum state into a predetermined coherent superposition of states is obviously a more demanding task than transfer to a single quantum state.
The price to pay is some loss of robustness: the relative intensity on the trailing slopes of the pulses needs to be controlled.

\subsection{Control of nonadiabatic losses: ultrahigh fidelity \label{Sec: nonadiabatic}}

In the early applications of STIRAP, its efficiency   was barely scrutinized because an accuracy of  over 90\%  sufficed for most purposes.
However, because of its robustness to decoherence STIRAP was  quickly recognized as a promising control tool for quantum information processing.
The latter demands very high fidelity, with admissible errors usually below $10^{-4}$.
Such small errors are difficult  to achieve with STIRAP
  as it approaches unit efficiency only asymptotically when the pulse areas increase.
Moreover,  very large laser intensities  may breach various assumptions:  other states will be coupled to the three states of STIRAP, and multiphoton ionization or dissociation may be appreciable.
 Various scenarios   have been considered for reducing the nonadiabatic coupling and hence achieving ultrahigh efficiency of STIRAP.
The following paragraphs describe some of these.

 \subsubsection{Nonadiabatic transitions \label{Sec: nonadiabatic transitions}}

The behavior of STIRAP away from the adiabatic limit has been discussed by several authors \mycite{Vit97loss,Fle96,Kob98a,Sun14}.
In particular, it was found that for smooth pulses, the nonadiabatic error vanishes exponentially with {the pulse area} \mycite{Elk95}.
This behavior is similar to the one predicted for two-level systems \mycite{Dyk62,Dav76}.
For Gaussian pulses, however, this is not the case and there is a power-law dependence\mycite{Lai96,Vit96,Dre98}.

\subsubsection{Pulse shaping \label{Sec: parallel Stirap}}

One approach to   reducing the nonadiabatic transitions in STIRAP is based on an approach that uses the Dykhne-Davis-Pechukas (DDP) method
 \mycite{Dyk62,Dav76} for estimation of the transition probability in a two-state system.
The DDP method relies on the so-called transition points defined as the (complex) zeroes of the quasienergy $\epsilon(t) = \sqrt{\Omega(t)^2 + \Delta(t)^2}$.
The transition probability is given by a sum over contour integrals in the complex plane from the origin to these transition points.
\mytcite{Gue02,Lac08} noticed that the two-state transition probability is minimized if the quasienergy does not have transition points, e.g. if $\epsilon(t)=$\,const.

\begin{figure}[tb]
\includegraphics[width=0.75\columnwidth]{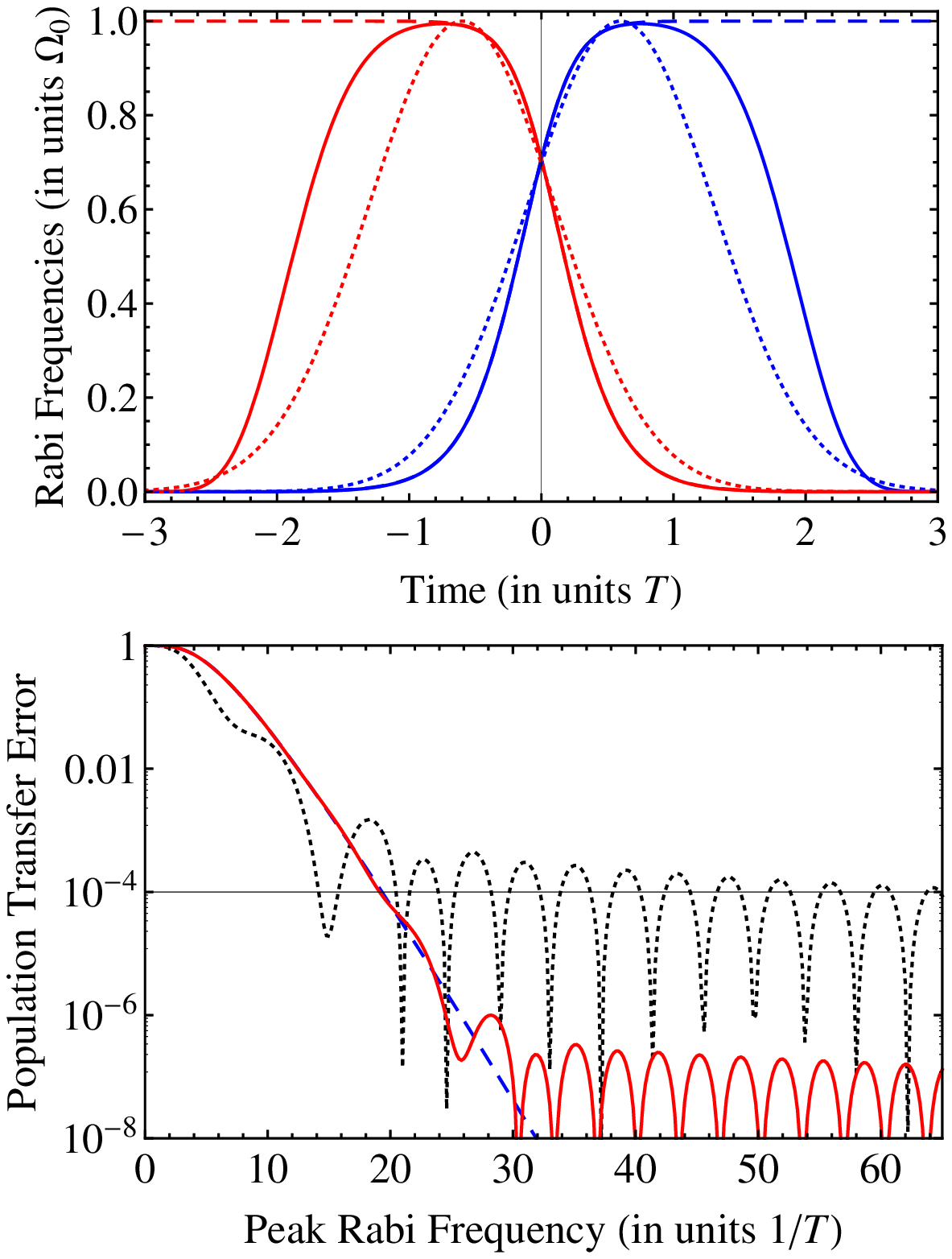}
\caption{\coloronline  
 DDP-optimized STIRAP by \mytcite{Vas09}.
\emph{Top}: pulse shapes.
Dotted curves: Gaussian pulses,
 dashed curves (fully overlapping with the solid curve in the pulse overlap region): $\Omega_{P}(t) = \Omax  \sin[\pi f(t)/2]$ and  $\Omega_{S}(t) = \Omax  \cos[\pi f(t)/2]$,
 solid curves: $\Omega_{P}(t) = g(t) \Omax  \sin[\pi f(t)/2]$ and  $\Omega_{S}(t) = g(t) \Omax  \cos[\pi f(t)/2]$,
 with $f(t) = 1 / (1+\e^{-4t/T})$ and $g(t) = \e^{-(t/2T)^6}$.
\emph{Bottom}: Population transfer error in STIRAP for the pulse shapes in the top frame.
\locationa{Fig. 2 of  \mytcite{Vas09} \bk }
}
\label{Fig:Vasilev2009-1}
\end{figure}

\mytcite{Vas09} used the DDP approach to optimize STIRAP by using the reduction of STIRAP to effective two-state systems on exact resonance and for large one-photon detuning.
In either case, the DDP-optimized pulse shapes must satisfy the relation
 $\Omega_{P}(t)^2 + \Omega_{S}(t)^2 = \text{const}$,
i.e., the eigenvalues remain constant, see Eq.~\eqref{STIRAP-eigenvalues}.
For   pulses of finite duration (as contrasted with Gaussians), this condition can be fulfilled exactly by the shapes shown by the dashed curves in Fig.~\ref{Fig:Vasilev2009-1} (top).
The solid curves in the same figure show more realistic pulse shapes, for which this condition is fulfilled only during the pulse overlap, which is when the dynamics occur.
%
An example is shown in the lower frame of Fig.~\ref{Fig:Vasilev2009-1}.
Here the Gaussian pulse shapes fail to reduce the population transfer error below the benchmark  value    $10^{-4}$ until   the pulse areas become very large.
By contrast, the DDP-optimized shapes 
    bring the error below this value   for much smaller pulse areas.
\mytcite{Bak16} further optimized this approach.
\mytcite{Che12} proposed another optimization of the \pump  and \Stokes pulse shapes using invariant-based inverse engineering.
\mytcite{Du16}  experimentally realized such a pulse-shape-optimized STIRAP with a large middle-state detuning with ${}^{87}$Rb atoms in a magneto-optical trap.
Finally, \mytcite{Dri09} proposed to use both pulse shaping and detuning chirping  in such a way that all three adiabatic energies of the  \L~system  were parallel;  they   called  this ``parallel STIRAP''.
This approach, however, places considerable population in the middle state \s2.

The trade-off of these pulse-shaping versions of STIRAP is clear: adiabaticity is improved at the expense of pulse shaping.
Such procedures lose a key advantage of STIRAP -- independence of the pulse shapes.
Therefore, the experimental feasibility of these proposals depends on the availability of pulse-shaping techniques for  the particular implementation.

\subsubsection{Shortcuts to adiabaticity \label{Sec: shortcuts}}

An alternative approach to reducing nonadiabatic losses uses  an additional field,   applied on the transition $\s1 \fromto \s3$ to form a loop linkage (a triangle or $\Delta$, rather than a  \L) of the three states \mycite{Una97a}.
They termed this approach ``control of diabatic losses''.
However, it turned out that   the amplitude of this additional field must be equal to $2\dot\mix(t)$ (up to a phase factor), i.e. for given \pump  and \Stokes  fields, it must have a specific time dependence and a   pulse area of $\pi$.
These constraints render the procedure impractical: If it is possible to apply a resonant $\pi$-pulse to the 1-3 transition, then this pulse will produce the desired population transfer without the need of the  \Stokes and \pump pulses.
Moreover, the additional field creates a closed interaction loop, which makes population transfer sensitive to the phases of the fields.

Similar proposals in two- and three-state systems have been made by \mytcite{Dem03,Dem05,Dem08} with the name  ``assisted adiabatic passage by counterdiabatic field'',
by \mytcite{Ber09} who used the term ``transitionless quantum driving'', and by \mytcite{Che10} who named it ``shortcut to adiabaticity''.
%

\subsubsection{Composite STIRAP \label{Sec: composite Stirap}}

\begin{figure}[tb]
\includegraphics[width=0.85\columnwidth]{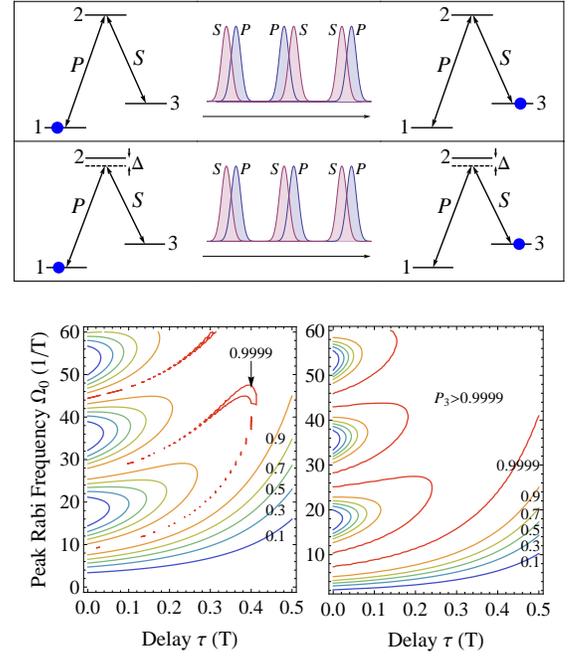}
\caption{\coloronline
Composite STIRAP.
\emph{Top}: The population is transferred from state $\s1$ to $\s3$ via a sequence of three $SP$ pulse pairs with appropriate relative phases.
When there is one-photon resonance ($\Delta = 0$), the ordering of the two pulse constituents   is reversed from pair to pair,
 while for nonzero $\Delta$  the ordering is the same for all pulse pairs.
%
\emph{Bottom}: Final population $P_3$ vs the pulse delay and the peak Rabi frequency for a single resonant STIRAP (left) and a sequence of five resonant  pulse pairs   (right) with sine-squared shapes of duration $T$,
 and  phases
$(\phi_{P,k},\phi_{S,k})$ given by
 $(0,\frac45\pi)(\pi,\frac85\pi)(\frac35\pi,\frac35\pi)(\frac85\pi,\pi)(\frac45\pi,0)$.
\location{Figs.~1 and 3 of  \mytcite{Tor13}.}
 }
\label{Fig:c-STIRAP}
\end{figure}

A rather different approach to optimization of STIRAP, which uses neither pulse shaping nor additional transitions, was proposed by \mytcite{Tor13}.
It uses the idea of composite pulses -- a coherent control technique, which is widely used in nuclear magnetic resonance \mycite{\NMR}.
A composite pulse is a sequence of pulses with well-defined relative phases, which are used as free control parameters to shape up the excitation profile in a desired manner.
In fact, similar ideas have been used in polarization optics    much earlier \mycite{\polarization}.

Recently, composite pulses have been successfully used in quantum optics \mycite{Tim08,Haf08,Tor11,Iva11}.
Nearly all work on composite pulses concerns resonantly driven two-state systems, mainly with rectangular pulses.
\mytcite{Tor11a} proposed to combine this technique with adiabatic passage -- composite adiabatic passage (CAP) -- which has been demonstrated experimentally by
\mytcite{Sch13}.

\mytcite{\composite} have extended these ideas to STIRAP.
Figure \ref{Fig:c-STIRAP}(top) illustrates how composite STIRAP operates.
The single pair of \Stokes  and \pump  pulses (SP) is replaced by a sequence of such pairs.
Figure \ref{Fig:c-STIRAP}(bottom) compares the efficiency for STIRAP (left) and composite STIRAP (right).
The ultrahigh efficiency region is expanded from a very tiny area in STIRAP to a large plateau in composite STIRAP, thereby making STIRAP suitable for high-fidelity quantum information processing.
This technique only requires an accurate phase control, which is possible in many experiments.

\subsection{Effects of decoherence \label{Sec:decoherence}}

A quantum system is always surrounded by an environment that is a source of decoherence.
Here we review  the detrimental effects of various types of decoherence on STIRAP.
STIRAP is  robust against some   causes of decoherence, e.g. irreversible population loss from the middle state $\s2$ and spontaneous emission within the system.
It is more sensitive to others, e.g. dephasing.

\subsubsection{Transition time \label{Sec:transition time}}

The   consequences of decoherence   are closely related to the  \emph{transition} time  $T_{\text{STIRAP}}$ in STIRAP.
\mytcite{Bor10} defined the transition time as the time it takes for the population $P_3$ to rise from $\epsilon$ to $1-\epsilon$.
For Gaussian pulses this  definition leads to
\be\label{transition time}
T_{\text{STIRAP}} = \frac{T^2}{\tau} \ln\sqrt{\frac{1-\epsilon}{\epsilon}}.
\ee
This time is to be distinguished from  the \emph{interaction} time, which is $T_{\text{interaction}} = 2T + \tau$,
where $T$ is the pulse width and $\tau$ is the pulse delay, and from the  \emph{overlap} duration.
Formula \eqref{transition time} reveals a   remarkable  feature: the transition time is inversely proportional to the pulse delay -- the longer the delay the shorter the transition time.
Of course, the pulse delay must stay within the limits discussed in Sec.~\ref{Sec:delay}.

\subsubsection{Irreversible population loss \label{Sec:loss}}

Because the evolution of the quantum system is never perfectly adiabatic and some transient population does visit the middle state $\s2$, a strong decay from this state may cause population loss.
The effects of irreversible population decay from state 2 to states outside the system has been studied by several authors
\mycite{Glu92,Fle96,Vit97loss}.
The most convenient way to model such loss is to add a negative imaginary term $-\frac12 \ii \Gamma$ in the Hamiltonian of Eq. \eqref{def-H}: $\Delta \to \Delta  -\frac12 \ii \Gamma$.
For small and medium decay rates $\Gamma$     the loss of transfer efficiency is dominated by decay of population that visits state $\s2$ due to imperfect adiabaticity.
For large $\Gamma$,    quantum overdamping \mycite{Sho06}  occurs,  decoupling     the entire system from the driving fields.
These two mechanisms lead to different dependence of the transfer efficiency on $\Gamma $: exponential at small $\Gamma$ and polynomial at large $\Gamma $.
Like the one-photon detuning linewidth $\Delta_{1/2}$, the loss ``linewidth'' $\Gamma_{1/2}$, at which $P_3$ drops to 1/2, is proportional to the squared peak Rabi frequency: $\Gamma_{1/2} \propto \Omax ^2$ \mycite{Vit97loss}.

\subsubsection{Dephasing \label{Sec:dephasing}}

The treatment of dephasing, or phase relaxation, requires solution of the more  complicated Liouville equation for the density matrix \mycite{Sho90},
\be\label{Liouville equation}
\ii \hbar \ddt \bm{\rho}\t =   \left[ \H\t, \bm{\rho}\t \right] - \ii \D\t,
\ee
where the dissipator $\D\t$ reads
\be
\D\t  = \hbar \left[\begin{array}{ccc}
0 & \gamma_{12} \rho_{12}\t & \gamma_{13} \rho_{13}\t \\
\gamma_{21} \rho_{21}\t & 0 & \gamma_{23} \rho_{23}\t \\
\gamma_{31} \rho_{31}\t & \gamma_{32} \rho_{32}\t & 0
\end{array}\right] .  \label{D}
\ee
Here $\gamma_{mn} = \gamma_{nm}$ are dephasing rates of the coherences $\rho_{mn}\t = \langle \psi_m|\hat{\rho}\t|\psi_n\rangle$, where $\hat{\rho}\t$ is the density operator.
Dephasing affects STIRAP by destroying the coherence between states $\s1$ and $\s3$, and thus leads to depopulation of the dark state.

\mytcite{Iva04} have derived    the adiabatic solution of the Liouville equation in the limit of strong dephasing,
\be\label{dephasing solution}
\rho_{11} = \rho_{22} = \tfrac13 - \tfrac13 e^{-\gamma_{13} \eta},\quad
\rho_{33} = \tfrac13 + \tfrac23 e^{-\gamma_{13} \eta},
\ee
with $\eta = \tfrac34 \int_{-\infty}^{\infty} \sin^2 2\mix\t dt $.
For Gaussian pulses, we have $\eta =3T^2/4\tau$, i.e. $\eta$ is proportional to the transition time, $T_{\text{STIRAP}}$.
Interesting conclusions follow from Eq.~\eqref{dephasing solution}.
For any value of $\gamma_{13}$, the final populations of states 1 and 2 are equal.
Moreover, the dephasing losses depend inversely on the pulse delay $\tau$.
Finally, the solution   is independent of the peak Rabi frequencies, hence increasing them does not reduce the   loss of efficiency.

In other studies of the detrimental effects of dephasing,  \mytcite{Dem02} have numerically explored STIRAP in liquid solutions by assuming that the three-state system is coupled to a classical bath.
\mytcite{Shi03} assumed a quantum bath and performed numerical simulations  using a quantum master equation.

\subsubsection{Spontaneous emission\label{Sec:spontaneous}}

 The    consequence of spontaneous emission within the  \L~system -- from  state \s2 to states $\s1$ and $\s3$ --  was studied by several authors \mycite{Ban91,Ban92,Iva05b}.
The description of this process requires equations for the density matrix \mycite{Kuh92,Bre02,Sca10}.
\mytcite{Iva05b} used adiabatic elimination of weakly coupled density matrix elements in the Liouville equation, from which an analytic approximation was derived.
For small-to-moderate decay rates STIRAP is not affected significantly by spontaneous emission because the middle state is unpopulated.
For strong decay rates STIRAP degenerates into incoherent optical pumping.
\mytcite{Sca11} studied this problem by using an effective Hamiltonian derived from a microscopic model.
They found that at zero environment temperature, the system-environment interaction acts as a pump toward the dark state.
Consequently, higher efficiency than in the phenomenological model \mycite{Iva05b} was obtained.

\subsection{Hyper-Raman STIRAP (STIHRAP) \label{sec-STIHRAP}}

\begin{figure}[tb]
\includegraphics[width=0.95\columnwidth]{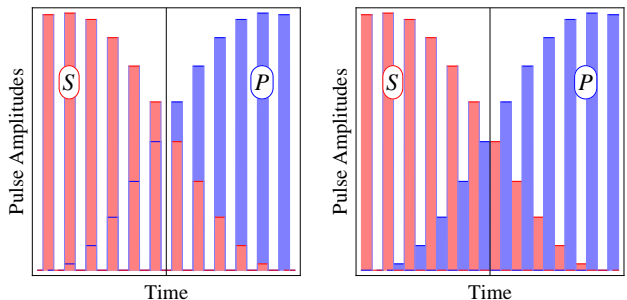}
 \caption{\coloronline
Sequences of coincident (left) and non-coincident (right) \Stokes and \pump pulse pairs used to produce piecewise-adiabatic passage (PAP).
\locationa{Fig. 1(c) of \mytcite{Sha07}.}
}
\label{fig-NV-pap}
\end{figure}

Access to high-lying electronic states of atoms or molecules requires ultraviolet or vacuum-ultraviolet (vuv) radiation, and it is tempting to consider using a multiphoton transition for this interaction.
For two-photon transitions, evaluation of the hyper-Raman interaction requires the frequency-dependent polarizability tensor ${\pmb \alpha}(\omega)$, whose
matrix elements between states $m$ and $n$ ($m,n= 1, 2, 3$) involve the product of two dipole-moment components summed over all possible intermediate states (including continua)\mycite{Yat98}.
The interaction energy associated with the polarizability involves the product of two electric-field amplitudes, leading 
 to two-photon Rabi frequencies, which are proportional to the intensities of the corresponding fields.
In addition, there occur dynamic Stark shifts,
that shift the diagonal Hamiltonian elements by $\H^{\rm Stark}_{nn}\t = \hbar [S_{nP}\t + S_{nS}\t ]$.
These shifts are proportional to the intensities too. 
It is these inevitable dynamic Stark shifts that  diminish the usefulness of population transfer by {\em stimulated hyper-Raman adiabatic passage} (STIHRAP)
\mycite{Gue98a,Gue98b,Yat98,Gue99}: the customary approach to improving adiabaticity by increasing the field intensities does not help because this introduces {Stark-shift-induced} two-photon detuning detrimental to the desired formation of the dark state.
Nevertheless, it has been possible to achieve population transfer by a combination of adiabatic and diabatic evolution \mycite{Boh01}, as in the technique of Stark-chirped rapid adiabatic-passage (SCRAP) \mycite{Ric00,Yat02c,Ran05}, albeit with population in state \s2.

\subsection{Piecewise adiabatic passage \label{Sec:PAP}  }

\mytcite{Sha07} proposed a technique, named  {\em piecewise adiabatic passage} (PAP), that produces STIRAP-like population transfer by trains of a large number of \Stokes  and \pump pulses.
The amplitudes of the \pump pulses  gradually increase, while those of the \Stokes  pulses gradually decrease, thereby forming global envelopes reminiscent of the \pump and \Stokes pulses in STIRAP, see Fig. \ref{fig-NV-pap}.
The interesting aspect of PAP is that the individual pulses can be shorter than the durations allowed by the adiabatic condition in STIRAP, which makes it possible to implement this STIRAP-like process with ultrashort pulses.
A similar concept was demonstrated experimentally in rubidium atoms by \mytcite{Zhd08}, who produced  a two-state level-crossing transition,  in which  both the field amplitude and the detuning chirp  change in steps.
This chirped-PAP was also used to create  superpositions of states in potassium atoms \mycite{Zhd09}.
However, attempts to implement the PAP-scheme in a molecule have hitherto failed \mycite{Bit12}.
A recent proposal \mycite{Sho16} suggests an implementation of a PAP analog in polarization optics, see Sec.~\ref{Sec:polarization}.

\mytcite{Ran12} proposed a related technique for producing complete population transfer  from state  \s1 to state  \s3  by a train of $N$ coincident \emph{PS} pulse-pairs, in which the maximum population in the middle-state \s2 is $\sin^2(\pi/4N)$.
In the limit of $N \gg 1$, it reduces to PAP, while for small $N$, it is similar to generalized  $\pi$  pulses.
\mytcite{Vai13} proposed a similar scheme, with a stepwise change of the fields, named ``digital adiabatic passage'' (DAP).

\section{STIRAP-like processes beyond three states \label{Sec:multi}}

\subsection{Multistate chains \label{Sec:chains}}


\begin{figure}[tb]
\includegraphics[width=0.90\columnwidth]{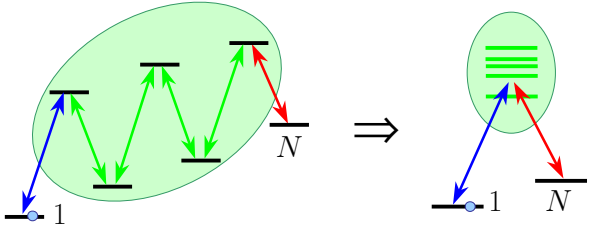}
\caption{\coloronline
\emph{Left}: Linkage pattern for $N$-state chain-STIRAP.
\emph{Right}: Effective set of parallel   \L~systems after diagonalization of the subsystem of middle states.
Multistate STIRAP is optimized  by tuning to a dressed middle state.
\locationa{Fig. 27 of \mytcite{Vit01a}.}
}
\label{Fig:chain}
\end{figure}

STIRAP has been implemented successfully in the chainwise-linked system of $N$ states, in which each state is connected only to its two neighbors,
$\s1 \fromto \s2 \fromto \s3 \fromto \cdots \fromto  N$.
Figure~\ref{Fig:chain} presents an example of such a chain
in a bent linkage pattern, which is found in manipulation of internal dynamics, in atom optics (Sec.~\ref{Sec-AtomOptics}), and other applications.
Ladder-like excitation is of considerable interest for producing dissociation or photoionization.

Adiabatic population transfer along a multistate chain by time-dependent fields has been described by numerous authors, and various STIRAP-like or STIRAP-inspired processes have been proposed and demonstrated in experiments.
Here we present the basic features of chain-STIRAP and describe a few important applications.

The RWA Hamiltonian of a multistate chain reads
\be
\label{Hchain}
\H = \frac{\hbar }{2}\left[\begin{array}{cccccc}
0 & \Omega_{1,2} & 0 & \cdots & 0 & 0 \\
\Omega_{1,2} & 2\Delta_{2} & \Omega_{2,3} & \cdots & 0 & 0 \\
0 & \Omega_{2,3} & 2\Delta_{3} & \cdots & 0 & 0 \\
\vdots & \vdots & \vdots & \ddots & \vdots & \vdots \\
0 & 0 & 0 & \cdots & 2\Delta_{N-1} & \Omega_{N-1,N} \\
0 & 0 & 0 & \cdots & \Omega_{N-1,N} & 0
\end{array}\right] ,
\ee
where the detunings are on the diagonal, and the Rabi frequencies $\Omega_{j,k}$ between states $j$ and $k$ are the off-diagonal elements.
The condition for two-photon resonance in STIRAP is replaced by the condition for ($N-1$)-photon resonance between the first and last states of the chain, as evident from the zeros in the first and last diagonal elements.
The intermediate states may,   in general,  be   nonresonant.
When all transitions are resonant, such multistate chains behave differently for odd and even numbers of states.

\subsubsection{Resonantly driven chains \label{Sec:chains-resonance-odd}}

A necessary condition for STIRAP-like population transfer in multistate chains is the existence of an eigenstate of the Hamiltonian, which connects the initial state $\s1$ to the final state $N$ of the chain.
For an \emph{odd} number of states  ($N=2n+1$), such a multilevel adiabatic-passage (AP) state  exists when all the lasers are on resonance with their corresponding transitions.
It still exists if the even-numbered states in the chain are detuned from resonance by the same detuning \mycite{Sho91,Mar91,Smi92}:
 $\Delta_{2j+1} = 0$ $(j=0,1,2,\ldots, n)$, and $\Delta_{2k} = \Delta$ $(k=1,2,\ldots,n)$.
Then the Hamiltonian of Eq.~\eqref{Hchain} has a zero eigenvalue.
The corresponding eigenstate $\AT$ is a time-dependent coherent superposition of only the odd-numbered states in the chain ($\s1$, $\s3$, $\ldots $, $ {2n+1}$).
In a bent linkage as in Fig.~\ref{Fig:chain}, this state contains only the lower states, with no contribution from the upper states, which makes it similar to the dark state in the \L~system. 

 For example, in a five-state chain the zero-eigenvalue   eigenvector of Eq. \eqref{Hchain} is a superposition
of states $\s1, \s3$ and $\s5$ only
\mycite{Mor83,Hio88,Sho91,Mar91,Smi92},
\begin{align}
\AT (t) &=  [ \Omega_{2,3}(t)\Omega_{4,5}(t)\psi_1-\Omega_{1,2}(t)\Omega_{4,5}(t)\psi_{3} \nonumber\\
&\quad +\Omega_{1,2}(t)\Omega_{3,4}(t)\psi_{5}] / {\cal N}(t),
\label{MultilevelDarkState}
\end{align}
where  ${\cal N}(t)$ is a normalization factor.
  If states \s2 and \s4 are fluorescent 
   then $\AT\t $ is a dark state, trapping population in states \s1, \s3 and \s5.

\begin{figure}[tb]
\includegraphics[width=0.80\columnwidth]{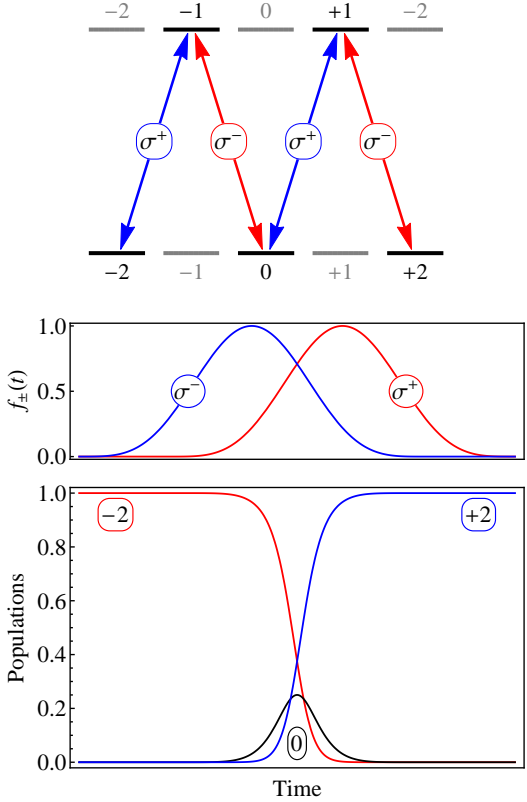}
\caption{\coloronline
\emph{Top}: Letter-M linkage pattern for five-state STIRAP between the magnetic sublevels of two degenerate levels, ground and excited, with angular momenta $J_g = J_e = 2$.
 The transitions are driven by a pair of laser pulses with right ($\sigma^+$) and left ($\sigma^-$) polarizations.
 \locationa{Fig. 30  of \mytcite{Vit01a}.}
\emph{Bottom}: Pulse shapes and populations for five-state STIRAP between the magnetic sublevels (indicated by numbers) in the top panel.
 }
 \label{Fig:M-chain}
 \end{figure}

A natural system for   applying multistate STIRAP is the chain formed by the   Zeeman sublevels of two degenerate levels.   Let the angular momentum be   $J$ for the lower level and $J$ or $J-1$ for the upper level,
  and take the two sequential pulses to be  polarized with opposite right ($\sigma^+$) and left ($\sigma^-$) circular polarizations, as shown in Fig.~\ref{Fig:M-chain}.
The Rabi frequencies $\Omega_{1,2}(t)$ and $\Omega_{3,4}(t)$ will follow the time dependence $f_+(t)$ of the $\sigma^+$ pulse, while the Rabi frequencies $\Omega_{2,3}(t)$ and $\Omega_{4,5}(t)$ will follow the time dependence $f_-(t)$ of the $\sigma^-$ pulse.
The relative amplitudes of the Rabi frequencies are determined by   Clebsch-Gordan coefficients \mycite{Sho90}.
For example, for  two levels with  angular momenta $J_g=J_e=2$, as shown in Fig.~\ref{Fig:M-chain}(a), the AP state reads
\be\label{dark-M22}
\AT\t = \frac{ \sqrt{3} f_-(t)^2\psi_1 + \sqrt{2} f_-(t)f_+(t)\psi_3 + \sqrt{3} f_{+}(t)^2\psi_5}{\calN(t)}.
\ee
If the system is prepared initially in the $M_g=-J$ ground sublevel, then a STIRAP-like transfer to the $M_g=+J$ sublevel can be achieved by applying the $\sigma^-$ pulse (\Stokesx) before the $\sigma^+$ pulse (\pumpx), because then the AP state has the asymptotic behavior $ \psi_1 \stackrel{t\to  -\infty }{\longleftarrow } \AT(t)  \stackrel{  t \to+\infty }{\longrightarrow } \psi_N$, with $N=5$.
The excited sublevels $\s2$ and $\s4$ remain unpopulated, even transiently, if the process is adiabatic.
However, the intermediate ground state $\s3$ acquires some transient population, as seen in Fig.~\ref{Fig:M-chain}.
In this particular example this does not lead to population loss because state $\s3$ is not a lossy state.

Multistate STIRAP in such angular-momentum   chains has been demonstrated experimentally and used extensively
 \mycite{Pil93,Val94,Gol94a,Gol94b,The98,Fea96,Fea98,God99,Web99} in the context of atom optics,  see Sec. \ref{Sec-AtomOptics}.

The chains with an \emph{even} number of states behave very differently than the ones with an odd number of states.
For $N=2n$, we have
$\det \H = (-1)^{n}\Omega_{1,2}^{2}\Omega_{3,4}^{2}\ldots \Omega_{N-1,N}^{2}\neq 0$, and hence $\H(t)$ does not have a zero eigenvalue.
More importantly, $\H(t)$ does not possess an AP state between the end states of the chain, $\s1$ and $ N$.
Hence a STIRAP-like population transfer is impossible.
Instead, the populations experience generalized Rabi oscillations \mycite{Ore92,Vit98a,Ban91b}.
%

\subsubsection{The off-resonance case  \label{Sec:chains-off-resonance}}

When the intermediate states in a chain are off resonance, while the two end states are still on ($N-1$)-photon resonance, chains with odd and even numbers of states behave similarly \mycite{Vit98a}.
There is no  zero eigenvalue and dark state of the Hamiltonian of Eq.~\eqref{Hchain}, but an AP state $\AT(t)$ may exist.
It has been shown \mycite{Vit98a}  that the condition for existence of an AP state is ${\cal D}^{(2,N-2)}{\cal D}^{(3,N-1)} > 0$, where
 ${\cal D}^{(j,k)}$ is the determinant of the matrix obtained by keeping the rows and columns from $j$-th to $k$-th in $\H$. 
For pulse-shaped couplings and nonzero detunings, this condition reduces to $\Delta_{2}\Delta_{N-1} > 0$.

For $N=4$, an AP state exists only when the middle detunings have the same sign, $\Delta_{2}\Delta_{3}>0$.
When $\Delta_{2}\Delta_{3}<0$, or one of them is zero, there is no AP state.
%
For $N=5$, an AP state exists if $(\Delta_{2}\Delta_{3}-\Omega_{2,3}^{2})(\Delta_{3}\Delta_{4}-\Omega_{3,4}^{2})>0$.
Thus if all detunings are nonzero, an AP state   exists if $\Delta_{2}\Delta_{4}>0$.
If $\Delta_{2}=0$, an AP state exists for $\Delta_{3}\Delta_{4}<0$.
If $\Delta_{4}=0$, an AP state  exists for $\Delta_{2}\Delta_{3}<0$.
If $\Delta_{3}=0$, an AP state  exists irrespective of the values of $\Delta_{2}$ and $\Delta_{4}$, which agrees with the result in Sec.~\ref{Sec:chains-resonance-odd}.
If $\Delta_{2}=\Delta_{4}=0$, an AP state exists irrespective of $\Delta_{3}$.

%
\begin{figure}[tb]
\includegraphics[width=0.99\columnwidth]{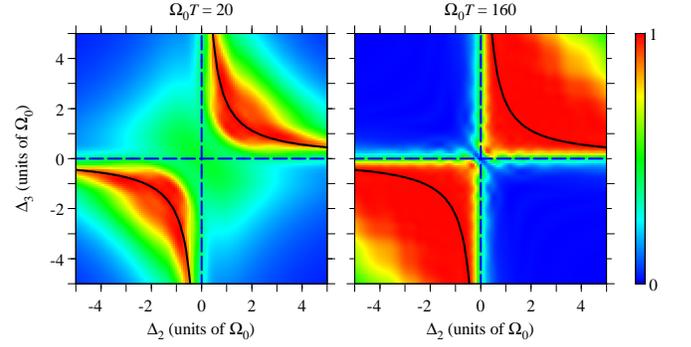}
\caption{\coloronline
Numerically calculated transfer efficiency for chain-STIRAP with $N=4$ states vs the two middle-state detunings $\Delta_2$ and $\Delta_3$.
The \pump and \Stokes pulse shapes are Gaussian, with peak Rabi frequency $\Omega_0$, pulse width $T$, and delay $\tau=T$,
  while the Rabi frequency of the pulse coupling the middle states $\s2$ and $\s3$ is constant and equal to $3\Omega_0$.
Left frame: $\Omega_0 = 20/T$; right frame: $\Omega_0 = 160/T$.
The solid curves show the dressed-state resonances, i.e. the energies of the dressed middle states.
\locationa{Fig. 4 of \mytcite{Vit98b}.} 
}
\label{Fig:chain-STIRAP-4states}
\end{figure}

Figure \ref{Fig:chain-STIRAP-4states} shows adiabatic population transfer in a four-state system versus the detunings from the two intermediate states $\s2$ and $\s3$.
High transfer efficiency is achieved only if $\Delta_2\Delta_3>0$ (first and third quadrant), while almost no transfer is possible for $\Delta_2\Delta_3<0$ (second and fourth quadrant).
As the pulse areas increase (from left to right frame), the high transfer efficiency regions in the first and third quadrants grow too.

We note that for nonzero middle detunings, the AP state   has, in general, nonzero components from all states involved, including the lossy excited states.
Hence  in the lossy regime     complete   population  transfer is impossible.

Recently, \mytcite{Kam13} proposed to use off-resonant four-state STIRAP-like population transfer between the two fine structure components of the metastable D-state in  alkaline-earth-metal ions.
A suitable spatial arrangement of the three laser fields makes it possible to exactly cancel the first-order Doppler shift.

\subsubsection{Straddle-STIRAP \label{Sec:straddle}}

\begin{figure}[tb]
\includegraphics[width=0.70\columnwidth]{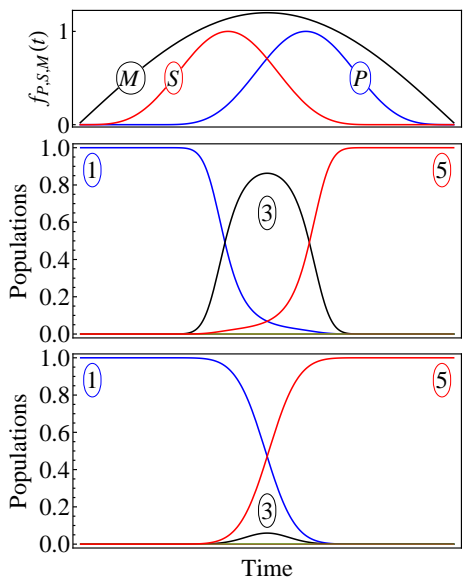}
\caption{\coloronline
Straddle-STIRAP in a chain with five states.
\emph{Top}: shapes of the \emph{P},  \emph{S}  and middle (\emph{M}) pulses.
All middle-state detunings are zero.
\emph{Middle}: Weak middle couplings: $\Omega_{2,3}^{0} = \Omega_{3,4}^{0} = 0.2 \Omega_0$.
\emph{Bottom}: Strong middle couplings: $\Omega_{2,3}^{0} = \Omega_{3,4}^{0} = 2 \Omega_0$.
}
\label{Fig:straddle-STIRAP}
\end{figure}

The problem of the nonzero middle-state populations can be alleviated by using ``straddle-STIRAP'' \mycite{Mal97}.
In straddle-STIRAP, all couplings between the middle states of the chain are at all times  much larger than the couplings for the first and last transitions -- the \pump and \Stokes fields.
Figure \ref{Fig:straddle-STIRAP} demonstrates the operation of straddle-STIRAP in a five-state chain.
Increasing the middle couplings reduces the populations of the middle states.
This approach has been experimentally demonstrated by \mytcite{Dan10}.

\subsubsection{Optimization of multistate STIRAP: dressed-state picture \label{Sec:dressed}}

A dressed-state approach of \mytcite{Vit98b} provides a particularly intuitive picture of STIRAP in multistate chains.
The $N-2$ coupled middle states can be considered as a field-dressed subsystem, which can be diagonalized, thereby replacing the bare states with dressed states, see Fig.~\ref{Fig:chain}.
The chain linkage is therefore replaced by two levels (the initial and final levels) both coupled to a set of closely spaced dressed states.
 The properties of this dressed subsystem can be controlled by the parameters of the dressing pulses (intensities and frequencies).
By tuning the \pump and \Stokes lasers to one of the dressed eigenstates, $\Phi_k$, the complex multistate dynamics is reduced to an effective  \L~system
$\psi_1\fromto \Phi_{k}\fromto \psi_{N}$, thereby facilitating efficient STIRAP-like transfer. 

This dressed-state approach provides further insight into straddle-STIRAP.
Indeed, if the dressing middle fields are strong, then the  splittings between the dressed energies are large.
Hence these states are far off resonance and receive little population.

The dressed picture also reveals the difference between odd and even number of states on resonance. 
For odd-integer $N$, one of the dressed states (the one with the zero eigenvalue) is always on resonance with the \pump and \Stokes lasers.
For even-integer $N$, the \pump and \Stokes lasers are tuned in the middle between two adjacent dressed energies and the ensuing interference between the different adiabatic paths leads to Rabi-like oscillations.
Hence staying on resonance is the best choice for odd-integer $N$, while the only possibility to achieve adiabatic population transfer for even-integer  $N$ is to choose appropriate nonzero middle-state detunings.

\subsubsection{Multiple intermediate states \label{sec-multi-lambda}}

The system of Fig.~\ref{Fig:chain} (right), in which the middle state 2 is replaced by a set of states, is interesting not only in the context of the preceding discussion by in its own right.
The presence of such states may open {multiple} transition paths $\s1\to\s3$, the interference between which may ruin the population transfer.
\mytcite{Cou92} were the first to consider this problem: they assumed  two intermediate states and equal couplings $\Omega_{P}(t)$ to state $\s1$ and equal couplings $\Omega_{S}(t)$ to state $\s3$.
\mytcite{Vit99b} studied the general case of $N$ intermediate states and arbitrary couplings.
They found that the dark state  remains a zero-eigenvalue eigenstate of the Hamiltonian only when the ratio $\Omega_{P,k}(t)/\Omega_{S,k}(t)$ between the couplings from each intermediate state $k$ to $\s1$ and $\s3$  does not depend on $k$.
Then the multi-\L~system behaves almost as a single  \L~system and STIRAP-like transfer $\s1\to\s3$ is possible, with almost no transient population in any intermediate state.
When this condition is not fulfilled there is no dark state but   there may exist,  for specific conditions on the single-photon detuning,  an AT state (albeit with contributions from the intermediate states).
It is most appropriate to tune the \pump and \Stokes lasers either just below or just above all intermediate states because then an adiabatic link $\s1\to\s3$ always exists, the transfer is more robust and the transient middle-state populations can easily be suppressed.

\subsection{Nearly degenerate states \label{sec-nearly-degenerate}}

In real physical systems, such as atoms and molecules, there are multiple states in the vicinity of the three states of the  \L~system.
These states may be present due to fine and hyperfine structure, Zeeman sublevels, or closely spaced   rovibrational levels in molecules.
They may interfere and impede, or even inhibit, STIRAP.

In a system of multiple states, the adiabatic energies may exhibit a very complicated pattern when plotted as a function of time.
The main hindrance to STIRAP is the emergence of narrow avoided crossings between the dark-state energy and its nearest neighbors, which may block the adiabatic path from state $\s1$ to state $\s3$.
Below we review the problems that emerge in two common situations: STIRAP between magnetic sublevels and STIRAP in a dense web of states.

\subsubsection{STIRAP between magnetic sublevels \label{sec-magnetic sublevels}}

\begin{figure}[tb]
\includegraphics[width=0.96\columnwidth]{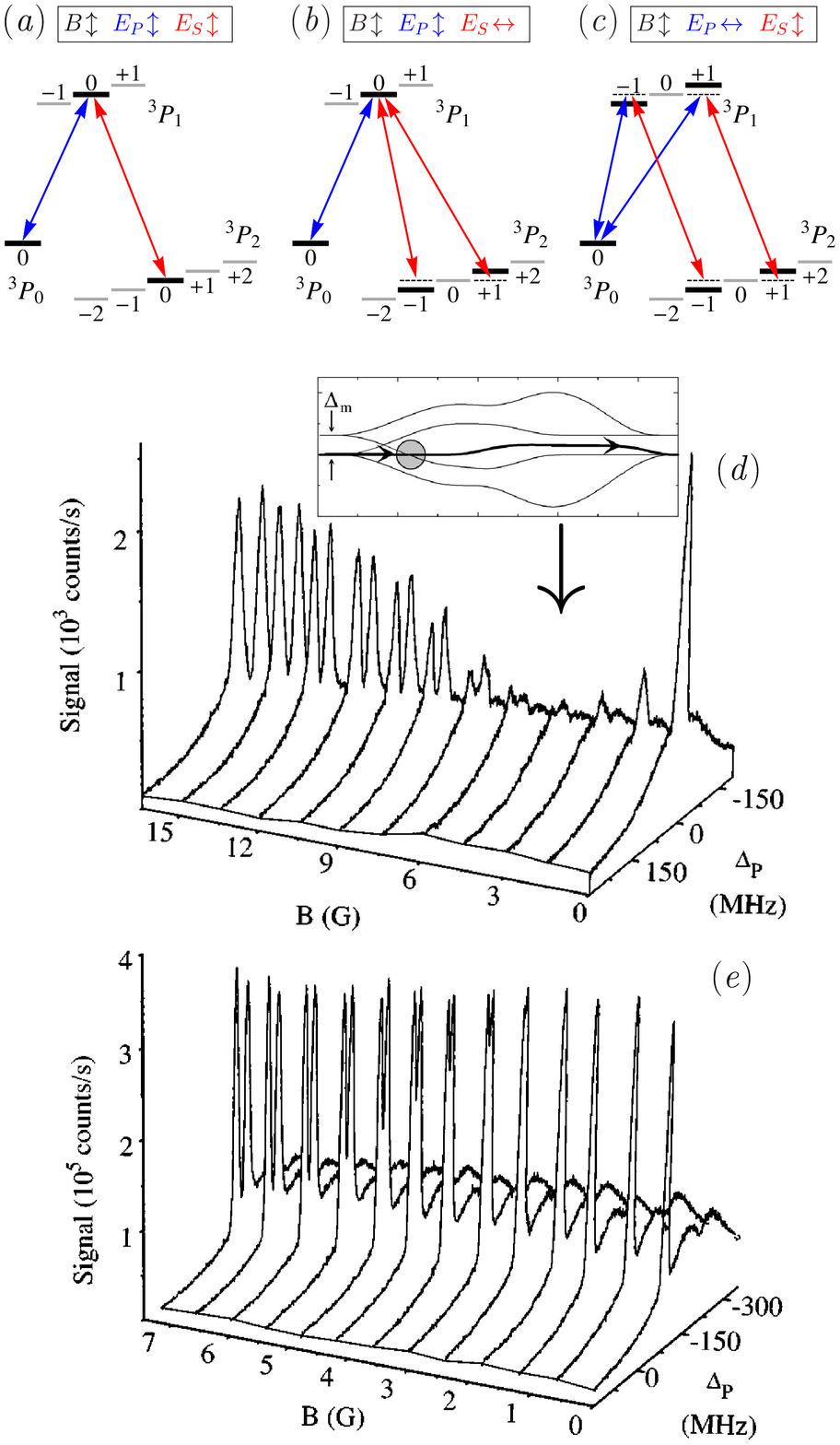}
\caption{\coloronline
(a)-(c) Linkage patterns between the magnetic sublevels of the levels $^3P_0$, $^3P_1$ and $^3P_2$ in Ne$^*$ for various choices of \pump and \Stokes polarizations with respect to the direction of the magnetic field $B$.
\locationa{Fig.~12 of \mytcite{Ber98}.}
(d) and (e) Population transfer efficiency in Ne$^*$ vs the detuning of the \pump laser field for a set of values of the magnetic field $B$ and the polarization choice of panel (c).
The \Stokes laser frequency is held fixed on resonance in frame (d) and off resonance  by 200 MHz in frame (e).
The inset in frame (d) shows the narrow avoided crossing appearing in the eigenenergies for moderate magnetic fields, which blocks the adiabatic path between the initial and target states. 
$\Delta_m$ is the Zeeman splitting.
\locationa{Figs. 14 and 15 of \mytcite{Mar96}.}
}
\label{Fig:Martin}
\end{figure}

A detailed theoretical and experimental study of STIRAP between magnetic sublevels   was carried out by \mytcite{Sho95b,Mar95,Mar96} in Ne$^*$ atoms.
That work demonstrated many of the problems present also in polyatomic molecules with their high density of energy levels, when the \Stokes  and/or \pump laser may couple several levels.
The level scheme in this experiment,  shown in Fig.~\ref{Fig:Martin} (top), involves the magnetic sublevels of $J=0$, $1$, and $2$ levels, and thus there are $9$ sublevels in total.
The population is initially in sublevel $J=0,M=0$.
A uniform magnetic field $\B$ is used to remove the Zeeman degeneracy, and to set the quantization axis.
The optical selection rules allow to select the sublevels participating in the dynamics by an appropriate choice of the laser polarizations with respect to the direction of $\B$.
Three special cases are shown in Fig. \ref{Fig:Martin}.
When the \pump and \Stokes fields are linearly polarized along $\B$, the selection rule $\Delta M=0$ applies.
Then only the three $M=0$ sublevels are coupled by the laser fields, cf. Fig.~\ref{Fig:Martin}(a): we have an ideal three-state  \L~linkage.
When the \pump polarization is parallel and the \Stokes polarization perpendicular to $\B$, four states are coupled, cf. Fig.~\ref{Fig:Martin}(b).
Conversely, when the \Stokes polarization is parallel and the \pump polarization perpendicular to $\B$, five states are coupled, cf. Fig.~\ref{Fig:Martin}(c).
Finally, in the general case of {arbitrary polarizations}, all nine magnetic sublevels are coupled.

Figure \ref{Fig:Martin} shows implementation of STIRAP in this system as a function of the \pump detuning $\Delta_{P}$ and the magnetic field $\B$.
In panel (d), the \Stokes detuning is set to zero with respect to the    transition frequency of the degenerate ($\B=0$) ${}^3P_2\fromto $\ ${}^3P_1$ transition, while the \pump detuning is scanned across the resonance.
For weak $\B$, a single peak in the   target-state population is observed near $\Delta_{P}=0$, because the $M=+1$ and $-1$ sublevels are too close to be resolved.
For strong $\B$, the Zeeman splitting increases and a symmetric two-peaked structure emerges, indicating population transfer to $M=+1$ or $-1$ sublevels.
A dramatic drop of efficiency occurs at moderate values of $\B$,  identified by \mytcite{Mar96} as due to the emergence of a narrow avoided crossing [inset   above panel (d)] in the adiabatic energy diagram, which blocks the adiabatic path between the initial and final states.
This connectivity obstacle can be removed by detuning the \Stokes laser from resonance, as evident in panel (e) of Fig.~\ref{Fig:Martin}.

\begin{figure}[tb]
\includegraphics[width=0.85\columnwidth]{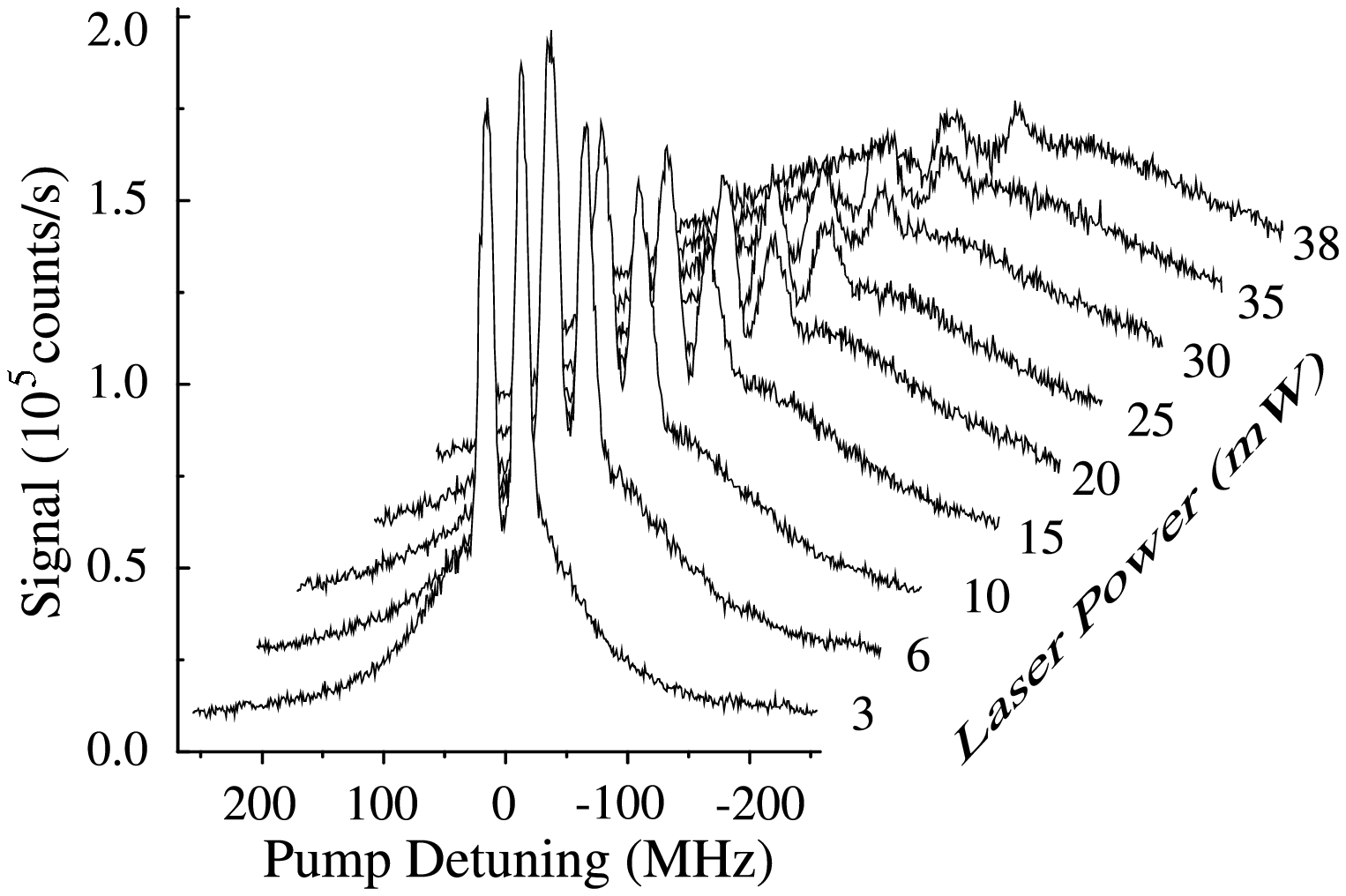}
\caption{
Population transfer in Ne$^*$ versus \pump detuning.
\location{Fig. 20  of \mytcite{Mar96}.}
\Also{Fig. 17 of \mytcite{Ber98} and Fig. 26 of \mytcite{Vit01a}.}
}
\label{Fig:Ne-power}
\end{figure}

Figure \ref{Fig:Ne-power} reveals another   possible problem in the implementation of STIRAP.
It shows an example,  in which efficient population transfer takes place for small Rabi frequencies but fails at higher laser power.
This is contrary to the adiabatic condition in STIRAP, which prescribes to increase the Rabi frequencies.
The reason for the breakdown is again the blocking of the adiabatic path
 as in Fig.~\ref{Fig:Martin}, a problem that can be cured, again, by detuning the lasers off their single-photon resonances.

To conclude, the  success of STIRAP in nearly degenerate systems depends on the existence of an adiabatic path that connects the initial and final states.
Although in most   cases careful analysis of the possible emergence of narrow avoided crossings blocking the adiabatic path is necessary,
 in general one-photon resonances should be avoided (but the two-photon resonance is still needed).

Finally, we point out that some other aspects of the influence of multiple nearly degenerate final states in STIRAP have been explored theoretically \mycite{Ban94,Kob98b}.

\subsubsection{STIRAP in a dense web of molecular states  \label{sec-Jakubetz}}

\begin{figure}[tb]
\includegraphics[width=0.70\columnwidth]{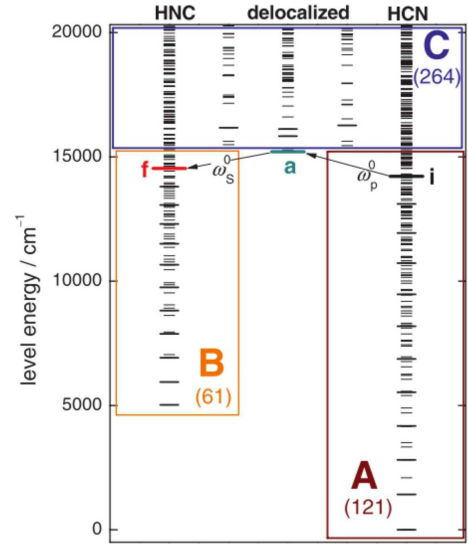}
 \caption{\coloronline
The 450 lowest $J = 0$ states of the HCN/HNC system
 used by \mytcite{Jak12} for studying STIRAP-like population transfer.  The horizontal grouping represents a coarse-grained
partitioning according to the localization of the states:
group A collects levels in the HCN well,  group B collects those in the HNC well, and group C are  high-lying states.
\location{Fig. 1 of \mytcite{Jak12}.}
}
\label{fig-Jak12-1}
\end{figure}

 \mytcite{Jak12} carried out a systematic numerical investigation of the effects of additional ``background'' states.
 Using  the double-well potential of the molecular HCN/HNC system, the level structure of which is shown in Fig. \ref{fig-Jak12-1}, he used an extensive parameter space of \Stokes and \pump pulses and hierarchies of  $(3 +N)$-state systems ranging in complexity from the basic three-state STIRAP
 $(N=0) $ up to  a dense web of linkages, $N=446$, drawn from the  lowest $J=0, K=0$ vibrational levels of \mytcite{Bow93}.%
\footnote{The simulation, intended to model general properties of a complex web of linkages, used vibrational dipole moments, treating rotational motion as sublevel averages in a rate-equation manner that ignored the prohibition on rotational $ 0 \fromto 0$ transitions for electromagnetic radiation.}
This study  showed that for pulse lengths of a few picoseconds the robustness of STIRAP disappeared as soon as the additional background couplings   exceeded about one tenth of the two basic STIRAP couplings, an  effect attributed, in part,  to the multitude of single- and multiphoton transitions that contribute to the elaborate linkage pattern.
Similar studies were presented by \mytcite{Dem02,Dem06},  who used a much smaller set of states and were therefore led to conclude that STIRAP may be possible for such a system.

\subsection{Tripod STIRAP \label{Sec:tripod}}

\begin{figure}[tb]
\includegraphics[width=0.90\columnwidth]{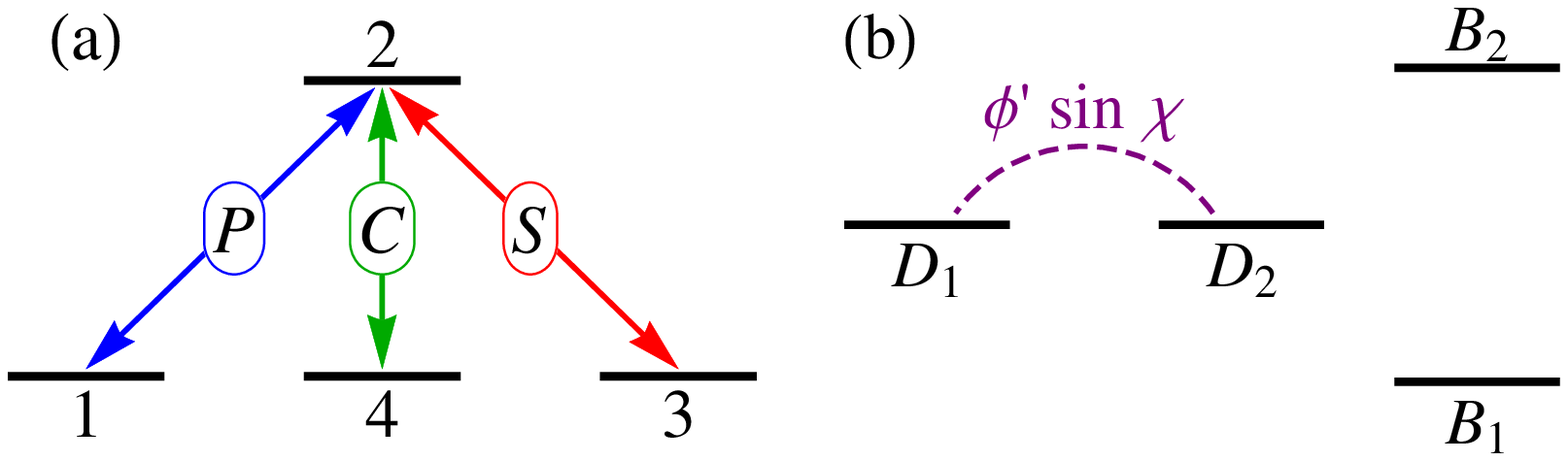}
\caption{\coloronline
$(a)$ The tripod linkage augments the \pump  and \Stokes fields of the $1 \fromto 2 \fromto 3$   \L~linkage of STIRAP   by an additional state $\s4$ coupled to the middle state $\s2$ by a control pulse $C$.
 $(b)$ The transformed tripod system in the adiabatic basis has two   bright states $B_1$ and $B_2$ unconnected with two degenerate dark states $D_1$ and $D_2$, one of which is initially populated.
In the adiabatic limit all   couplings can be neglected except the one between the two  dark states.
 }
\label{Fig:tripod}
\end{figure}

The four-state system in which three of the states are linked, by three separate fields, to a single state, has become a popular system in quantum physics.
 The linkage pattern may appear as the letter Y (two of the levels have higher energy), inverted Y (two ground states), or a tripod (three ground states and a single excited state).
In either of these, one may consider a three-state main chain, from which the fourth state forms a branch.
The effects of branches upon a main chain have been discussed for steady fields [Chap. 21 of \mycite{Sho90}] and pulsed excitation \mycite{Una98,Kob98a,Kob98b,Kob98c}
 including a method for suppression of  unwanted branches \mycite{Gen13}. 

In the tripod   extension of STIRAP, proposed by
\mytcite{Una98,Una99}, 
 the three-state  \L~linkage    gains   an additional state $\s4$, coupled to the intermediate state $\s2$ by a third, control laser $C$ with Rabi frequency $\Omega_{C}(t)$, cf. Fig.~\ref{Fig:tripod}.
The crucial difference from the   \L~linkage is that the tripod system has two, rather than one, zero-energy dark states.
The latter 
  have been found already by \mytcite{Cou92}.
In fact, the existence of two dark states in such a system follows from the Morris-Shore decomposition
\mycite{Mor83}.
This degeneracy has  important consequences, as will be explained below.
 Adiabatic evolution  leads to a coherent superposition of states, rather than to a single state.
This superposition can be controlled by the ordering of the pulses, the time delay between them, and the control pulse strength \mycite{Una98,Una99}.

 \subsubsection{Tripod linkage \label{sec-tripod}}

For simplicity, we consider only the case of exact   single-photon  resonance.
  With the states ordered as in Fig.~\ref{Fig:tripod} the RWA Hamiltonian of the tripod system   reads
\be\label{Htripod}
\H(t) = \frac{\hbar }2\left[\begin{array}{cccc}
0 & \Omega_{P}(t) & 0 & 0 \\
\Omega_{P}(t) & 0 & \Omega_{S}(t) & \Omega_{C}(t) \\
0 & \Omega_{S}(t) & 0 & 0 \\
0 & \Omega_{C}(t) & 0 & 0
\end{array}\right] .
\ee
It has two nonzero eigenvalues,
$\varepsilon_{B_1}(t) = -\varepsilon_{B_2}(t)=\case12   \Orms (t)$,
where $\Orms(t) \equiv \sqrt{\Omega_{P}(t)^2+\Omega_{S}(t)^2+\Omega_{C}(t)^2}$, and two zero ones, $\varepsilon_{D_1}(t) = \varepsilon_{D_2}(t)=0$.
The corresponding eigenstates\mycite{Una98} 
  fall into two classes: two degenerate null-eigenvalue dark states that have no component of state 2,
 \bse \label{Phi-tripod}
\begin{align}
\Phi_{D_1} &= \psi_1\cos\mixa - \psi_3\sin\mixa\cos\mixb - \psi_4\sin\mixa \sin\mixb ,  \label{D1} \\
\Phi_{D_2} &= \psi_3\sin\mixb - \psi_4\cos\mixb, \label{D2}
\end{align}
\ese
with the two time-dependent mixing angles
\be\label{tripod-angles}
\tan \mixa (t)=\frac{\Omega_{P}(t)}{\sqrt{\Omega_{S}(t)^2 + \Omega_{C}(t)^2}},\quad
\tan \mixb (t)=\frac{\Omega_{C}(t)}{\Omega_{S}(t)},
\ee
and two bright states (with components of state 2), whose explicit form is unimportant here.
%
For $\Omega_{C}(t)=0$  the mixing angle $\mixa (t)$ reduces to the mixing angle $\mix(t)$ used in STIRAP.
Then the second mixing angle is zero, $\mixb(t) = 0$, and state $\Phi_{D_1}(t)$ reduces to the usual dark state.
The  two degenerate dark states  form a decoherence-free qubit,   a useful tool for quantum information processing [Sec.~\ref{Sec:QInfo}].

\subsubsection{Tripod adiabatic evolution \label{sec-tripod-adiabatic-evolution}}

Because the two bright states have energies $\pm\case12\hbar \Omega (t)$, the nonadiabatic transitions between the dark and bright states can be suppressed by using large pulse areas, and the dynamics can be confined within  the dark states $\Phi_{D_1}(t)$ and $\Phi_{D_2}(t)$.
For counterintuitively ordered \pump and \Stokes pulses, we have the initial condition $\mixa (-\infty) = 0$ and hence  $\Phi_{D_1}(-\infty)=\psi_1$.
However, because $\Phi_{D_1}(t)$ and $\Phi_{D_2}(t)$ are degenerate the nonadiabatic coupling between them cannot be suppressed, even in the adiabatic limit.
The nonadiabatic coupling between  $\Phi_{D_1}(t)$ and $\Phi_{D_2}(t)$ is
 $-\ii\langle \dot{\Phi}_{D_1}(t) \vert \Phi_{D_2}(t) \rangle = \ii\dot{\mixb}(t) \sin\mixa(t)$,
 and it causes a transition between them with probability \mycite{Una98}
\be\label{beta}
P_{D_1\to D_2} = \sin^2 \beta , \qquad \beta = \int_{-\infty}^{\infty} \dot{\mixb}\t \sin\mixa\t dt .
\ee
If the system starts in state $\s1$ and the evolution is adiabatic, it will end  in a superposition of the dark states,
\be\label{initially-in-Phi1}
\Psi\t  \stackrel{t\to +\infty }{\longrightarrow } \Phi_{D_1}(+\infty )\cos \beta - \Phi_{D_2}(+\infty )\sin \beta .
\ee

Because $\mixa (\infty)$ and $\mixb (\infty)$ depend on the asymptotic values of the ratios of the Rabi frequencies, the mixing angle $\beta$ can be controlled merely by the pulse   ordering.
Thus by suitably choosing this   ordering, and the relative Rabi frequencies (for coincident pulses), one can create any desired superposition of the three ground states $\s1$, $\s3$ and $\s4$ in a decoherence-free fashion.
Three special cases are  listed below, in each of which the \Stokes and \control pulses precede (but overlap) the \pump pulse,  so that $\mixa (\infty) = \pi/2$ and   hence both dark states $\Phi_{D_1}(\infty)$ and $\Phi_{D_2}(\infty)$ are superpositions of $\s3$ and $\s4$,
\bse
\bea
\label{tripod-case1}
\Psi (\infty ) =& -\psi_3\sin \beta - \psi_4\cos \beta \ \ & (S-C-P), \\
\label{tripod-case2}
\Psi (\infty ) =& -\psi_3\cos \beta + \psi_4\sin \beta \ \ & (C-S-P), \\
\label{tripod-case3}
\Psi (\infty ) =& -\tfrac1{\sqrt{2}} (\psi_3 + \psi_4)\qquad \quad & (C\equiv S - P).
\eea
\ese
The relative phase in the created superpositions can be altered by changing the relative phases of the laser fields.

The above examples demonstrate that tripod-STIRAP provides a great deal of freedom in controlling   quantum superpositions.
Even more opportunities arise in more complex multistate systems that have two or more degenerate dark states, which are treated with use of the decomposition method of \mytcite{Mor83}; cf.  \mytcite{Iva08,Kyo06,Kyo07a,Sho11,Sho13,Una01c}.

\subsubsection{Experimental tripod demonstration \label{sec-tripod-demo}}

\begin{figure}[tb]
\includegraphics[width=0.75\columnwidth]{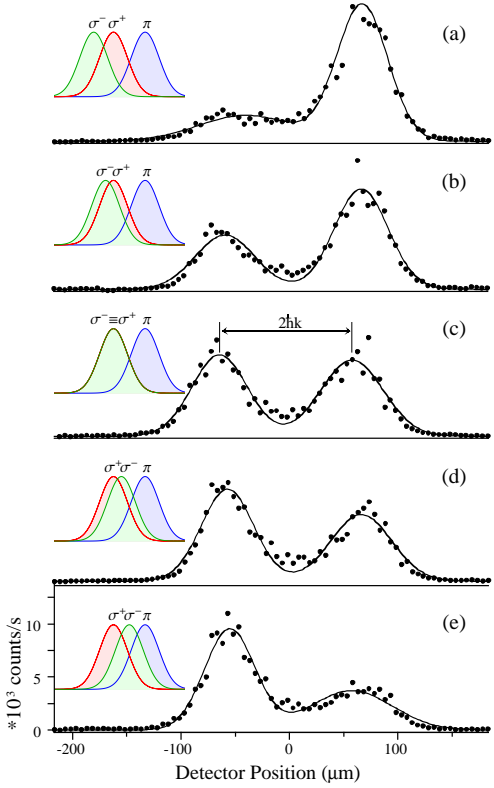}
 \caption{\coloronline
Momentum distribution for various possible orderings for the \pump ($\pi$), \Stokes ($\sigma^+$) and \control ($\sigma^-$) pulses in tripod-STIRAP.
\locationa{Fig.  2  of \mytcite{The99}.}
}
\label{Fig:tripod-exp}
\end{figure}

Tripod-STIRAP was demonstrated by \mytcite{The99} in a beam of Ne$^*$ atoms crossing three suitably arranged laser beams at right angles.
The initially populated state ${}^3P_0,M=0$ (state $\s1$) is coupled by a $\pi$-polarized \pumpx-laser field to an excited state ${}^3P_1, M=0$ (state $\s2$), which in turn is coupled via $\sigma^+$ ($S$) and $\sigma^-$ ($C$) laser fields to two magnetic sublevels of level ${}^3P_2$, $M=-1$ (state $\s3$) and $M=+1$ (state $\s4$), respectively; see Fig.~\ref{Fig:Martin}(b).
The atoms encounter the \pump laser last, while the timing of the \Stokes and \control laser beams is varied by displacing their axes.
In this way tunable superpositions of the magnetic sublevels $M=\pm1$ of state ${}^3P_2$ are created.
Because the $\sigma^+$ and $\sigma^-$ beams propagate in opposite directions, the momentum transfer to states $M=+1$ and $-1$ has opposite signs, resulting in coherent beam splitting by momentum $2\hbar k_S$.
This splitting results in two peaks separated by (122$\pm $2) $\mu$m, see Fig.~\ref{Fig:tripod-exp}.
When the $\sigma^-$ beam precedes the $\sigma^+$ beam [Fig.~\ref{Fig:tripod-exp}(a)], the $M=+1$ sublevel is predominantly populated, in agreement with Eq.~\eqref{tripod-case1}.
On the contrary, when the $\sigma^+$ beam precedes the $\sigma^-$ beam [Fig.~\ref{Fig:tripod-exp}(e)], it is the $M=-1$ sublevel that is predominantly populated, in agreement with Eq.~\eqref{tripod-case2}.
When the axes of the two beams coincide, a 50:50 beam splitting is achieved, as predicted by Eq.~\eqref{tripod-case3}.

Tripod STIRAP was further used by \mytcite{Vew03} in a modified version of the experiment by \mytcite{The99} described above, to create coherent superpositions of the sublevels $M=\pm 1$, see Sec.~\ref{sec-superposition}.

\subsection{Two-state STIRAP \label{sec-two state STIRAP}}

\begin{figure}
\includegraphics[width=0.80\columnwidth]{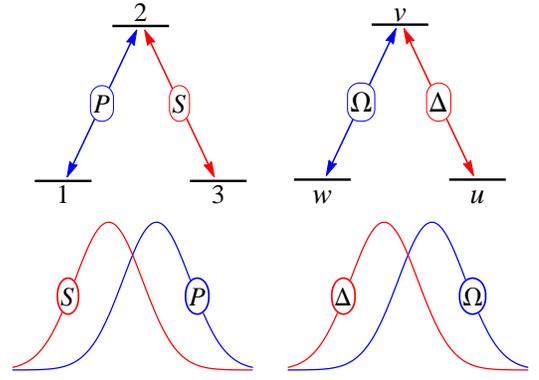}
\caption{\coloronline
Formal equivalence of linkages (top) and pulses (bottom) for STIRAP in a three-state system (left) and two-state STIRAP (right).
\locationa{Fig. 1 of \mytcite{Vit06}.}
}
\label{Fig:2s-STIRAP}
\end{figure}

Two-state STIRAP is an adiabatic technique for creating a maximally coherent superposition   in  a two-state system \mycite{Vit06,Yat02c}.
It is based on a formal analogy between the STIRAP equations and the Bloch equations for a two-state system \mycite{All87}.
By redefining the components of the state vector one obtains a torque equation similar to the Bloch equations \mycite{Fey57},
\be\label{torque}
\frac{d}{dt} \mathbf{B} = \mathbf{Q} \times \mathbf{B},
\ee
where $\mathbf{B} = [u,v,w]$ is the two-state Bloch-vector, and $\mathbf{Q} = [-\Omega_P, 0, \Omega_S]^T$ is the torque vector.
The two-state Bloch-vector components correspond to the three-state probability amplitudes through the formulas
 $ u\t = -C_3\t$, $v\t = -\ii C_2\t$, and $ w\t = C_1\t$,
as illustrated in Fig.~\ref{Fig:2s-STIRAP}.
Apart from a factor of 2, the   two-state detuning $\Delta\t$ corresponds to the \Stokes pulse, and the two-state Rabi frequency $\Omega\t$ to the \pump pulse.
The    condition for initial population in the lower state of the two-state system, $w(-\infty )=-1$, becomes $C_1(-\infty )=-1$ in two-state STIRAP.
The pulse  ordering pertinent to STIRAP requires, for two-state STIRAP, a time-dependent (pulsed) detuning $\Delta\t$ that precedes the Rabi frequency $\Omega\t$,   see  Fig.~\ref{Fig:2s-STIRAP}.
This  sequence 
  will rotate the tip of the Bloch vector from the south pole to the equator,   creating thereby a  final state described by the values $| u|  =1$, $v=w=0$, 
  corresponding to the three-state amplitudes $|C_3| = 1, C_2 = C_1 = 0$.
States with $w= 0$   are states of \emph{maximal coherence}. 

The ``dark state'' superposition    produced by two-state STIRAP   is not really a \emph{{quantum} state} of the system, but a sum of the inversion $w\t$ and the coherence $u\t$,
\be
d\t = w\t\cos \mix \t + u\t\sin \mix \t,
\ee
with $\mix \t = \arctan [\Omega \t/\Delta \t]$.
When the detuning pulse $\Delta \t$ precedes the coupling pulse $\Omega \t$ the mixing angle $\mix \t$ has the same asymptotics as in STIRAP; hence   the initial and final conditions  are  $d(-\infty )=w(-\infty )$ and $d(\infty)=u(\infty )$.
 Because the adiabatic passage is robust,   the  Bloch vector rotation is also robust: it depends only weakly on the
  extent of the overlap of the two pulses and the peak values of $\Delta \t$ and $\Omega \t$.
The adiabatic condition is similar to the one in STIRAP: it requires large coupling pulse area and large detuning area,
$\left\vert \int_{-\infty}^{\infty} \Omega\t dt \right\vert  \gg 1$, $\left\vert \int_{-\infty}^{\infty} \Delta\t dt \right\vert \gg 1$.

Various techniques exist for producing pulse-shaped detuning $\Delta (t)$   that is delayed in time with respect to a pulse-shaped Rabi frequency of the excitation pulse.
One can use a pulse shaper for pico- or femtosecond pulses, an acoustooptic modulator  for microsecond pulses, or a pulsed magnetic field (to Zeeman shift the state energies).
Alternatively, one can use the dynamic Stark shift induced by a strong off-resonant pulse to modify the energies of the two states \mycite{Yat02c}.

\begin{figure}
\includegraphics[width=0.95\columnwidth]{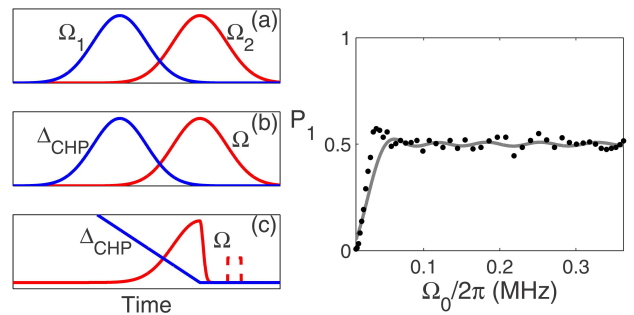}
\caption{\coloronline Experimental demonstration of two-state STIRAP with a ${}^{40}$Ca$^+$ ion in a spherical Paul trap.
Left-hand frames show pulses:
$(a)$ Two Gaussian pulses for Rabi frequencies as used in STIRAP.
$\Omega_1$ is the  \Stokes   field, $\Omega_2$ is the \pump field.
$(b)$ Two Gaussian pulses for use with detuning $\Delta_{\rm CHP}$ and Rabi frequency $\Omega$ of two-state STIRAP.
$(c)$ The linearly-chirped detuning and partial-Gaussian pulse used in the experiment.
Right-hand frame: the observed transition probability.
Two-state-STIRAP creates an equal superposition of states $4S_{1/2}$ and $3D_{5/2}$. 
\location{Figs. 1 and 3  of \mytcite{Yam08}.}
}
\label{Fig:Yamazaki-2s-STIRAP}
\end{figure}

Two-state STIRAP was demonstrated by \mytcite{Yam08} in an experiment with a trapped ${}^{40}$Ca$^+$ ion.
Figure \ref{Fig:Yamazaki-2s-STIRAP} shows the signature of the created superposition.
Instead of a pulsed detuning (middle) they used a linear chirp ending at the time of maximum of the Rabi frequency (bottom).
Because only the behavior of $\Delta(t)$ and $\Omega(t)$ in the overlap region matters, this change did not matter.
The experimental results confirm that the populations of the created superposition are very robust to changes of the peak Rabi frequency $\Omega_0$, as predicted.

This intriguing analogy between STIRAP and chirped two-state excitation is a rare example
of an instance when one can learn something about a simpler system (the two-state system) by using the knowledge about a more complex system (the three-state system).

\subsection{Population transfer involving a continuum \label{Sec:continuum}}

\subsubsection{STIRAP via a continuum \label{sec-continuum-coherence}}

Our understanding of the photoionization continuum has evolved in the last decades from being viewed as an incoherent irreversible drain of population described by Fermi's golden rule to a medium that can, under certain conditions, support coherent processes.
Continuum coherence  is essential in autoionization \mycite{Fan61,Fan68,Rza81} where it leads to the so-called laser-induced continuum structure (LICS) caused by the destructive interference of two ionization channels, which can suppress photoionization at a certain wavelength.

\mytcite{Car92,Car93} were the first to propose replacing the middle state in STIRAP by a continuum of states, arguing that STIRAP is largely insensitive to the properties of the middle state, and that such a scheme would offer a continuous range of energies.
They  modeled this continuum as an infinite set of equidistant discrete states, taken to the limit of infinitesimal energy separation, each equally strongly coupled to the two bound states -- a discretized quasi-continuum [Sec.~16.1 of \mycite{Sho90}].
They found that the dark state exists with this model and that, with the counterintuitive pulse sequence,  it would  allow  complete population transfer between the two bound states.

This conclusion turned out to be a consequence of using an inadequate model \mycite{Nak94}.
 In a real continuum, with a nonzero Fano parameter \mycite{Fan61,Kni90} and Stark shifts, population transfer would be greatly reduced.
\mytcite{Car95,Car96,Yat97,Pas97,Vit97c} subsequently found that significant partial transfer may still be feasible.
 \mytcite{Nak96,Pas98} further suggested that a STIRAP-like process can take place via an autoionizing state.
\mytcite{Una98b,Una00a} discussed a tripod-type linkage with the intermediate level lying in the continuum.
All schemes for population transfer via continuum make use of features similar to LICS \mycite{Kni90,Hal98,Yat99a}.

\begin{figure}[tb]
\includegraphics[width=0.45\columnwidth]{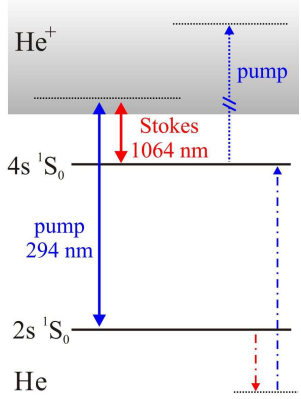}
\caption{\coloronline
Energy levels for the demonstration of  STIRAP via a continuum in He$^*$ atoms.
The  \pump and \Stokes lasers couple the initial state $2s\ {}^1S_0$ and the target state $4s\ {}^1S_0$ to the same ionization continuum.
\location{Fig. 1 of \mytcite{Pet05}.} 
\label{Fig:He*-levels}
}
\end{figure}

Despite these limitations \mytcite{Pet05} demonstrated continuum-STIRAP in He$^*$ atoms with 6\% efficiency, see Fig.~\ref{Fig:He*-levels}.
Later, \mytcite{Pet07}  improved the efficiency to 23\%.
The main limit to higher efficiency was identified to be  the irreversible ionization of the target level $4s\ {}^1S_0$ by the \pump laser.

\subsubsection{STIRAP-like transfer into a continuum \label{sec-into-continuum}}

\begin{figure}[tb]
\includegraphics[width=0.55\columnwidth]{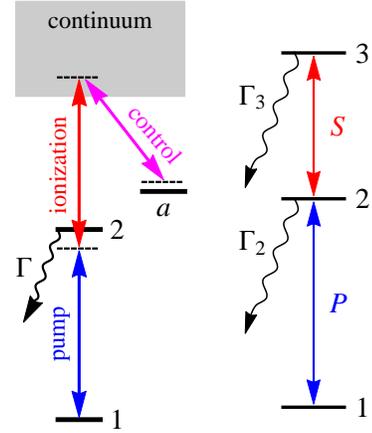}
 \caption{\coloronline
\emph{Left}: Linkage scheme for STIRAP-like population transfer into a continuum enhanced by an ancillary state $a$ -- an example of LICS.
\locationa{Fig. 4 of   \mytcite{Ran07}.}
\emph{Right}: Linkage structure for steering  population flow through lossy states.
\locationa{Fig. 1 of \mytcite{Gar05}.}
}
\label{fig-NV-lics-loss}
\end{figure}

Instead of replacing the middle state $\s2$ by a continuum \mytcite{Sha94,Fri96,Var96,Var99}
 proposed replacing the target state $\s3$ by a  photoionization or dissociation continuum, as shown in  Fig.~\ref{fig-NV-lics-loss}(left).
The objective differs dramatically from that of   STIRAP via a continuum: instead of trying to avoid population loss, now the goal is to maximize this loss in the form of ionization or dissociation.
As in STIRAP, the aim is to minimize the transient population of the middle state $\s2$ in order to avoid population loss via spontaneous emission to other discrete states.
\mytcite{Tha06a} extended this idea to multiple continua, e.g. the control of the photodissociation channels of the process CH$_3(v)$+I$^*(^2P_{1/2})$ $\leftarrow$ CH$_3$I $\rightarrow$ CH$_3(v)$+I$(^2P_{3/2})$ \mycite{Tha06b}.

Unlike STIRAP via a continuum a dark state cannot be formed between the initial state 1 and a continuum of final states.
To overcome this difficulty, \mytcite{Ran07}  proposed to use a LICS  by embedding an ancillary state $a$ into the continuum by a third laser, as seen in Fig.~\ref{fig-NV-lics-loss}(left).
This LICS-STIRAP allows one to produce almost complete photoionization, with negligibly small population losses from state $\s2$.

\subsubsection{Control of loss channels \label{sec-control-loss}}

The principle of STIRAP has been applied, in theoretical work and experiments \mycite{Gar05,Gar06},
to control the population flow in a molecular ladder of three states in Na$_2$, consisting of a nondecaying initial state \s1 and rapidly decaying middle and upper states  \s2 and \s3, see Fig.~\ref{fig-NV-lics-loss}(right).
The sink of the population flow out of the system is controlled by the pulse timing.
When the \pump pulse precedes the \Stokes pulse, the population loss occurs primarily through state 2.
When the two pulses coincide in time, some population reaches state 3 and decays from there.
When the \Stokes pulse precedes the \pump pulse, as in STIRAP, almost the entire population passes through state 3.
An analytic description of this process has been presented by \mytcite{Yat06}.

\section{STIRAP  in atomic and molecular physics \label{sec-AMO}}

Atoms and molecules have provided the major platforms for STIRAP, with the intent of  producing selective, robust and complete population transfer.
Very soon it was realized that the coherent excitation process had additional benefits, such as    redirection of atomic and molecular velocities.
The following section discusses some examples.

\subsection{Neutral atoms \label{Sec:atoms}}

\subsubsection{Population transfer in atoms \label{Sec:atoms-population-transfer}}

Originally, STIRAP was used in atoms primarily for coherent momentum transfer in order to build coherent beam splitters and mirrors for atom interferometry, see Sec.~\ref{Sec-AtomOptics} below for more details.
A brief overview here begins with the first demonstration of STIRAP in neutral atoms presented by 
\mytcite{Pil93}, who achieved a little over 50\% efficiency of population transfer between magnetic sublevels of the ground hyperfine level $F=4$ in ${}^{133}$Cs.
\mytcite{Wei94,Wei94PRA,Gol94a,Gol94b,Fea96,Fea98,God99,Web99} applied STIRAP in the same atom with nearly 100\% efficiency.
\mytcite{Law94a} demonstrated STIRAP in metastable He$^*$, and \mytcite{Kul97}  used STIRAP in the same atom for subrecoil laser cooling in 1D, 2D and 3D.
\mytcite{Ess96} demonstrated STIRAP in ${}^{87}$Rb.
\mytcite{Mar96} implemented STIRAP in ${}^{20}$Ne$^*$, while \mytcite{The98}  implemented tripod STIRAP in the same atom.
Rempe and co-workers, and Kuhn and co-workers have used vacuum-STIRAP with ${}^{85}$Rb and ${}^{87}$Rb atoms in numerous experiments for producing single photons, entangled photon pairs, shaped photons, etc. (see Sec.~\ref{Sec:vacuum-STIRAP} for further details and references).

Further applications include:
\begin{itemize}

\item  population transfer in a ladder system to the 5D$_{1/2}$ state of ${}^{85}$Rb atoms with high efficiency \mycite{Sup97,Sni12};

\item  transfer to the 44D$_{5/2}$  Rydberg state in cold  ${}^{85}$Rb \mycite{Cub05};

\item  monitoring the population evolution in the participating levels during the transfer process in ${}^{87}$Rb \mycite{Gea07};

\item  transfer by microwave fields between Zeeman sublevels 
 of  ${}^{133}$Cs for a fountain-based primary frequency standard \mycite{Cha05};

\item  application of fractional STIRAP in a vapor cell to prepare a coherent superposition of the $F = 1$ and $F=2$ hyperfine levels in the ground state of  ${}^{87}$Rb \mycite{Obe07} or transfer between hyperfine levels in an ultracold sample of those atoms \mycite{Du14};

\item  demonstration of transfer in Na vapor using pulses of 7 ps duration with the homogeneous linewidth being sufficiently broad to transfer the entire thermal ensemble to the $5s$ level \mycite{Hic15};

\item proposals for deterministic creation \mycite{Pet13} and extraction \mycite{Pet15} of a single Rydberg atom in an atomic ensemble.

\end{itemize}

\subsubsection{Coherent superposition states \label{sec-superposition}}

\begin{figure}[tb]
\includegraphics[width=0.80\columnwidth]{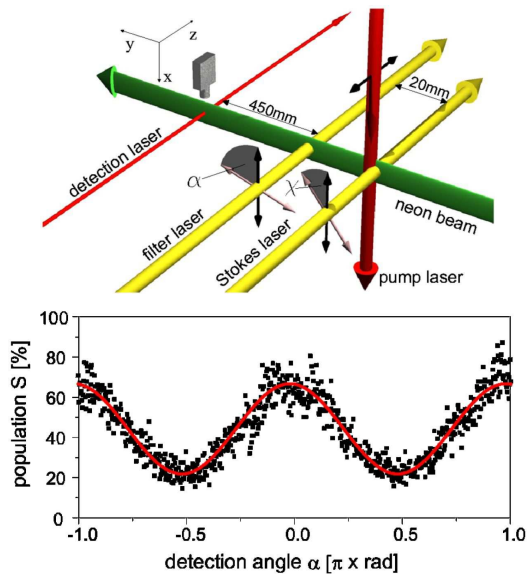}
\caption{\coloronline
Creation and measurement of a coherent superposition of magnetic sublevels in Ne$^*$ atoms: beams geometry (top) and signal (bottom).
The linear polarization direction of the Stokes laser is rotated by an angle $\chi$ with respect to the $x$ axis.
The polarization direction of the filter laser used for detection is rotated by an angle $\alpha$.
 \location{Figs.~2 and 3 of \mytcite{Vew03}.}
 }
\label{Fig:Vewinger2003}
\end{figure}

\mytcite{Vew03,Vew07a,Vew07b,Vew10,Hei06} used tripod-STIRAP and variations of it to prepare coherent superpositions of magnetic sublevels in metastable ${}^{20}$Ne$^*$ atoms.
They 
 created a coherent superposition of the magnetic sublevels $M=\pm1$ of the level ${}^3P_2$ with a well-defined relative phase.
Figure \ref{Fig:Martin}(b) shows the relevant sublevels in this experiment.
The initially populated state 
${}^3P_0$ is coupled by $\pi$-polarized light (pump) to the intermediate state 
${}^3P_1$, which in turn is coupled to the sublevels $M=\pm 1$ of level 
 ${}^3P_2$ by $\sigma^\pm$ polarized light (Stokes lasers $S_\pm$) produced by a single $\pi$ polarized laser beam perpendicular to the \pumpx-laser beam,
 see Fig.~\ref{Fig:Vewinger2003}(top).
The Stokes polarization direction was rotated by an angle $\chi$ with respect to the $x$ axis, and this angle was mapped onto the phase of the created superposition of sublevels $M=\pm 1$.
The measurement of the superposition was performed by mapping the superposition parameters onto the populations of the $M$ states of ${}^3P_2$ by a filter laser
 with polarization rotated at an angle $\alpha$ to the $x$ axis. 
This field depleted the population, subject to the relevant optical selection rules (the surviving population is the read-out signal).
The light-induced-fluorescence signal observed by a subsequent unpolarized probe laser is proportional to $\cos^2(\chi-\alpha+\phi)$ ($\phi$ is an arbitrary phase, e.g., from an external magnetic field).

We also note several theoretical proposals for creation of coherent superpositions of degenerate sublevels by STIRAP \mycite{Kis01,Kis02b,Kar03,Tha04,Kis04b,Kis05}.

\begin{figure}[tb]
\includegraphics[width=0.80\columnwidth]{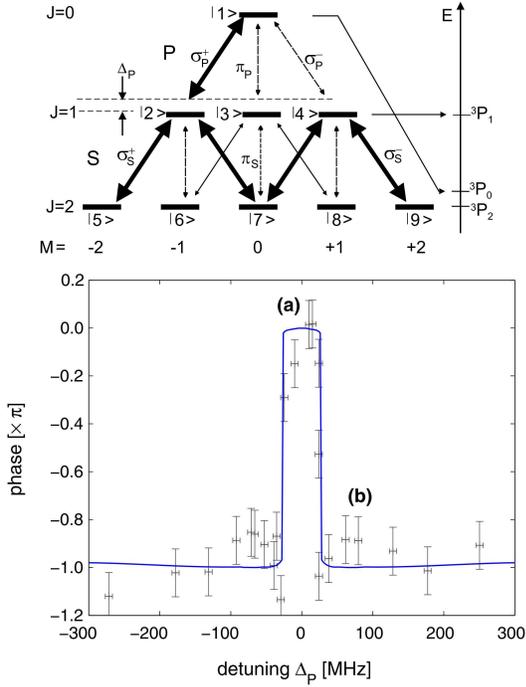}
\caption{\coloronline
Level scheme (top) and signal (bottom) for creation and measurement of a coherent superposition of the magnetic sublevels $M=-2$, 0 and $+2$ of the metastable level  ${}^3P_2$ in Ne$^*$ atoms.
\location{Figs.~1  and 5 of \mytcite{Hei06}.}
}
\label{Fig:Heinz2006}
\end{figure}

\mytcite{Hei06} introduced and demonstrated another method for control of the phase of a coherent superposition by changing the carrier frequency of the driving lasers.
The method exploits two distinctly different mechanisms: STIRAP and CPT.
The level scheme and the beam geometry are shown in Fig.~\ref{Fig:Heinz2006}.
The initially populated state ${}^3P_0$ in ${}^{20}$Ne$^*$ is coupled by $\sigma^+$-polarized light (\pumpx) to the intermediate state ${}^3P_1$, which in turn is coupled to the sublevels of level ${}^3P_2$ by $\sigma^\pm$ polarized light (\Stokes lasers) produced by a single $\pi$ polarized laser beam parallel to the \pump beam. When both the \pump and \Stokes lasers are on resonance, the population is transferred by STIRAP to the coherent superposition
\be
\ket{\Psi\t} = -s_{-2}\t \ket{-2} - s_{0}\t \ket{0} + s_{+2}\t \ket{+2}.
\ee
When \Stokes is on resonance, but the \pump field is detuned by more than 10 MHz (which is the two-photon linewidth of STIRAP), STIRAP is disabled and the population is optically pumped into the superposition
\be
\ket{\Psi \t} = c_{-2}\t\ket{-2} - c_{0}\t \ket{0} + c_{+2}\t \ket{+2}.
\ee
The probability amplitudes $s_m\t$ for STIRAP and $c_k\t$ for CPT are real and positive; their specific values are unimportant here.
The relative phase between the amplitudes $s_{-2}\t$ and $s_0\t$ for STIRAP is zero, while the relative phase between $c_{-2}\t$ and $c_0\t$ for CPT is $\pi$.
Figure \ref{Fig:Heinz2006} demonstrates such a phase switching between the values 0 (for $|\Delta_P| < 10$ MHz, where STIRAP dominates) and $\pi$ (for $|\Delta_P| > 10$ MHz where CPT dominates).

Subsequently, \mytcite{Vew07a} proposed and \mytcite{Vew07b} experimentally demonstrated additional implementations of extensions of STIRAP and tripod-STIRAP, for robust deterministic preparation of superpositions of two or three magnetic atomic sublevels and measurement of their relative amplitudes and phases  \mycite{Vew10}.

\subsection{Atom optics \label{Sec-AtomOptics}}

Coherent atomic excitation involves photon absorption and emission and hence it is always accompanied by transfer of photon momenta to the atoms. This momentum change   is the basis for laser cooling of atoms \mycite{Chu98,Coh98,Phi98,Met12}.
Momentum transfer induced by optical beams is essential for the design of
 mirrors, beam splitters and atom interferometers -- a subject termed {\em atom optics} \mycite{Ada94}.
An atomic beam splitter splits the atomic wavefunction into a coherent macroscopic superposition of two wavepackets propagating in different spatial directions.
An atomic mirror deflects these wavepackets so that the split matter waves can be brought together to interfere, thereby forming an atomic interferometer.
This interference is possible only if the atomic beam splitters and mirrors are coherent.
Because STIRAP features efficient, robust and dissipationless coherent transfer of population and momentum,  it was quickly recognized as a perfect tool for atom interferometry \mycite{Mar91}.

\subsubsection{Atomic mirrors \label{Sec-AtomMirror}}

\begin{figure}[tb]
\includegraphics[width=0.99\columnwidth]{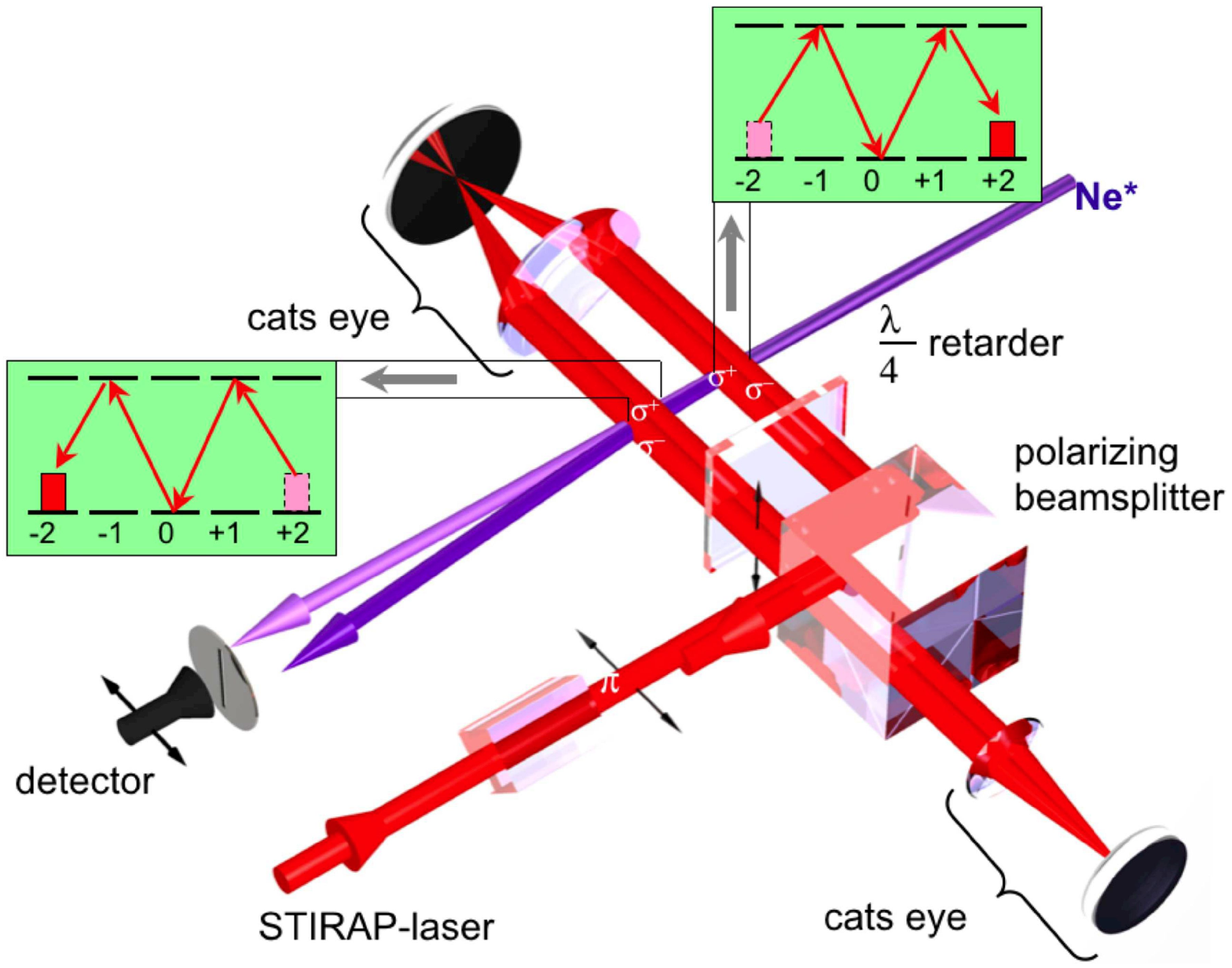}
 \caption{\coloronline Experimental setup for the ${}^{20}$Ne$^*$ atomic mirror  of \mytcite{The98}.
\locationa{Fig.~26 of  \mytcite{Ber98}.}
 }
\label{Fig:Bergmann1998-setup}
\end{figure}

A very convenient system for multiphoton coherent momentum transfer is the chainwise transition formed between the magnetic sublevels of two degenerate levels.
A pair of counterpropagating laser pulses with opposite circular polarizations $(\sigma^+,\sigma^-)$ acts upon a ground level with an angular momentum $J_{g}$  and an excited level with an angular momentum $J_{e}=J_{g}$ or $J_{g}-1$.
Figure ~\ref{Fig:M-chain} shows an example for such a chain for   $J_{g}=J_{e}=2$.
When this system is prepared in one of the end ground sublevels, e.g. $M=J_{g}$, and is driven adiabatically by two laser pulses
in the counterintuitive order ($\sigma^+$ before $\sigma^-$), then multistate STIRAP transfers population from $M=J_{g}$ to $M=-J_{g}$.
If the two fields propagate in the same direction, the momentum kicks from the absorption and stimulated emission processes cancel each other.
If the two pulses propagate in opposite directions, as shown in Fig.~\ref{Fig:Bergmann1998-setup}, then the absorption of a $\sigma^-$ photon and the stimulated emission of a $\sigma^+$ photon each impart to the atom a momentum $\hbar k$ in the direction of the $\sigma^-$ pulse.
 Therefore, in this case the multistate STIRAP is accompanied by a transfer of momentum $4\hbar k$ in the direction of the $\sigma^-$ pulse, a deflection that acts as an atomic mirror.

\begin{figure}[tb]
\includegraphics[width=0.70\columnwidth]{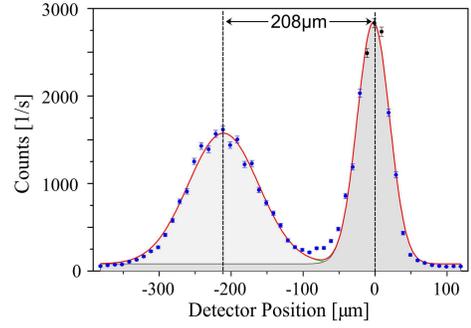}
\caption{\coloronline
Deflection of a beam of ${}^{20}$Ne$^*$ atoms due to transfer of eight photon momenta after double adiabatic passage from the $M = 2$ sublevel to $M = -2$ and back to $M = 2$.
The narrow undeflected original distribution is observed due to the presence of ${}^{22}$Ne isotope atoms that are insensitive to the light.
The width of the deflected beam is broader because the deflection angle depends on the velocity of the particles.
\location{Fig.  7 of \mytcite{The98}.}
 }
\label{Fig:Theuer1998-results}
\end{figure}

Figure \ref{Fig:Bergmann1998-setup} illustrates the setup in the Ne$^*$ experiment of \mytcite{The98}.
A beam of metastable ${}^{20}$Ne$^*$ atoms, prepared in the $M=+2$ magnetic sublevel of the metastable ${}^{3}P_2$ level crossed two zones of two displaced but partly overlapping circularly polarized cw laser beams in the counterintuitive order ($\sigma^+\sigma^-$).
In the first interaction zone the population was adiabatically transferred to the $M=-2$ sublevel of the $^{3}P_{2}$ level with nearly 100\% efficiency, without residing at any time in the decaying upper level ${}^{3}D_{2}$.
Because the two laser beams were counterpropagating each atom received a total momentum of $4\hbar k$ in the direction of the $\sigma^{-}$ beam.
Then the atoms encountered a second interaction zone, with beams ordered oppositely to the first zone ($\sigma^-\sigma^+$), which enabled a second multistate STIRAP transfer that brought the atoms back into the initial state with the transfer of   another momentum kick of $4\hbar k$ in the direction of the $\sigma^-$ beam.
Figure \ref{Fig:Theuer1998-results} shows the experimental results after the second STIRAP.

STIRAP-based atomic mirrors in similar Zeeman chains have been demonstrated in a number of other experiments.
 \mytcite{Pil93,Val94,Gol94a} achieved momentum transfer of $8\hbar k$, with about 50\% efficiency in the single-pass multistate STIRAP between the $M_{F}=-4$ and $M_{F}=4$ Zeeman sublevels in the hyperfine transition $F_{g}=4\leftrightarrow F_{e}=4$ of the D$_{2}$ line of ${}^{133}$Cs atoms.
 \mytcite{Law94a} demonstrated momentum transfer of $4\hbar k$ with 90\% efficiency after double STIRAP ($M=-1\to M=1\to M=-1$) between the ground-state sublevels in the $2^{3}S_{1}\leftrightarrow 2^{3}P_{0}$ transition of He$^{\ast }$ by using circularly polarized lasers.
They also demonstrated momentum transfer of $6\hbar k$ with 60\% efficiency after a triple pass with linearly polarized lasers.

\subsubsection{Atomic beam splitters and interferometers \label{Sec-AtomBeamSplitter}}

Fractional STIRAP (Sec.~\ref{sec-fractional}) with counterpropagating \emph{P} and \emph{S} fields has been a popular tool for creation of atomic beam splitters, because the creation of a coherent superposition of two atomic states is accompanied by splitting of the initial momentum distribution into two (or more) momenta distributions.
A suitable combination of beam splitters and mirrors make a matter-wave interferometer.

\begin{figure}[tb]
\includegraphics[width=0.65\columnwidth]{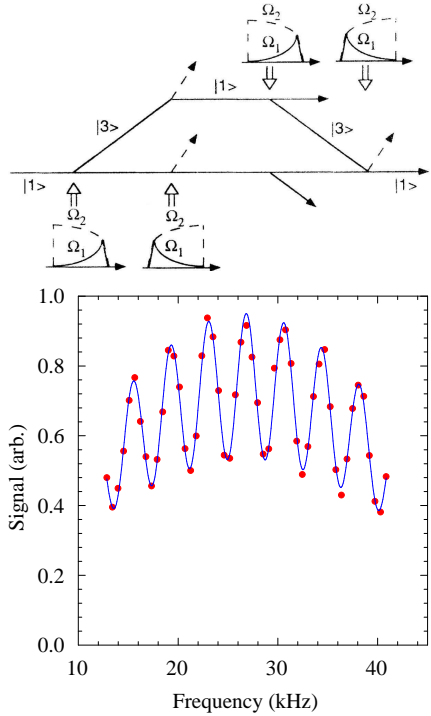}
\caption{\coloronline
\emph{Top}: Scheme for an atomic interferometer using STIRAP with ${}^{133}$Cs atoms.
  The double arrows show the propagation axis of the light {and the solid lines are the atomic trajectories.}
\emph{Bottom}: Interference fringes {in the population of the ${6S_{1/2},F=3,M_{F}=0}$ sublevel vs the Raman frequency difference of the \emph{P} and \emph{S} pulses.}
\locationa{Figs.~2 and 4 of  \mytcite{Wei94}.}
 }
\label{Fig:Weitz1994}
\end{figure}

 \mytcite{Wei94,Wei94PRA} demonstrated the first STIRAP-based atomic interferometer on the transition between the two ${}^{133}$Cs hyperfine ground states ${6S_{1/2},F=3,M_{F}=0}$ and ${6S_{1/2},F=4,M_{F}=0}$ via the excited state $6P_{1/2},F=3$   (or 4), $M_{F}=1 $.
They used a  $\sigma^{+}\sigma^{+}$ polarization configuration, a choice that makes the transfer insensitive to magnetic fields -- an essential property for precision interferometry.
This atom interferometer has the Bord\'{e}'s four-$\pi /2$ geometry \mycite{Bor89,Rie91} and involved four sequential atomic beam splitters, as shown in Fig.~\ref{Fig:Weitz1994} (top).
Each of the beam splitters used half-STIRAP with counterpropagating laser pulses {with 95\% efficiency.}
The observed interference {fringes are} shown in Fig.~\ref{Fig:Weitz1994} (bottom).

Burnett  and coworkers \mycite{Fea96,Fea98,God99,Web99}
  have built atomic interferometers with ${}^{133}$Cs atoms, with both $\sigma^+ \sigma^-$ and $\sigma^+ \pi$ polarizations.
 They used a combination of microwave fields for ground-state manipulations and laser fields for momentum transfer by STIRAP.
 First, a $\pi /2$ microwave pulse was used to create a  superposition of the $F=3,M=0$ and $F=4,M=0$ sublevels.
Then another $\pi /2$ pulse was applied to induce Ramsey fringes.
Between the $\pi/2$ pulses,
 two orthogonally propagating and partly overlapping pulses of $\sigma^+$ and $\pi$ polarizations in the  $\sigma^+ \pi$ case transferred the population from $F=4,M=0$ to $F=4,M=4$ in an eight-photon STIRAP via the sublevels of the upper 6P$_{1/2},F=4$ level.
 Then STIRAP was reversed and the population was returned back to $M=0$.
   This atom interferometer was used  for measuring the Berry phase \mycite{Web99} and the temperature of an atomic ensemble \mycite{Fea98}.

\subsubsection{Coherent manipulation of laser-cooled and trapped atoms \label{sec-laser-cooled}}

\begin{figure}[tb]
\includegraphics[width=0.85\columnwidth]{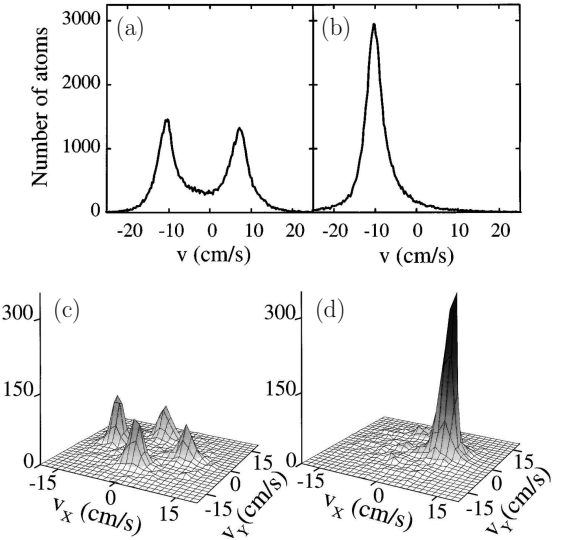}
\caption{
Momentum transfer in subrecoil laser cooling peaks in the experiment of  \mytcite{Kul97}.
\emph{Top}: 1D laser cooling.
\emph{Bottom}: 2D laser cooling.
\location{Figs.~1 and 2 of  \mytcite{Kul97}}
}
\label{Fig:Kulin1997}
\end{figure}

STIRAP has been used to coherently manipulate the atomic wavepackets resulting from subrecoil laser cooling by velocity-selective coherent population trapping (VSCPT)
\mycite{Asp88, Asp89,Kas91a,Kas91b,Kas92,Law95,Chu98,Coh98,Phi98}.
In 1D the momentum distribution of atoms cooled by VSCPT has two peaks, at $+\hbar k$ and $-\hbar k$ (both narrower than the photon recoil momentum $\hbar k$), which correspond to the two components of the dark state.
By gradually lowering the amplitude of one of the components of the standing wave used for VSCPT, which amounts to half-STIRAP, \mytcite{Ess96}
 coherently transferred ${}^{87}$Rb atoms  into a single momentum state, still with a subrecoil momentum spread.
In another landmark experiment,  \mytcite{Kul97} used adiabatic transfer of He$^*$ atoms into a single wavepacket in 1D (with nearly 100\% efficiency), 2D (90\% efficiency) and 3D (75\% efficiency).
Figure \ref{Fig:Kulin1997}   shows results of this experiment in 1D and 2D.
A distribution characterized initially by   two (1D) and four (2D) peaks [frames (a,c)] is transferred to a single momentum peak [frames (b,d)].

For such wavepacket manipulation, the coherence of the two initial momentum components is crucial.
Hence, this operation can be used to prove the coherence of the two momentum peaks at $\pm\hbar k$ \mycite{Ess96}.

\subsubsection{Measurement of weak magnetic fields \label{Sec-Larmor}}

\begin{figure}[tb]
\includegraphics[width=0.75\columnwidth]{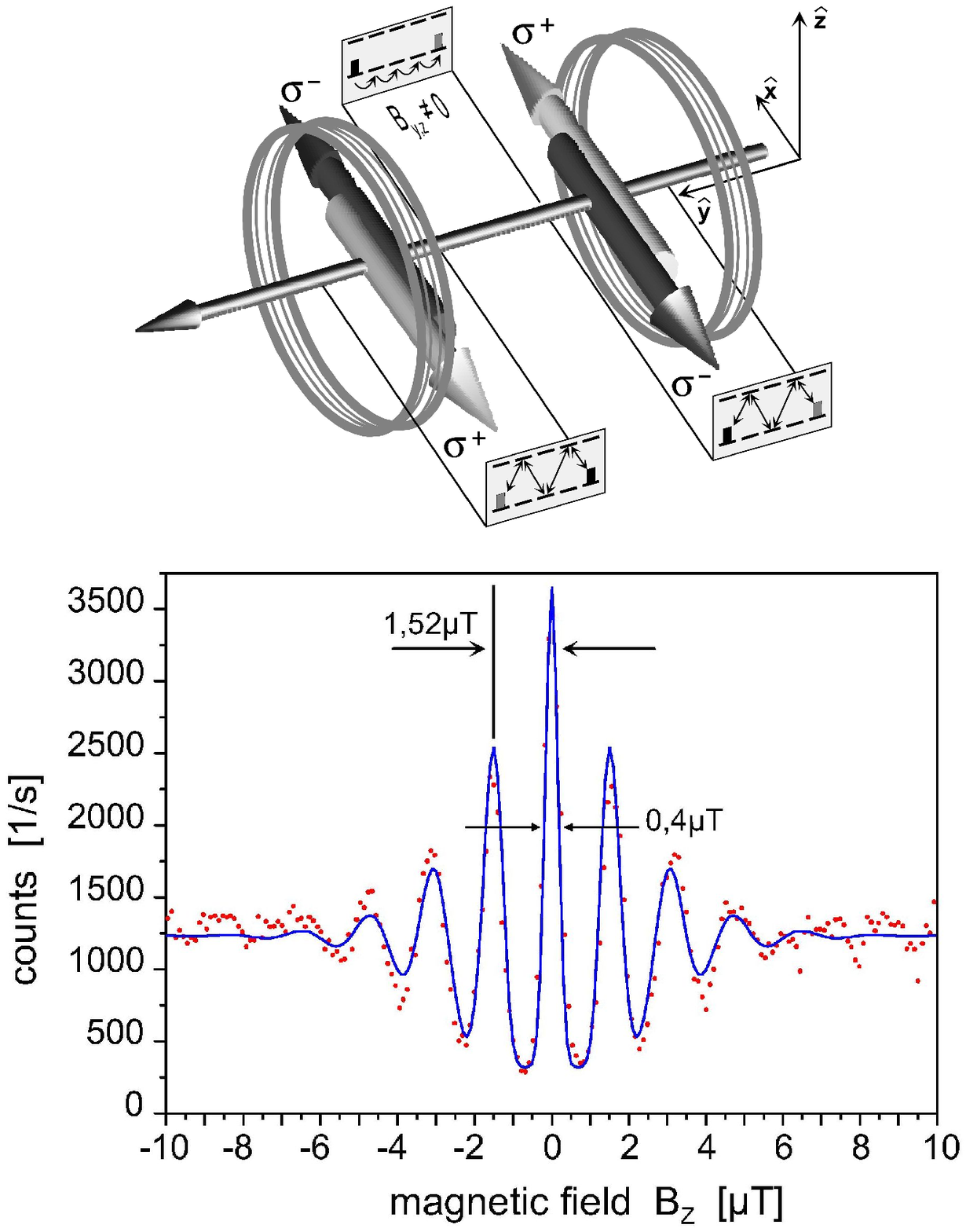}
\caption{\coloronline Larmor velocity filter:
 experimental setup (top) and variation of the flux of deflected Ne$^*$ atoms with the magnetic field (bottom).
\location{Figs.~10 and 12 of  \mytcite{The98}.}
\Also{Fig. 34 of \mytcite{Vit01a} and Fig. 19 of \mytcite{Vit01b}. }
}
\label{Fig:Theuer1998-larmor}
\end{figure}

The atomic beam deflection by coherent momentum transfer was used by   \mytcite{The98}    to design a technique termed a ``Larmor velocity filter'' for measuring    small magnetic fields along the axis of the atomic beam.
The scheme consists of two STIRAP zones through which the beam travels,
see Fig.~\ref{Fig:Theuer1998-larmor} (top).
In the first zone, Ne$^*$ atoms were prepared in the $M=2$ sublevel of the $^{3}P_{2}$ metastable state and transferred to the $M=-2$ sublevel.
In the second zone, the atoms that had remained in the $M=-2$ sublevel were transferred back to the initial $M=2$ sublevel.
The magnetic field in the region between the two zones caused Larmor precession, thereby altering the   populations of  magnetic sublevels and affecting the momentum transfer in the second zone.
The resulting narrow-peaked interference pattern, an example of which is shown in Fig.~\ref{Fig:Theuer1998-larmor}(bottom), permits measurement of weak magnetic fields.

\subsection{Single-atom cavity quantum electrodynamics \label{Sec:vacuum-STIRAP}}

\subsubsection{Cavity-STIRAP}

The first application of STIRAP beyond laser-driven atoms and molecules in free space was proposed by   \mytcite{Par93} in cavity quantum electrodynamics (QED).
They   proposed to create coherent superpositions of photon number (Fock) states of the cavity mode by mapping a coherent superposition of Zeeman atomic sublevels onto the cavity field.
\mytcite{Lan00} extended this idea to two degenerate cavity modes of orthogonal polarizations.

\begin{figure}[tb]
\includegraphics[width=0.99\columnwidth]{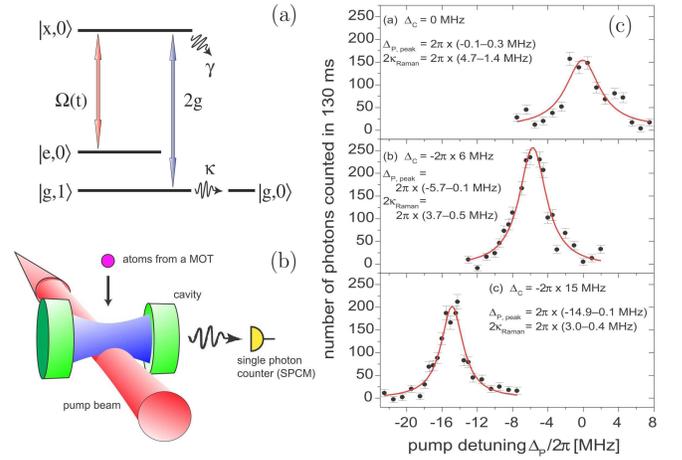}
\caption{\coloronline
(a) Scheme for vacuum-STIRAP.
 The labels $g,e,x$ refer to atomic levels and 0,1 refer to the cavity photon number.
 The initial state is $\ket{e,0}$ and the final state, producing the outgoing photon, is $\ket{g,1}$.
 The pulsed \pump field is $\Omega\t$, the \Stokes field is the vacuum coupling $2g$.
%
(b) Experimental setup used for demonstration of vacuum-STIRAP.
(c) Number of counted photons vs \pumpx-field detuning for different cavity detunings $\Delta_C$ (corresponding to $\Delta_S$).
Maximum at $\Delta_P=\Delta_C$ signals vacuum-STIRAP.
\location{Fig.~1 of  \mytcite{Vas10} and Figs. 4 and 5 of  \mytcite{Hen00}.}
}
\label{Fig:Hennrich-1}
\end{figure}

In cavity-STIRAP (or vacuum-STIRAP), a laser beam excites one branch of the Raman transition (usually \pumpx) of a single atom, while the cavity vacuum stimulates the emission of the photon on the other branch (usually \Stokesx), Fig.~\ref{Fig:Hennrich-1}(a).
The quantized field of the single-mode cavity provides the \Stokes coupling (a vacuum-Rabi frequency) denoted by $g(t)\sqrt{n+1}$, where $n$ is the number of photons in the cavity mode and $g(t)$ is the coupling strength in vacuum ($n=0$).
The \pump laser beam is focused inside the cavity but slightly below the cavity axis, as shown in Fig.~\ref{Fig:Hennrich-1}(b).
Therefore the \pump and \Stokes fields are seen as a counterintuitive pulse sequence by an atom falling through the cavity.

The dynamics is described by the combined atom-photon states $\ket{\psi,n}=\ket{\psi}\ket{n}$.
In RWA, only three such atom-field states are coupled: $\ket{\psi_1,n}$, $\ket{\psi_2,n}$, $\ket{\psi_3,n+1}$.
The dark atom-field state corresponds to energy $E_n=\hbar n \omega$, with $\omega$ being the cavity mode; it reads
\be
\ket{E_n} = \frac{2g(t)\sqrt{n+1}\ket{\psi_1,n} - \Omega_P(t)\ket{\psi_3,n+1}}{\sqrt{4(n+1)g(t)^2 + \Omega_P(t)^2}}.
\ee
In the adiabatic limit, complete decoherence-free transfer $\ket{\psi_1,n}\to\ket{\psi_3,n+1}$ is achieved, without populating the decaying excited state $\ket{\psi_2,n}$ at any time.
Because of the quantized cavity field the usual adiabatic condition becomes $\Omega_P T_P \gg 1$  for the pump field, and $2g_{\max} T (n+1) \gg 1$ for the Stokes field.
For $n=0$ (empty cavity initially), a single-photon state is created out of the vacuum after the atom passes through the cavity.
If the atom arrives in a coherent superposition of Zeeman sublevels then cavity-STIRAP will produce a coherent superposition of Fock states.
The transfer of coherence from an atom  to a field mode is reversible; likewise, it allows the mapping of cavity field onto atomic ground-state coherence,  which has been suggested as a method for measuring cavity fields \mycite{Par95}.


\begin{figure}[tb]
\includegraphics[width=0.75\columnwidth]{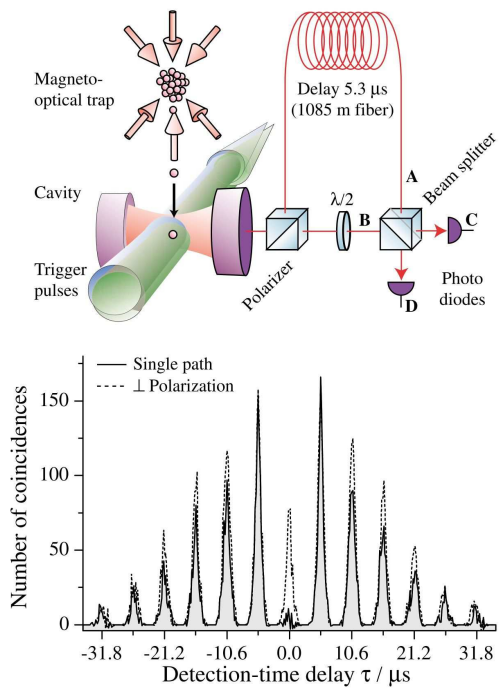}
\caption{\coloronline
\emph{Top}: An atom-cavity system emits unpolarized single photons via vacuum-STIRAP.
 A photon traveling along path A is delayed so that it impinges on the BS simultaneously with a subsequent photon traveling along path B.
\emph{Bottom}: Number of coinciding photodetections in the two output ports vs the time difference between the detections.
When only a single path is open antibunching is observed due to Hanbury-Brown-Twiss interference (solid curve).
When both paths are open but have perpendicular polarizations
no interference occurs and the
beam splitter randomly directs the photons to the photodiodes (dashed line).
\location{Fig.~2 of   \mytcite{Leg04}.}
}
\label{Fig:Legero2004a}
\end{figure}

\subsubsection{Generation of single photons and photon networks}

Cavity-STIRAP has been used in numerous experiments by Rempe, Kuhn and co-workers.
Cavity-STIRAP with a single cold rubidium atom passing through a high-finesse cavity was demonstrated experimentally by \mytcite{Hen00}.
Figure \ref{Fig:Hennrich-1} shows the idea of the experiment and the observed efficiency of single-photon generation versus the pump detuning with peaks on two-photon resonance.
This experiment did not operate as a single-photon source because its continuous driving simply mapped the (Poissonian) atom statistics to the photons.
\mytcite{Kuh02}  constructed a deterministic single-photon source in a similar experiment by using a pulsed driving and a pulsed recycling.
This made it possible to produce, on demand, a stream of single-photon pulses from the same atom, see Fig.~\ref{Fig:Legero2004a}.
\mytcite{Leg04} observed quantum bits of two single photons generated by vacuum-STIRAP.
\mytcite{Hen05} studied the statistics of the photons emitted by a single atom and observed the transition from antibunching to bunching.
 \mytcite{Wil07PRL} produced a stream of single photons with alternating circular polarization by alternately exposing the atom to laser pulses of two different frequencies and polarizations.

 \mytcite{Wil07S} built the basic element of a distributed quantum network: an atom-photon quantum interface that entangles a single atom with a single photon and maps the quantum state of the atom onto a second single photon, thereby producing an entangled photon pair.
 \mytcite{Rit12} proposed a quantum network architecture based on single atoms embedded in optical cavities.
 Quantum connectivity between the nodes is achieved by the coherent exchange of a single photon,
 with vacuum-STIRAP playing the central role in the sending, receiving, storing and releasing the photons.
A proof-of principle experiment demonstrated the transfer of an atomic quantum state and the creation of entanglement between two identical nodes in separate laboratories connected by an optical fiber of 60 m length.
\mytcite{Nol13} demonstrated teleportation of quantum bits between two single atoms in distant laboratories with a fidelity of 88\%.
%

\subsubsection{Shaping photons}

\begin{figure}[tb]
\includegraphics[width=0.95\columnwidth]{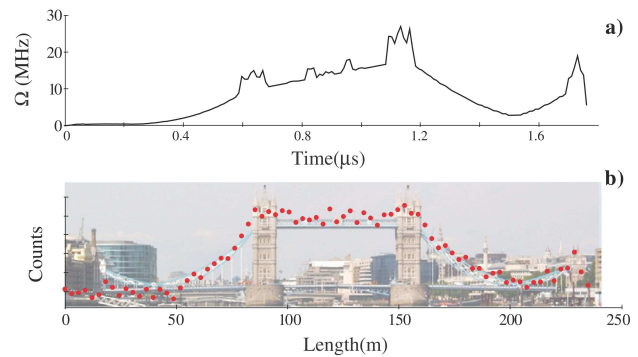}
\caption{\coloronline
 (a) Control-laser-field shape, which produces a photon shape (b)  mimicking the Tower Bridge in London.
\locationa{Fig.~6 of  \mytcite{Nis11}.}
}
\label{Fig:Kuhn}
\end{figure}

\mytcite{Vas10} developed a vacuum-STIRAP based technique, which produces single photons of arbitrary pre-defined shape by tailoring the shape of the pump laser pulse.
The control of single-photon shapes is an important tool for use in  quantum teleportation and quantum memories, which are essential elements in quantum cryptography and quantum computing.
The desired photon shape is imposed a priori and because it is proportional to the probability amplitude $c_g(t)$ of state $\ket{g,1}$ [Fig.~\ref{Fig:Hennrich-1}(a)], the Schr\"odinger equation becomes an equation for the laser-field shape $\Omega(t)$.
   \mytcite{Nis11} demonstrated this in a proof-of-principle experiment, cf. Fig.~\ref{Fig:Kuhn}, and then in another experiment to create photonic qubits, qutrits and ququads -- photons divided into 2, 3 and 4 time bins \mycite{Nis13}.
 \mytcite{Dil12} provided also an analytic solution for the temporal shape of the control laser field needed for the inverse problem: capture a single photon with one atom coupled to an optical cavity.

\subsection{Trapped ions  \label{sec-trapped-ions}}

\begin{figure}[tb]
\includegraphics[width=0.60\columnwidth]{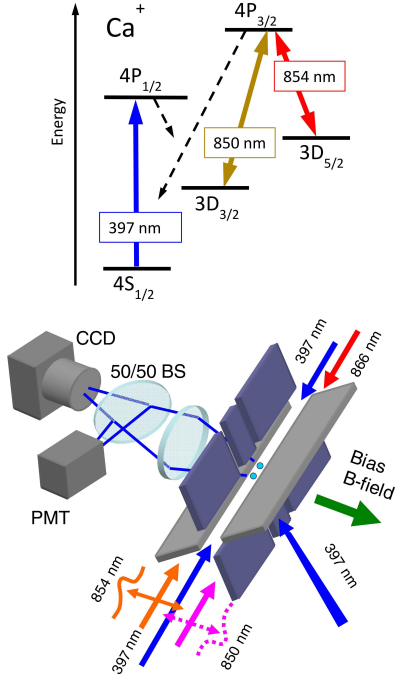}
\caption{\coloronline  
\emph{Top}: Level scheme  used for STIRAP in ${}^{40}$Ca$^+$ ions.
Population is transferred between the sublevels of the metastable levels 3D$_{3/2}$ and 3D$_{5/2}$ via the excited sublevels of 4P$_{3/2}$.
\emph{Bottom}: Experimental setup of ${}^{40}$Ca$^+$ ions trapped in a segmented linear Paul trap.
The detection is performed via light-induced fluorescence collected by a CCD camera and a photomultiplying tube (PMT).
\locationa{Fig.~2 of  \mytcite{Sor06}.}
}
\label{Fig:Sorensen2006}
\end{figure}

\begin{figure}[tb]
\includegraphics[width=0.90\columnwidth]{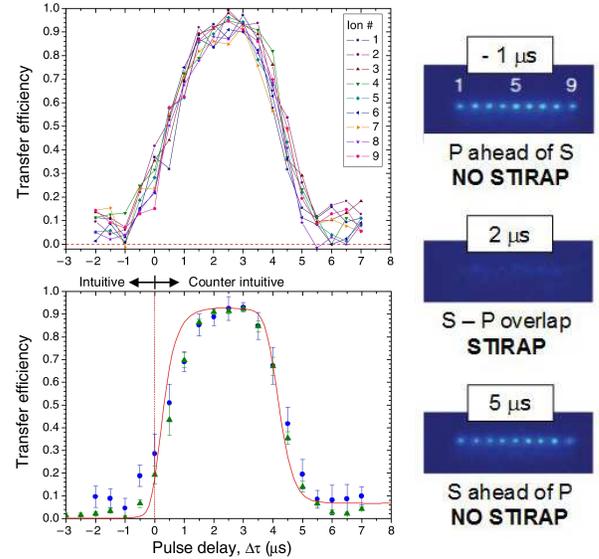}
\caption{\coloronline
\emph{Left}: Population transfer efficiency in the system of Fig.~\ref{Fig:Sorensen2006} vs the pulse delay for a string of nine trapped ions.
with results for the individual ions (upper left) and the average over all ions (lower left).
\emph{Right}: CCD images of the fluorescence from the middle level 4P$_{3/2}$ for the string of nine ions for various time delays between the \pump and \Stokes pulses.
 The absence of fluorescence signals high transfer efficiency.
The maximum transfer occurs for delays between 2 and  3 $\mu$s.
 \locationa{Fig.~3 of  \mytcite{Sor06}.}
 }
\label{Fig:Sorensen2006-3}
\end{figure}

STIRAP has also become a popular tool for coherent control of trapped ions.
STIRAP was used for efficient qubit manipulation \mycite{Sor06} and detection \mycite{Mol07} of ${}^{40}$Ca$^+$ ions trapped in a segmented linear Paul trap.
 Population transfer between the metastable levels 3D$_{3/2}$ and 3D$_{5/2}$ proceeds via the 4P$_{3/2}$ level, see Fig.~\ref{Fig:Sorensen2006}.
The \pump and \Stokes pulses were provided by lasers with wavelenghts of  850 nm and 854 nm, respectively, while the other lasers indicated in Fig.~\ref{Fig:Sorensen2006} were used for preparation and detection.
Population transfer efficiency of over 90\% was achieved, see Fig.~\ref{Fig:Sorensen2006-3}.
 What makes this figure remarkable 
  are the upper frame and the CCD images on the right, which show that each individual ion undergoes successful STIRAP.

\begin{figure}[tb]
\includegraphics[width=0.90\columnwidth]{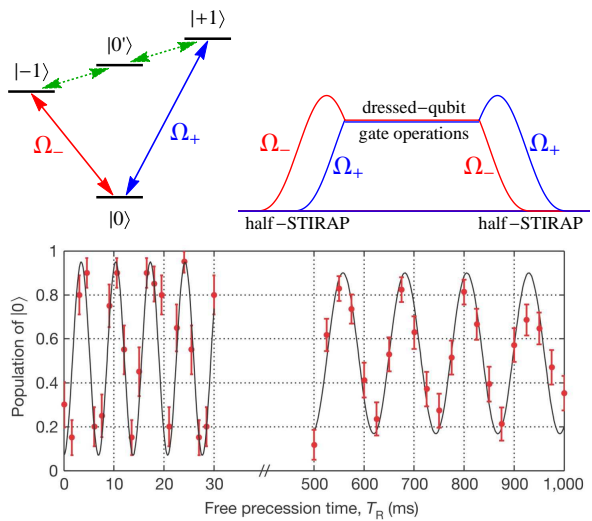}
\caption{\coloronline
\emph{Upper left}: Linkage pattern used  in the experiment by  \mytcite{Tim11}.
\emph{Upper right}: Pulse timing.
The dressed state $\ket{D}$ (or $\ket{B}$) is prepared by half-STIRAP with two microwave fields starting from sublevel $\ket{+1}$.
Then the amplitudes of the two microwave fields are held equal and constant, while the dressed qubit is driven by an rf field coupling the transitions  $\ket{0^\prime}\fromto\ket{-1}$ and $\ket{0^\prime}\fromto\ket{+1}$.
In the end, another half-STIRAP is applied for the detection stage.
\emph{Lower panel}: Ramsey fringes verifying the coherence of the dressed qubit $\{\ket{0^\prime},\ket{D}\}$.
\locationa{Figs.~1, 2, and 3 of  \mytcite{Tim11}.}
}
\label{Fig:Timoney2011}
\end{figure}

Half-STIRAP 
 was used recently in two experiments that demonstrated manipulation of a dressed-state qubit formed of hyperfine sublevels of a ${}^{171}$Yb$^+$ ion in a linear Paul trap \mycite{Tim11,Web13}, cf. Fig.~\ref{Fig:Timoney2011}. 
Instead of bare atomic states the qubit is constructed from state $\ket{0^\prime}$ and the dressed state $\ket{B} = (\ket{+1} + \ket{-1})/\sqrt{2}$ or $\ket{D} = (\ket{+1} - \ket{-1})/\sqrt{2}$.
The dressed qubit is insensitive to magnetic field fluctuations, and its coherence exceeds the coherence time of the bare-state qubit by three orders of magnitude, from the milliseconds to the seconds range,  cf. the Ramsey fringes in Fig.~\ref{Fig:Timoney2011} (bottom).
 Half-STIRAP was used twice: to populate the dressed state $\ket{D}$ (or $\ket{B}$) initially and then to drive it back to one of the bare states for detection in the end.
Between the two half-STIRAP zones, \mytcite{Tim11} manipulated the dressed qubit $\{\ket{0'},\ket{D}\}$ (or $\{\ket{0'},\ket{B}\}$) by driving both transitions $\ket{0^\prime}\fromto\ket{-1}$ and $\ket{0^\prime}\fromto\ket{+1}$ by a rf field.
\mytcite{Web13} made use of a stronger magnetic field and the ensuing larger second-order Zeeman shift to drive $\ket{0}\fromto\ket{+1}$ only (and hence avoid the phase-sensitive closed loop formed when driving both rf transitions).
Recently,  \mytcite{Ran15}   extended this method to prepare and detect  all three magnetic-insensitive dressed states of the system, thereby forming a magnetic-field insensitive qutrit.
\mytcite{Tim11} and \mytcite{Coh15} also described protocols for constructing conditional two-qubit gates with the decoherence-free qubits.
In another development,  \mytcite{Wei15}  used rf pulses to perform  ground-state cooling of  a dressed-state qubit, reaching $\ket{n=0}$ population of 88\%, which allowed them to drive Rabi oscillations between vibrational levels $\ket{0}$ and $\ket{1}$.

Very recently   \mytcite{Geb16}  used STIRAP for manipulation and detection of the motional phonon number associated with two hyperfine ground states of a trapped ${}^{25}$Mg$^+$ ion.
The insensitivity of STIRAP to the value of the couplings made it possible to remove a single phonon independent of the initial (thermal of Fock) motional state, an objective that is impossible with a resonant $\pi$ pulse.
They used the technique to measure the wavefunction overlap of excited and ground motional states, which is an important ingredient of force sensing.

\subsection{Molecules \label{Sec:Molecules}}

Molecules provided the initial physical platform for demonstration of STIRAP by Bergmann and co-workers, and STIRAP still remains a popular tool for their studies.
Both \textsc{cw} and pulsed lasers were used in the early experiments, which were motivated by the desire to study chemical reactions dynamics.
In the last decade, STIRAP has become an enabling technology for transferring  ultracold Feshbach molecules into their rovibrational ground state, see Sec.~\ref{sec-ultracold molecules}.

\subsubsection{Early experiments with molecular states \label{Sec:Molecules-1990}}

 After preliminary results \mycite{Gau88} the first comprehensive experimental demonstration of STIRAP was achieved by Bergmann and coworkers in Na$_2$ \mycite{Gau90} in a crossed-beams geometry.
 The experiment achieved nearly complete population transfer from the initial level ($v=0,J=5$) to the final level ($v=5,J=5$) of the molecules in their electronic ground state $X\ ^{1}\Sigma _{g}^{+}$.
 Because Na$_2$ has relatively strong transition moments, only moderate laser intensities of about 100 W/cm$^{2}$, produced by \textsc{cw} laser beams mildly focused to a spot diameter of a few hundred $\mu $m into the molecular beam, were needed to guarantee large pulse areas.

\begin{figure}[tb]
\includegraphics[width=0.57\columnwidth]{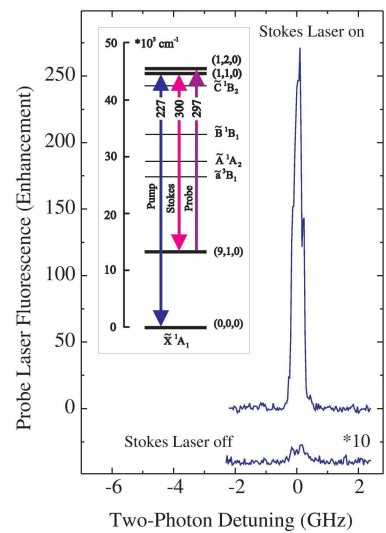}
\caption{\coloronline
Experimental demonstration of STIRAP in SO$_2$ with ultraviolet transitions (227 nm for the  pump field and 300 nm for the  Stokes field), with pulse durations of 2.7 ns for the  pump pulse and 3.1 ns for the  Stokes pulse.
Typical laser intensities were 10 MW/cm$^2$, yielding Rabi frequencies of about $10^{10}$ s$^{-1}$.
The curves show the transfer efficiency vs. the two-photon detuning with the  Stokes laser on and off.
Inset: Relevant energy-level scheme and linkage pattern.
 \locationa{Figs. 1, 2  and 3 of \mytcite{Hal96}.}
}
\label{Fig:Halfmann1996}
\end{figure}

The first implementation of STIRAP with pulsed lasers was demonstrated with nanosecond pulses in the electronic ground state of ${}^{14}$N${}^{16}$O molecules \mycite{Sch93,Kuh98}, see Fig.~\ref{Fig:Schiemann1993}.
The most complex molecule, in which STIRAP has been successfully implemented to populate very high-lying vibrational levels, is SO$_2$ \mycite{Hal96}.
The large density of levels results in much smaller transition dipole moments compared to atoms or diatomic molecules,
 which demand much higher laser power to enforce adiabatic evolution.
Figure \ref{Fig:Halfmann1996} shows nearly 100\% efficiency of population transfer from the rotational level $3_{03}$ of the vibrational ground state $(0,0,0)$ to the same rotational level $3_{03}$ of the $(9,1,0)$ overtone in the electronic ground state $X\ {}^1A_1$ via the vibrational level $(1,1,0)$ of the excited electronic state $C\ {}^1B_2$.
Figure \ref{Fig:Halfmann1996} displays the relevant level diagram and the probe-laser-induced fluorescence from the final state when only the pump pulse was present, and when both the pump and Stokes laser pulses were applied in the counterintuitive order.
The signal increased by more than two orders of magnitude in the latter case.

\subsubsection{Formation of ultracold molecules \label{sec-ultracold molecules}}

\begin{figure}[tb]
\includegraphics[width=0.80\columnwidth]{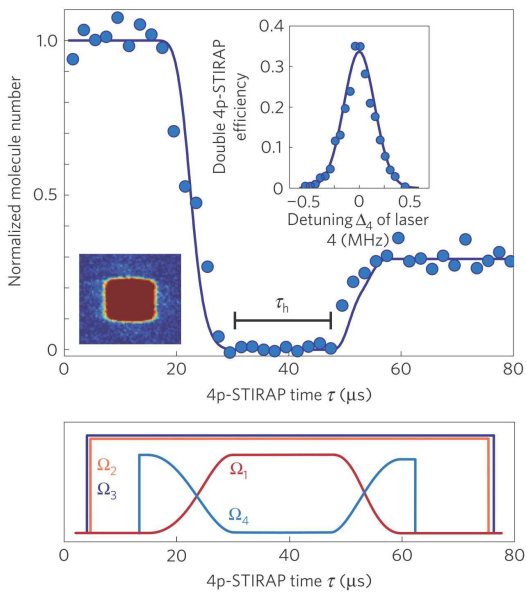}
\caption{\coloronline
Four-photon straddle-STIRAP transfer to the rovibronic ground state $\ket{v = 0, J = 0}$ and back in ultracold Cs$_2$ molecules in the experiment of
 \mytcite{Dan10}.
\emph{Top}: Transfer efficiency vs time (main frame) and vs detuning $\Delta_4$ of the laser on the last transition (inset).
\emph{Bottom}: Temporal pulse shapes.
 \location{Fig.~3 of  \mytcite{Dan10}.}
}
\label{Fig:Danzl2010}
\end{figure}

Laser cooling has had a major impact on atomic physics by making possible atom and ion trapping, quantum degenerate gases, and quantum information processing with atoms and ions.
Laser cooling methods, however,  are not generally applicable to molecules due to their rovibrational structure and the ensuing absence of closed two-level transitions.
Hence a major route to ultracold molecular gases is by association of ultracold atoms.
Soon after the first creation of Bose-Einstein condensate (BEC) in the gas phase it was proposed to use two-color photoassociation with counterintuitively ordered pulses \mycite{Jav99,Var99,Mac00,Dru03,Kuz09}.
The challenges in the application of STIRAP to atoms$\to$molecule conversion is the smallness of the free-bound dipole matrix elements.
It has been suggested \mycite{Mac00} that such a conversion could still be possible thanks to Bose-enhancement of the free-bound couplings.

In fact, \mytcite{Sag05} demonstrated photoassociation of ultracold Rb and Cs atoms to form RbCs molecules in $v = 0$  of the electronic ground state using stimulated emission pumping (SEP), an incoherent process \mycite{Kit81,Bar13}.
At the same time, \mytcite{Win05} documented dark resonances in a superposition of Rb BEC and degenerate gas of Rb$_2$ ground-state molecules, thereby providing strong evidence that achieving coherent population transfer is possible.
 \mytcite{Win07} were the first to report STIRAP-based transformation of  Rb$_2$ Feshbach molecules to chemically-stable bound molecules, although yet with substantial vibrational excitation.

A series of experiments followed, in which, starting from Feshbach states formed in atomic BECs by suitable magnetic tuning, the system was transferred by  STIRAP into

\begin{itemize}

\item the vibrational level $v'' = 73$ of the electronic ground state of  ${}^{133}$Cs$_2$ \mycite{Dan08};

\item the rovibrational ground level  of the lowest triplet or singlet electronic state of   $^{40}$K$^{87}$Rb \mycite{Ni08};

\item the rovibrational ground level of the triplet state of  ${}^{87}$Rb$_2$ \mycite{Lan08}.

\end{itemize}

Later, \mytcite{Dan10} used an optical lattice with two atoms per site, to transfer population into the lowest energy level, including hyperfine energy, of ${}^{133}$Cs$_2$.
For detection via absorption imaging, the molecules were excited again to the weakly bound molecular state and dissociated for detection.
Two versions of STIRAP were used: two sequential two-photon three-level STIRAPs, achieving about 60\% efficiency,  and a single-pass five-state straddle-STIRAP, achieving 57\% efficiency, shown in Fig.~\ref{Fig:Danzl2010}.

\begin{figure}[tb]
\includegraphics[width=0.60\columnwidth]{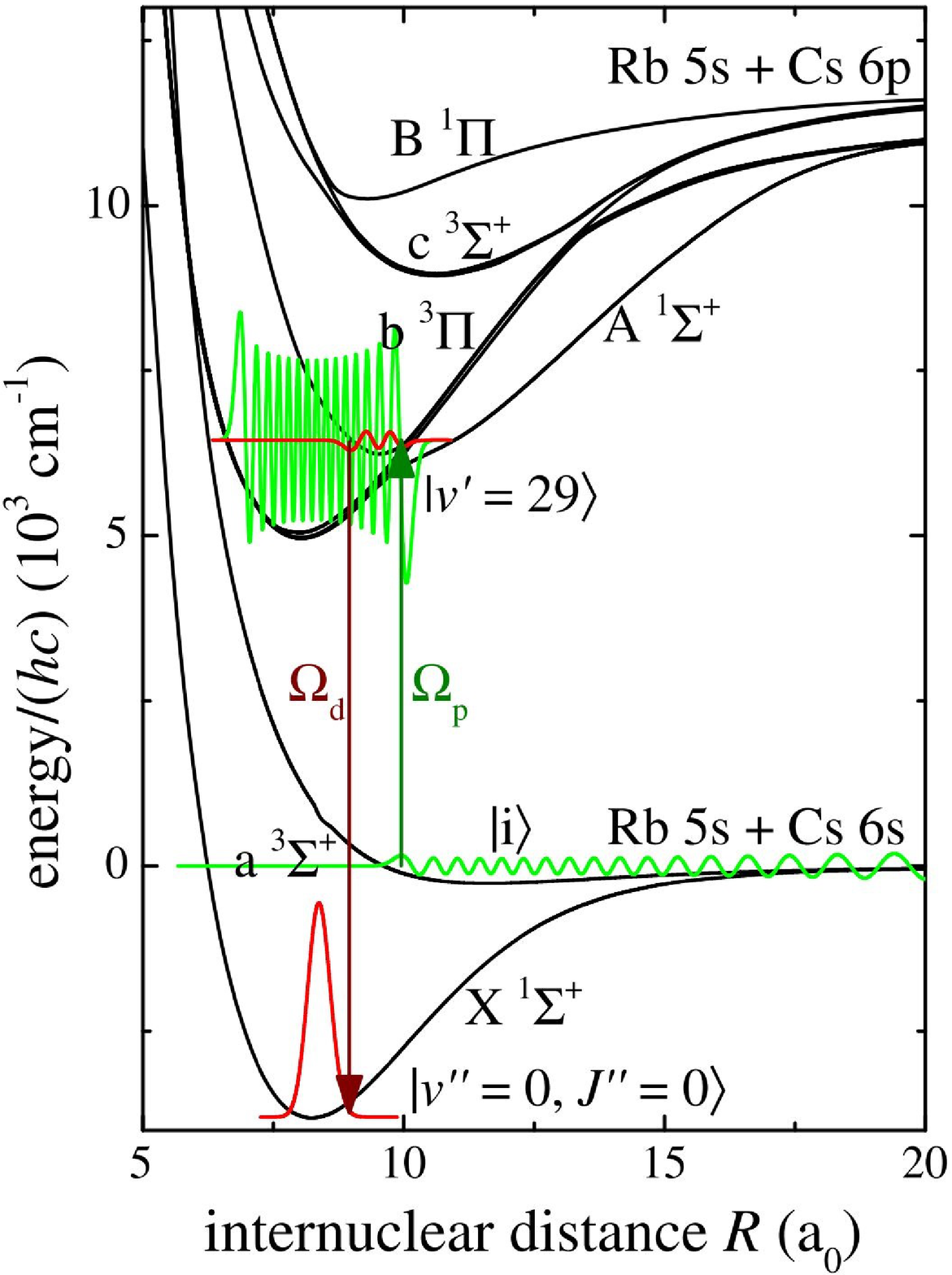}
\caption{\coloronline
 STIRAP scheme used by \mytcite{Tak14} for forming   RbCs molecules in their lowest rovibrational singlet ground state.
The transfer from the Feshbach state $\ket{i}$ to the rovibrational ground-state level
$\ket{v'' = 0,J'' = 0}$  involves the $\ket{v' = 29}$ level belonging to the $b ^3\Pi (\Omega = 1)$ electronically excited state.
The red and green lines show the wavefunctions that are coupled by the STIRAP  lasers \pump and \Stokes  with Rabi frequencies $\Omega_p$  and $\Omega_d$, respectively.
\location{Figs.~1 of  \mytcite{Tak14}.}
 }
\label{Fig:Takekoshi2014-a}
\end{figure}

\begin{figure}[tb]
\includegraphics[width=0.90\columnwidth]{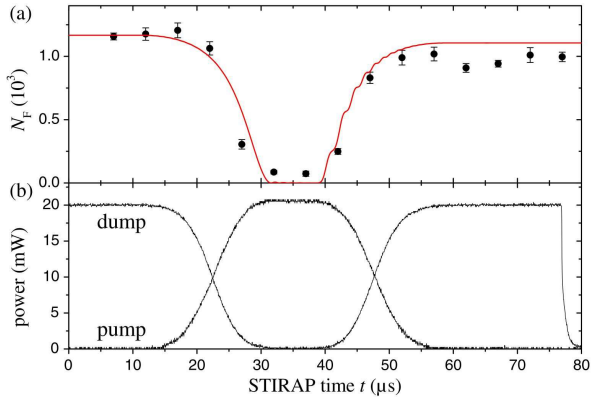}
\caption{\coloronline Formation of RbCs molecules in their lowest rovibrational singlet ground state by STIRAP in the experiment of  \mytcite{Tak14}.
(a) Population histories of the Feshbach state.
 After the first STIRAP process, the Feshbach molecules disappear.
A second STIRAP process drives the population from the lowest-lying molecular level back to the Feshbach state for detection.
 The one-way STIRAP efficiency is 90\%.
(b) Timing of the laser pulses.
 \location{Fig.~2 of  \mytcite{Tak14}.}
 }
\label{Fig:Takekoshi2014-b}
\end{figure}

 An approach using photoassociation of atoms in a magneto-optical  trap prior to the STIRAP transfer into the  rovibronic ground state of  $^{41}$K$^{87}$Rb was used by \mytcite{Aik10}.
Photoassociation directly through STIRAP, starting from pairs of  ${}^{84}$Sr atoms in the ground state of the wells of an optical lattice was successfully demonstrated by \mytcite{Ste12}.
The ${}^{84}$Sr$_2$  molecules were formed in a vibrational level close to the dissociation limit.

Recently \mytcite{Tak14}    reported the creation of ultracold dense samples of ${}^{87}$Rb${}^{133}$Cs molecules in their rovibrational and hyperfine-singlet ground state, i.e. in the absolutely lowest molecular state.
Figure \ref{Fig:Takekoshi2014-a} shows the excitation scheme and
Fig. \ref{Fig:Takekoshi2014-b} shows an example of STIRAP transfer.
At about the same time,  \mytcite{Mol14}  achieved similar results with the same molecule.
The two experiments produced a similar number of molecules in the rovibrational ground state (about 1000).
The difference was in the trap geometry.
 \mytcite{Mol14}   used a 3D optical trap, while  \mytcite{Tak14}   used a lattice of 2D pancake-shaped traps.
As a consequence, the latter authors  could reach higher efficiency (90\%), while the former authors could apply a larger electric field, thereby achieving measurement of the ground-state dipole moment with smaller uncertainties and realization of larger laboratory-frame electric dipole moments (up to 0.35 Debye).

Very recently  \mytcite{Par15}   reported the creation of an ultracold dipolar gas of fermionic ${}^{23}$Na${}^{40}$K molecules in their absolute rovibrational and hyperfine ground state by using a similar Feshbach-STIRAP approach.
Because this molecule is stable against two-body chemical reactions \mycite{Zuc10}, a relatively long lifetime (more than 2.5 seconds) was reported.
By applying a homogeneous electric field, a dipole moment of up to 0.8 Debye was achieved.
Results of ultracold ${}^{23}$Na${}^{87}$Rb were also reported \mycite{Guo16}.

Dipolar molecules near absolute zero may enable improved precision measurements of fundamental constants \mycite{Bar14, Mol16}
as well as new schemes in quantum computing \mycite{DeM02b} and quantum simulation of condensed matter materials \mycite{Bar12}.

\subsubsection{Collision dynamics \label{sec-chemical-reaction}}

Soon after the  invention of STIRAP, Bergmann and co-workers applied it to  the study  of several chemical reaction  dynamics.
 \mytcite{Dit92} studied the reaction Na$_2(v'')$ + Cl $ \to$  NaCl + Na$^*$ and monitored the variation of the total rate of Na(3p) formation in a crossed-beam experiment.
Changes of the sodium D-line emission were observed as the vibrational excitation of the Na$_2$ molecules was varied.
Although the majority of the data from this experiment were taken using vibrational excitation by Franck-Condon pumping \mycite{Sho11}, some crucial data were obtained with vibrational excitation by STIRAP.

In another experiment, \mytcite{Kul96} studied the dependence of the negative ion formation through dissociative electron attachment in the process Na$_2(v'') $ + e $\to$ Na + Na$^-$ on the vibrational excitation.
STIRAP was used to determine the location of the crossing between the potential energy curves for the
Na$_2$ +  e and  Na + Na$^-$ systems, which was found to lie between the $v''$ = 11 and $v''$= 12 levels of Na$_2$.
\mytcite{Kei99} explored  attachment of low-energy electrons to vibrationally excited sodium dimers in a supersonic molecular beam.
STIRAP was used to  vibrationally excite   these molecules.
In another early application of STIRAP, \mytcite{Kau01} studied the dependence of the rate of the dissociative attachment process
 Na$^{**}$ + Na$_2(v'') \to$ Na$^+ +$ Na + Na$^-$ on the vibrational excitation of the Na$_2$ molecule.

\subsubsection{Adiabatic passage by light-induced potentials \label{Sec:APLIP}}

\mytcite{Gar98} [see also \mytcite{Sol00a,Sol00b}] proposed to apply the STIRAP ideas  to wavepacket dynamics, in the transfer of an electronic wavepacket  between the ground vibrational states of two displaced molecular potentials of Na$_2$ in a process termed adiabatic passage by light-induced potentials (APLIP).

In order not to be limited by  the Frank-Condon principle (the overlap between the initial and final wavepackets is very small), the pulse durations must be longer than the vibrational time scale.
APLIP shares many STIRAP features, such as efficiency and robustness to parameter variations.
However, the two-photon resonance condition cannot be satisfied in APLIP.
APLIP transfers the wavepacket through a ``valley'', which emerges in the light-induced potential.

Various extensions of APLIP have been proposed.
\mytcite{Kal00} suggested APLIP with chirped pulses.
\mytcite{Rod00} extended APLIP to excited vibrational states, \mytcite{Gon06} to polyatomic molecules with intramolecular couplings among the vibrational modes, and \mytcite{Cha09} to molecular photodissociation.
\mytcite{Suo14} extended APLIP to multistate systems.

\subsubsection{Further applications in molecules \label{Sec:hole-STIRAP}}

\mytcite{Kra01b} pointed out that STIRAP can transfer holes, as well as electrons, between three molecular orbitals.
Of special interest is the case when electron and hole STIRAP's coexist: electron-STIRAP transfers an electron from a lower full orbital to a higher empty one via an empty middle one ($1_e\fromto 2_h \fromto 3_h$), while hole-STIRAP transfers a hole from a higher empty orbital to a full lower one via a full middle one ($3_h\fromto 2_e \fromto 1_e$).
The competition between these two processes leads to controllable bifurcating processes in molecular systems.

Finally,  STIRAP has been used in proposals for laser-controlled molecular current routers \mycite{Tha07,Tha08,Tha09}.

\subsection{Bose-Einstein condensates \label{Sec:BEC}}

\begin{figure}[tb]
\includegraphics[width=0.70\columnwidth]{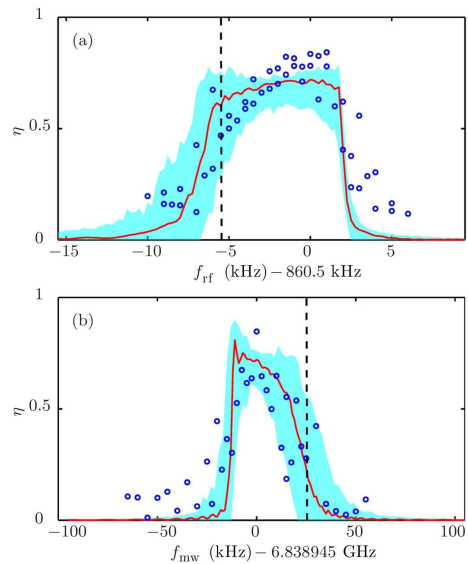}
\caption{\coloronline
STIRAP in ${}^{87}$Rb BEC: Transfer efficiency vs (a) two-photon detuning, and (b) one-photon detuning.
The dashed lines mark the (a) two-photon and (b) one-photon resonances.
The dots are experimental data, the solid lines are numerical simulation, and the cyan bands are numerical simulation with white magnetic-field noise included.
\location{Fig.~5 of \mytcite{Dup15}.}
}
\label{Fig:BEC}
\end{figure}

BEC features high atom densities, which introduce nonlinear terms in the Bloch equations due to interparticle interactions. 
The latter lead to resonance shifts and collisional losses, which pose some challenges to the implementation of STIRAP.
\mytcite{Dup15} reported thorough experimental study of STIRAP in ${}^{87}$Rb BEC magnetically trapped in the vicinity of an atom chip.
Population transfer with efficiency of up to 87\% took place in the transition between the hyperfine sublevels $F = 2, m_F = 2$ and $F = 2, m_F = 1$ via the lower state $F = 1, m_F = 1$ driven by two microwave pulses.
The \pump and \Stokes Rabi frequencies differed by a factor of 3, which, together with the effects of collisional losses and nonlinear shifts, led to asymmetric transition profiles, see Fig.~\ref{Fig:BEC} and Sec.~\ref{Sec: asymmetry}.

Theoretical activities on STIRAP in BEC have focused mainly on proposals for photoassociation of ultracold atoms into ultracold molecules \mycite{Jav99,Var99,Mac00,Dru03,Kuz09},  see Sec.~\ref{sec-ultracold molecules},
and spatial adiabatic passage between different potential wells \mycite{Gra06,Rab08,Nes09}, see Sec.~\ref{sec-CTAP}.
We point out an interesting theoretical proposal by \mytcite{Nan04} for STIRAP-inspired method for creation of a superfluid vortex in an oblate, axis-symmetric BEC by exposing it to two co-propagating laser pulses, one in the fundamental Gaussian mode and the other in a Gauss-Laguerre mode.
They numerically demonstrated complete transfer of the external angular momentum from the light field to the matter wave.

\subsection{Spatial adiabatic passage (SAP) \label{sec-CTAP}}

\begin{figure}[tb]
\includegraphics[width=0.95\columnwidth]{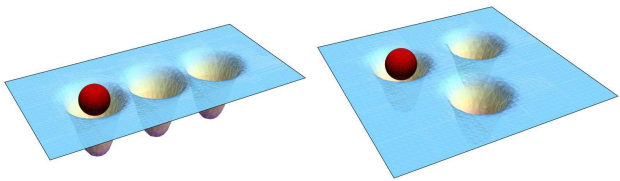}
 \caption{\coloronline
Spatial adiabatic passage of a particle between three potential wells.
Tunneling induces couplings between the wells.
\emph{Left}: a linear string of wells, corresponding to a three-state $\Lambda$ chain.
\emph{Right}: a two-dimensional set of wells corresponding to a loop system.
}
\label{fig-sap}
\end{figure}

The three equations of STIRAP for the probability amplitudes of the three states have been  adapted to the case of
  a particle than can be localized in three distinct potential wells (traps),  see Fig.~\ref{fig-sap}(left).
When two wells are sufficiently close to each other the particle matter wave  can tunnel between them,  at a rate that increases  as the trap separation diminishes.
{Moving the traps  or modifying the barrier height allows control of the tunneling probability.}
By adjusting the timing of separations in a chain of three traps one can, in principle,
reproduce the interaction sequence and adiabatic conditions {in analogy to}  STIRAP:  spatial motion in physical space replaces Hilbert-space motion to design a procedure initially known as\emph{ matter-wave STIRAP} \mycite{Eck04,Eck06} or {\em coherent tunneling by adiabatic passage} (CTAP) \mycite{Gre04}, and presently referred to as {\em spatial adiabatic passage} (SAP) \mycite{Men13,Men14,Men16}.

This technique was proposed by \mytcite{Eck04,Eck06}
as a robust tool for   transporting a single neutral atom between the outer traps of     a row of three optical-trap potentials.
Further interesting lines of development are presented in the recent review by  \mytcite{Men16}, discussing the transfer
\begin{itemize}
\item  of electrons in a chain of quantum dots \mycite{Gre04};

\item  in two-dimensional optical lattices \mycite{Mer07,McE10,Lon14}, as in Fig.~\ref{fig-sap}(right);

\item  of a hole (an empty site) in an array of three traps holding neutral atoms \mycite{Ben10};

\item  of electron spin states \mycite{Hol06,Hun13};

\item  in linear chains with more than three traps \mycite{Pet06};

\item  including the interatomic interaction in a BEC \mycite{Gra06,Rab08,Nes09};

\item by fractional STIRAP to coherently distribute  a BEC amongst three wells \mycite{Rab12};

\item of atoms between three \wgs~by double-STIRAP resulting in an atomic velocity filter \cite{Loi14}.

\end{itemize}

\section{STIRAP in quantum information \label{Sec:QInfo}}

Dues to  its inherent robustness to parameter errors and resilience to some types of decoherence,  STIRAP has emerged as a popular tool in quantum information.
Several examples of this development are briefly reviewed in this section.

\subsection{Single-qubit gates  \label{Sec:single_qubit}}

The pursuit of complete population transfer from one state to another, or creation of a coherent superposition by partial population transfer, starts from a \emph{specific} initial state.
A quantum gate requires a specified response of the qubit for \emph{any} initial condition.
The most general unitary transformation of a qubit reads
\be\label{SU(2)}
\mathbf{U} = \left[ \begin{array}{cc} a & -b^* \\ b & a^*  \end{array} \right],
\ee
where $a$ and $b$ are two complex (Cayley-Klein) parameters ($|a|^2+|b|^2 = 1$).
The construction of the SU(2) gate \eqref{SU(2)} means that the following transforms are performed:
\be\label{SU(2)-12}
\ket{1} \to a \ket{1} + b \ket{2}, \qquad
\ket{2}\to -b^* \ket{1} + a^* \ket{2}.
\ee
In a \emph{closed} two-state system, the fulfillment of the first of these guarantees the fulfillment of the second, and vice versa.
However, if the qubit is a subsystem of a larger system (with the so-called ancilla states), e.g. if the qubit is formed of the lower states $\ket{1}$ and $\ket{3}$ in the \L~system of STIRAP, the fulfillment of one of the transformations \eqref{SU(2)-12} does not guarantee the fulfillment of the other.
 For instance, the direct application of fractional STIRAP produces a coherent superposition of states $\ket{1}$ and $\ket{3}$ when the system is initially in state $\ket{1}$.
However, if the system is initially in state $\ket{3}$ and all fields are on resonance, fractional STIRAP would produce a superposition of all three states.
Hence it does not produce a qubit gate.

\begin{figure}[tb]
\includegraphics[width=0.80\columnwidth]{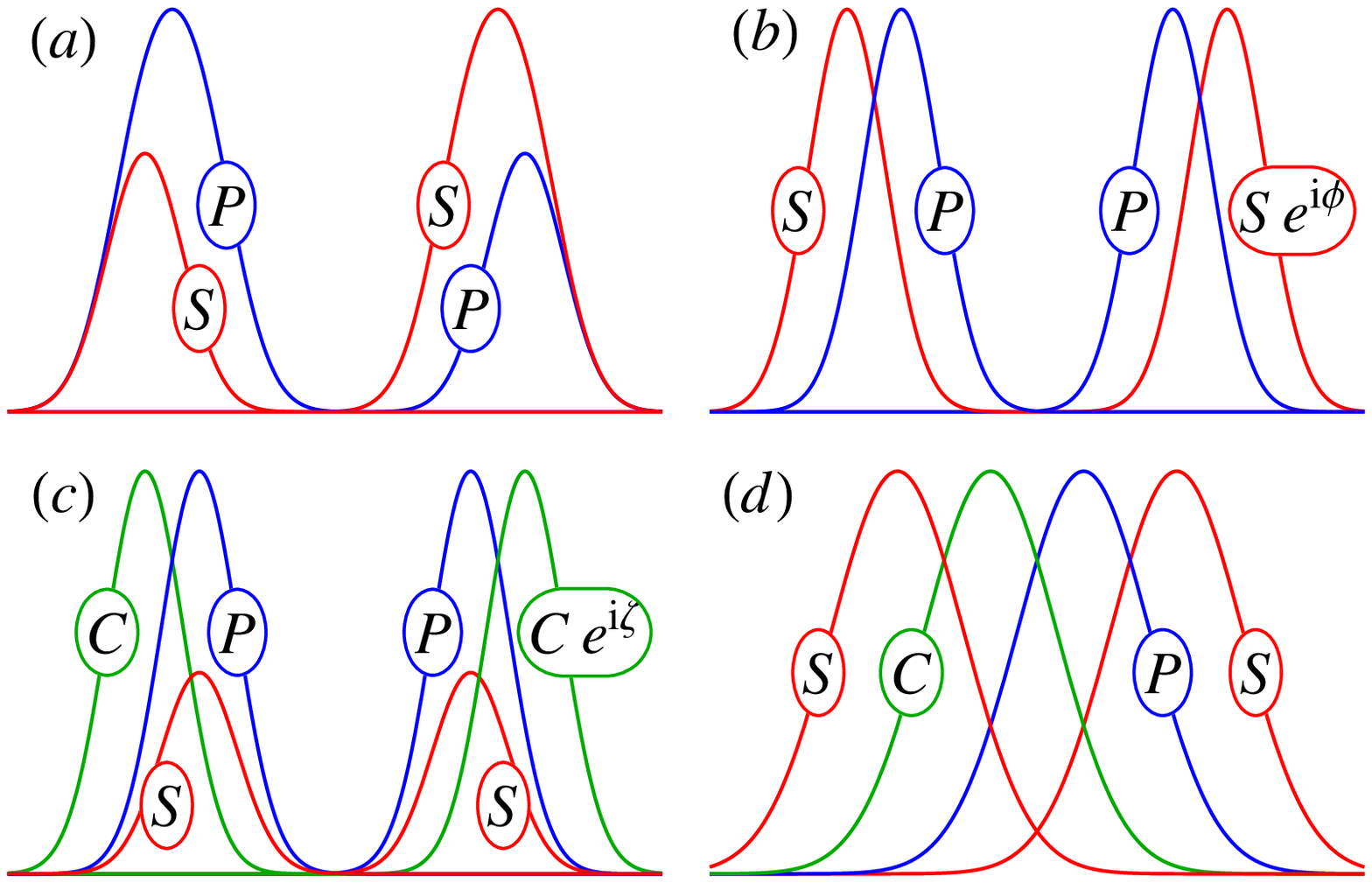}
\caption{\coloronline Pulse sequences for rotation gates with
(a)  two fractional-STIRAP processes, and (b) two STIRAP processes  in the $\Lambda$-system of Fig.~\ref{fig-links2dN-r}.
\locationa{Fig.~4 of   \mytcite{Lac06} and Fig.~2 of \mytcite{Rou13}. }~
(c) Pulse sequence for rotation gate created by two tripod-STIRAP processes, {with the linkage pattern of Fig.} \ref{Fig:tripod}.
\locationa{Fig.~1 of \mytcite{Kis02a}.}~
(d) Pulse sequence for observation of a geometric phase without a dynamical phase in a tripod system.
\locationa{Fig.~1 of \mytcite{Una04}.}
}
\label{Fig:timings}
\end{figure}

STIRAP can still be used to construct robust single-qubit gates.
One possibility is to use a large single-photon detuning.
Then, as described above (Sec.~\ref{sec-three-state}), the middle state $\ket{2}$ can be eliminated adiabatically and the   \L~ system is reduced to an effective two-state system of states $\ket{1}$ and $\ket{3}$.
In this case fractional STIRAP will act as an SU(2) gate for the qubit formed of states $\ket{1}$ and $\ket{3}$.
Alternatively, \mytcite{Lac06}  showed that robust rotation gates can be produced in a   \L~system by a sequence of two (inverted and regular) fractional STIRAP processes, see Fig.~\ref{Fig:timings}(a).
If the ratio $\Omega_P(t) / \Omega_S(t)$ tends to $\cot\alpha$ initially and to $\tan\alpha$ in the end, then this sequence produces a robust rotation gate of angle $2\alpha$.

\mytcite{Bet13} proposed to use double-STIRAP in a three-state ladder, with a pulse sequence such as the one in Fig.~\ref{Fig:timings} (b), but with nonzero single-photon detuning $\Delta$ in the first step and $-\Delta$ in the second step, in order to implement a rotation gate in trapped Rydberg atoms.
The sign flip in $\Delta$ reduces the dependence of the acquired phase on the  (uncertain) number of atoms.

\subsection{Geometric gates \label{Sec:geometric}}

Of particular interest to quantum information has been the tripod version of STIRAP because of its two dark states.
 \mytcite{Una99}
 recognized that the phase factors associated with the two dark states during the evolution are of non-Abelian nature,
 and the ensuing mixing angle between the two dark states is of geometric, or Pancharatnam-Berry, origin
 \mycite{Pan56a,Pan56b,Ber84,Wil84,Aha87}.
Indeed, this phase,
\begin{align}
 \beta &= \int_{-\infty}^{\infty} \dot{\mixb}(t) \sin\mixa(t)  dt = \int_{\mixb_i}^{\mixb_f} \sin\mixa  d\mixb \notag\\
  &= \oint \frac{\Op}{\sqrt{\Op^2+\Os^2+\Omega_{C}^2}}
   \frac{(\Os d\Omega_{C}  - \Omega_{C} d\Os)}{\Os^2+\Omega_{C}^2}
\end{align}
does not depend on time but only on the (closed) trajectory in the parametric space of  $\{\Op,\Os,\Omega_{C}\}$.
If the pulses have the same time dependence, this geometric phase will vanish.
However, if the pulses have different time dependences,  then the geometric phase is nonzero;
 moreover, it can be controlled by the pulse delays \mycite{Una98,Una99,Mol07}.
Following up on the work of \mytcite{Una99}, implementations of geometric gates have been proposed with trapped ions \mycite{Dua01}, Rydberg atoms \mycite{Mol08PRL}, and sodium dimers \mycite{Men07}.

\mytcite{Kis02a} proposed a robust rotation gate by application of two STIRAP processes in the tripod system of Fig.~\ref{Fig:tripod}.
The qubit is formed of states $\ket{1}$ and $\ket{3}$, while state $\ket{4}$ is an ancilla state.
The pulse sequence is shown in Fig.~\ref{Fig:timings} (c).
The couplings of the transitions $\ket{1}\fromto\ket{2}$ ($\Omega_P$) and $\ket{3}\fromto\ket{2}$ ($\Omega_S$) have the same time dependence but different amplitudes and phases: $\Omega_P(t) = f(t) \cos\xi$ and $\Omega_S(t) = \e^{\ii\eta}  f(t) \sin\xi$.
The coupling $\Omega_C(t)$ of the ancilla transition $\ket{4}\fromto\ket{2}$ is shifted in time with respect to the other two.
 The second $\Omega_C(t)$ pulse is phase shifted with respect to the first $\Omega_C(t)$ by a phase $\zeta$.
In the adiabatic limit, this sequence of pulses produces the unitary transformation
$\e^{-\ii\zeta/2} R_{\textbf{n}}(\zeta) = \e^{-\ii\zeta/2 -\ii \zeta \mathbf{n}.\bm{\sigma}/2}$,
 where $\mathbf{n} = (\sin2\xi \cos\eta,\sin2\xi \sin\eta,\cos2\xi)$ and $\bm{\sigma} = (\sigma_x,\sigma_y,\sigma_z)$ is a vector of Pauli's matrices.
The entire process is as robust as STIRAP and only good control of the relative laser phases is required.

\begin{figure}[tb]
\includegraphics[width=0.90\columnwidth]{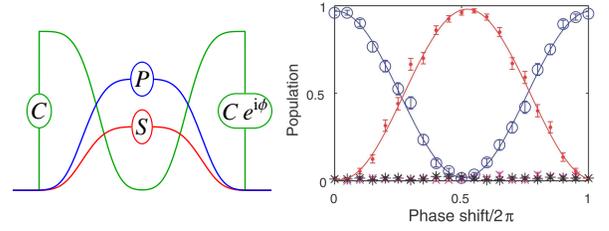}
\caption{\coloronline Geometric phase gate demonstration by \mytcite{Toy13}.
\emph{Left}: Pulse sequence.
\emph{Right}: Populations $P_1$ (blue hollow circles), $P_3$ (red filled circles), $P_4$ (magenta crosses) and $P_2$ (black asterisks).
\locationa{Figs.~1 and 2 of  \mytcite{Toy13}.}
}
\label{Fig:Toyoda2013}
\end{figure}

\mytcite{Toy13} demonstrated the Kis-Renzoni gate  in an experiment with trapped ${}^{40}$Ca$^+$ ions.
They used the $S_{1/2}-D_{5/2}$ electric-quadrupole transition, with the $S_{1/2}, M=-1/2$ sublevel serving as the common (``upper'') tripod state, and three $D_{5/2}$ sublevels: $M=-3/2,+1/2$ (forming the qubit) and $M=-5/2$ (ancilla state) coupled to the common state $S_{1/2}, M=-1/2$ by $P$, $S$ and $C$ fields, respectively.
The gate pulse sequence is shown in Fig.~\ref{Fig:Toyoda2013} (left); it is a modified version of the sequence of Fig.~\ref{Fig:timings}(c).
The gate operation is shown in Fig.~\ref{Fig:Toyoda2013} (right) where the population is seen to oscillate vs the phase $\phi$ between the qubit states $M=-3/2$ and $+1/2$, with only negligible population in the other two states.

\mytcite{Rou13} extended these ideas to an $N$-pod -- a fan linkage of $N$ lower states $\ket{k}$ coupled to a single excited state $\ket{e}$.
They showed that a double-STIRAP sequence, as the one in Fig.~\ref{Fig:timings} (a),  can produce a Householder reflection \mycite{Hou58} in the subset on $N$ lower states. Householder reflections are a powerful tool for construction of arbitrary quantum gates of qudits ($d$-state systems) \mycite{Iva06,Iva07,Iva08}.

In many implementations, the geometric phase is usually accompanied by a much larger dynamical phase, which makes its observation challenging.
\mytcite{Una04} showed that in the tripod system, the pulse sequence shown in Fig.~\ref{Fig:timings}(d), cancels the dynamical phase, thereby allowing to observe the unperturbed geometric phase.
\mytcite{Mol08PRA} proposed an implementation of a geometric phase gate by double-STIRAP in a   \L~system, with a time-dependent relative phase between the pump and Stokes fields (which is equivalent to time-dependent detunings).
Time-dependent detunings have been used also by \mytcite{Nak13}.
\mytcite{Das07} showed that, although a Berry phase cannot be accumulated in STIRAP in a resonant  \L~system, such a phase may emerge in the presence of decoherence.

\subsection{Entangled states \label{Sec:entangled}}

A number of authors have proposed using STIRAP and fractional STIRAP to construct many-qubit entangled states.
In two of the most ubiquitous quantum-information platforms -- trapped ions and trapped atoms -- STIRAP allows one to perform qubit manipulations without populating the noisy common bus mode, i.e. the  vibration mode shared by the trapped ions or the cavity mode shared by the trapped atoms.

\begin{figure}[tb]
\includegraphics[width=0.70\columnwidth]{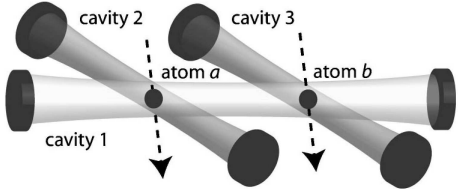}
\caption{
Two atoms passing through the intersections of three cavities produce a five-state sequentially-coupled chain of states. The axis of cavity 1 is spatially displaced with respect to the other axes, thereby producing counterintuitively-timed atom-cavity couplings.
\location{Fig. 1 of  \mytcite{Lar05}.}
}
\label{Fig:Larson2005-1}
\end{figure}

A few examples of the many theoretical proposals that use STIRAP and STIRAP-inspired schemes for creation of entangled states  include:

\begin{itemize}

\item
a method for creating entangled Bell states of two qubits \mycite{Nie10} using two pulse pairs with single-photon detunings of opposite signs \mycite{Una02,Gon02}, or by using the relative phase between the pulses \mycite{Mal04a,Mal04b,Mal04c} [an alternative using frequency chirps for two \mycite{Una01} and multiple qubits \mycite{Una02b} has been proposed too];

\item
 generation of two-particle \mycite{Bar00} and many-particle entangled states, such as the Greenberger-Horne-Zeilinger (GHZ) state \mycite{Nie10},  of dipole-dipole interacting Rydberg atoms by using the dipole blockade effect \mycite{Una02,Mol08PRL};

\item
a method to create many-particle entangled states of trapped ions \mycite{Una03} with controllable collective interactions of Lipkin-Meshkov-Glick type \mycite{Lip65};

\item
cavity-QED schemes that  map atomic Zeeman coherences onto photon states and generates entangled photon multiplets and atom-photon entanglement in a two-mode optical cavity \mycite{Lan00}, entanglement between atoms in coupled cavities \mycite{Che07}, two or three atoms, two photons, or an atom and a photon in a two-mode cavity \mycite{Bis03,Kis04a,Gar08,Amn05a,Amn05b,Amn10,Amn12}, an atom and BEC or two BECs \mycite{Che12b,Che14}, and Bell inequality tests for entangled photons \mycite{Bei00b};

\item
 a method to adiabatically transfer field states between two partly overlapping cavities via an atom passing through them \mycite{Mat01},
 and an extension to various cavity-field states in multiple cavities and multiple atoms \mycite{Lar05},  see Fig.~\ref{Fig:Larson2005-1};

\item
 a method to produce entangled Fock states in time, frequency, and space, using light storage of a single photon (or few-photon states) in the atomic coherence $\rho_{12}$ of two atomic states 1 and 2, followed by partial transfer of some of the population of state 2 to another state 3 by fractional STIRAP,
 and then partial retrieval of the initial coherence by the reading pulses \mycite{Wan04};

\item
``dissipation-assisted adiabatic passage'' for entangling atoms in a cavity, in which the presence of spontaneous decay corrects nonadiabatic errors \mycite{Mar03}; 

\item
  transfer the quantum state of two molecular dissociation fragments, whose internal and translational states are naturally entangled, to an entangled photon pair \mycite{Pet03};

\item
transportation of a qubit, operator measurements, and entanglement, in a one-dimensional array of quantum sites with a single sender and multiple receivers
\mycite{Gre06};

\end{itemize}

\begin{figure}[tb]
\includegraphics[width=0.80\columnwidth]{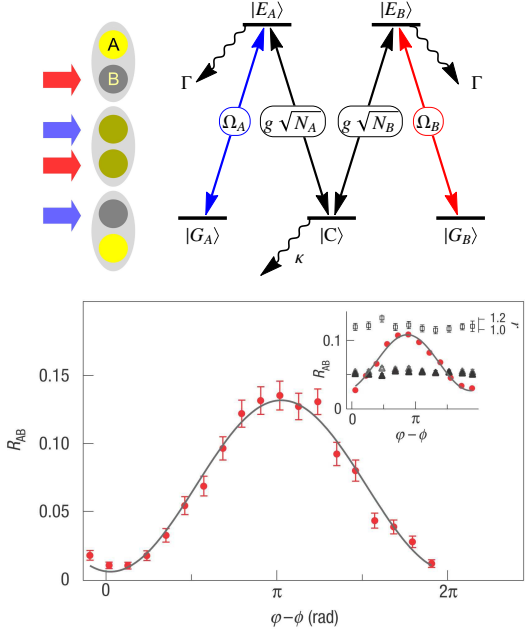}
\caption{\coloronline
\emph{Top}: Linkage pattern used for phase-coherent transfer of a spin wave between two atomic ensembles A and B (containing $N_A$ and $N_B$ atoms) in the experiment of \mytcite{Sim07}.
The write process populates state $\ket{G_A}$.
$g\sqrt{N_A}$ and $g\sqrt{N_B}$ are the collective couplings of the magnons to the optical resonator mode, whereas $\Omega_A$ and $\Omega_B$ are the laser-atom couplings.
The spin wave is transferred from A to B by a counterintuitive sequence of laser pulses (top left), with the pulse addressing B preceding the pulse addressing A, thereby leaving the lossy photonic mode $\ket{C}$ empty.
\emph{Bottom}: Phase coherence between the spin waves in the two ensembles after a partial transfer of a magnon. 
\locationa{Figs. 1 and 2 of \mytcite{Sim07}.}
}
\label{Fig:Simon2007}
\end{figure}

\mytcite{Sim07} experimentally demonstrated phase-coherent transfer by multistate STIRAP of a spin wave (quantized collective spin excitation, or magnon) from one ensemble of ${}^{133}$Cs atoms to another via an optical resonator serving as a quantum bus, see Fig.~\ref{Fig:Simon2007}.
Benefitting from the features of STIRAP, this bus was only virtually populated.
An entangled state with one excitation jointly stored in the two ensembles was deterministically created by fractional STIRAP.

\mytcite{Lin08} proposed a method for generation of arbitrary-sized Dicke states in a chain of trapped ions, which are equally-weighted coherent superpositions of collective states of qubits that share the same number of excitations.
The ion qubits are cooled to their vibrational ground state $\ket{0}$.
Then a vibrational Fock state $\ket{m}$ with $m$ phonons is prepared.
Next, the system is driven from this state to  the desired Dicke state by multistate STIRAP via a multiqubit dark state by two delayed pulses applied simultaneously on all $N$ ions, the first on the carrier transition, and the second on the red-sideband transition.

\begin{figure}[tb]
\includegraphics[width=0.80\columnwidth]{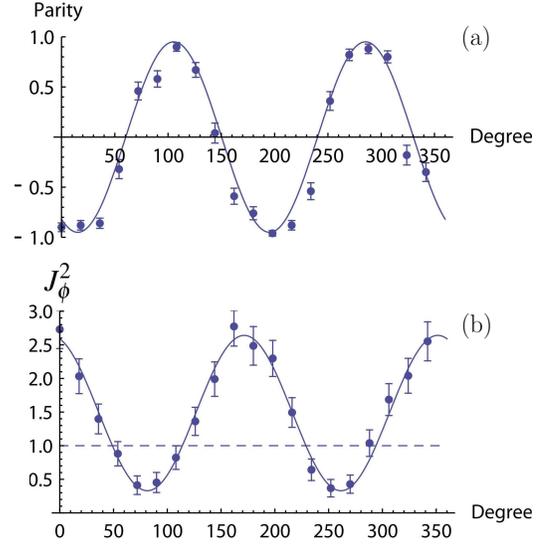}
\caption{\coloronline
(a) Parity oscillation of the two-ions Dicke state.
(b) Oscillation of the squared spin of the half-excited Dicke state of four ions. 
\location{Fig. 4 of \mytcite{Nog12}.}
}
\label{Fig:Noguchi2012-4}
\end{figure}

\mytcite{Nog12} experimentally demonstrated  a modified version of this proposal with global red- and blue-sideband pulses in chains of two and four trapped ${}^{40}$Ca ions.
Figure \ref{Fig:Noguchi2012-4} shows the parity oscillation for two ions (top) and the oscillation of the squared spin for four ions (bottom), with fidelities of 96\% and 86\%, respectively.

\mytcite{Lin08a} proposed another adiabatic method for creation of Dicke states of trapped ions -- by global addressing by a chirped pulse. This method has been demonstrated experimentally by \mytcite{Toy11}, and a modified version by \mytcite{Hum09}.

\subsection{Two-qubit gates \label{Sec:gates}}

STIRAP and fractional STIRAP have been the engines also in proposals for two-qubit quantum gates.
\mytcite{Pel95} proposed  a scheme for a two-qubit control-unitary gate with multilevel atoms in an optical cavity. 
The scheme has three steps:
(i) the state of the first qubit is transferred by STIRAP from the first atom to the non-qubit states of the second atom;
(ii) the four states of the second atom are manipulated by single-atom techniques;
(iii) the inverse of step (i) is performed.

\mytcite{Pac02} proposed two-qubit conditional phase gates of ions trapped in a cavity using a combination of STIRAP and environment-induced quantum Zeno effect.
The latter keeps the qubits in a decoherence-free subspace \mycite{Bei00}.
The method avoids both spontaneous emission, because of STIRAP, and cavity loss because no photon is present in the cavity at any time.
A modified version of this method \mycite{Pac04} creates a  two-qubit phase gate, with either dynamical or geometric phase, by using a common laser addressing of the two qubits in a single step.
\mytcite{Got04} proposed STIRAP-inspired implementations of one-, two- and three-qubit phase gates 
 of atoms in a single-mode optical cavity.

\mytcite{Mol07} used tripod-STIRAP to propose  a set of universal gates for quantum computing based on geometric phases: a one-qubit phase gate, a Hadamard gate, and a two-qubit phase gate.
\mytcite{San05} proposed a robust STIRAP-based SWAP gate of two atomic qubits in an optical cavity protected from cavity losses and atomic decoherence.
Similar approaches have been used to construct CNOT \mycite{San06} and control-unitary gates \mycite{Lac06}.

\subsection{Quantum algorithms \label{Sec:algorithms}}

\mytcite{Dae07,Dae08a} proposed to use multistate inverted fractional STIRAP for adiabatic implementation of Grover's quantum search algorithm, which finds a marked item in an unsorted list of $N$ items \mycite{Gro97}.
The search database is an ensemble of $N$ identical three-level atoms trapped in a single-mode cavity and driven by two lasers.
The marked atom has an energy gap between its two ground states.
Starting from an initial entangled state, inverted fractional STIRAP allows one to populate the marked state in time that scales as $\sqrt{N}$, thereby achieving the same speed-up as the discrete Grover algorithm.
\mytcite{Dae08b} proposed to use ``parallel-STIRAP'' (Sec.~\ref{Sec: shortcuts})  for the same purpose.

\section{STIRAP in solid state physics \label{sec-solid}}

In the first decade after its discovery  STIRAP was demonstrated exclusively in gas-phase atoms and molecules.
During the last decade, some solid-state systems have attracted significant attention as candidates for implementation of coherent light-matter interaction.
These developments have been largely motivated by promising applications to quantum information processing because solids offer appealing physical platforms for scalable quantum computing.
We begin our discussion with experimental implementations of STIRAP in rare-earth-metal-ion-doped dielectric crystals,
followed by a discussion of color centers in diamond, superconductors,  quantum dots, and semiconductors.

\subsection{Doped crystals \label{sec-doped}}

Crystals doped with  rare-earth-metal ions possess suitable properties for coherent interactions between light and matter:
 high density, robustness, scalability, narrow optical linewidths (unlike most other solid materials) and, in particular,  long  coherence times.
For this reason, doped solids have emerged as the leading candidate for optical data storage for quantum computing
 after successful implementations of various coherent control techniques \mycite{Ham97,Tur01,Lon05,Rip05}.
The EIT technique \mycite{Har90,Fle05} is especially important as it enables the deceleration of a light pulse in the doped crystal and the pulse storage in the atomic coherence, i.e. in a superposition of two atomic states.
Such states are readily available among the hyperfine states of the ground level of the dopant ions.
The stopped light is then released on demand by an inverted EIT process.

\begin{figure}[tb]
\includegraphics[width=0.75\columnwidth]{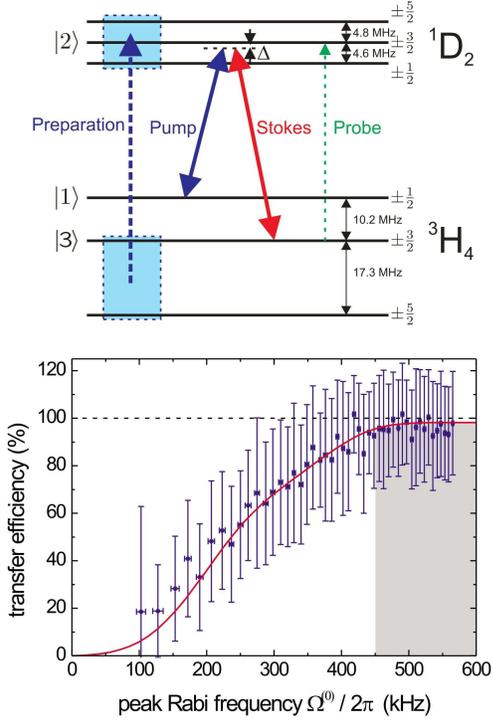}
\caption{\coloronline
Demonstration of STIRAP in Pr$^{3+}$:Y$_2$SiO$_5$.
\emph{Top}: Energy levels and linkages.
\emph{Bottom}: Population transfer efficiency vs peak Rabi frequency.
\location{Fig.~4 and 8 of  \mytcite{Kle08}.}
}
\label{Fig:Klein-1}
\end{figure}

\begin{figure}[tb]
\includegraphics[width=0.75\columnwidth]{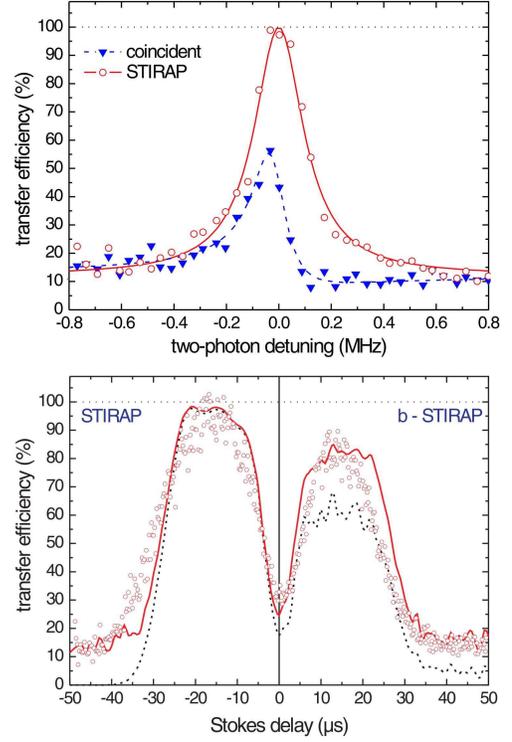}
\caption{\coloronline
Demonstration of STIRAP in Pr$^{3+}$:Y$_2$SiO$_5$.
\emph{Top}: Population transfer efficiency vs the two-photon detuning (\pump frequency fixed,  \Stokes frequency varied),
 with a peak on two-photon resonance.
\emph{Bottom}: Population transfer efficiency vs pulse delay.
The \emph{P} and \emph{S} fields are off resonance by the same detuning $\Delta = 2\pi \times 320$ kHz (see Fig.~\ref{Fig:Klein-1}), so that they are on two-photon resonance.
The dashed curve shows the data with the incoherent population transfer contribution excluded.
\location{Figs.~2 and 4 of  \mytcite{Kle07}.}
}
\label{Fig:Klein-2}
\end{figure}

The first experimental demonstration of STIRAP in doped solids was presented by   \mytcite{Got06,Got07} in  Pr$^{3+}$:Y$_2$SiO$_5$ crystal.
 \mytcite{Kle07,Kle08} conducted a thorough experimental study of STIRAP between hyperfine levels of praseodymium ions in a cryogenically cooled Pr$^{3+}$:Y$_2$SiO$_5$ crystal, see Fig.~\ref{Fig:Klein-1} (top).
Because of the   large inhomogeneous broadening of the medium (about 10 GHz),
a preparation step first depletes the populations of a group of energy levels near the target state by spectral-hole burning and optical pumping (thereby creating a ``spectral pit'').
The subsequent interaction involves only Pr$^{3+}$ ions with spectral features that fall within this spectral pit: their spectral properties   resemble those of atoms in a gas phase.
\mytcite{Kle07} demonstrated STIRAP between the degenerate sublevels $M = \pm 1/2$ and $M=\pm 3/2$ of the $^3$H$_4$ hyperfine level.
Figure \ref{Fig:Klein-1}(bottom) shows a characteristic feature of STIRAP -- the steady increase of the population transfer efficiency with the peak Rabi frequency until saturation (nearly 100\% transfer).
Fig.~\ref{Fig:Klein-2}(top) shows another characteristic feature of STIRAP -- the transfer efficiency vs the two-photon detuning, with
100\% STIRAP efficiency on two-photon resonance, contrasted to only 50\% efficiency with coincident pulses.

 \mytcite{Kle07,Kle08} have also observed bright-STIRAP, which uses intuitively-ordered pulses detuned from single-photon resonance (Sec.~\ref{sec-ordering}).
Its signature is found in the right part of Fig.~\ref{Fig:Klein-2}(bottom).
Its efficiency is lower than in STIRAP (left part) because bright-STIRAP uses a bright eigenstate of the Hamiltonian, which has a component of the decaying state \s2.

In another experiment, \mytcite{Ale08} conducted thorough study of STIRAP in Tm$^{3+}$:YAG crystal.
They achieved 90\% efficiency of population transfer by STIRAP, and 45\% with bright-STIRAP.

\subsection{Color centers in diamond \label{sec-diamond}}

\begin{figure}[tb]
\includegraphics[width=0.90\columnwidth]{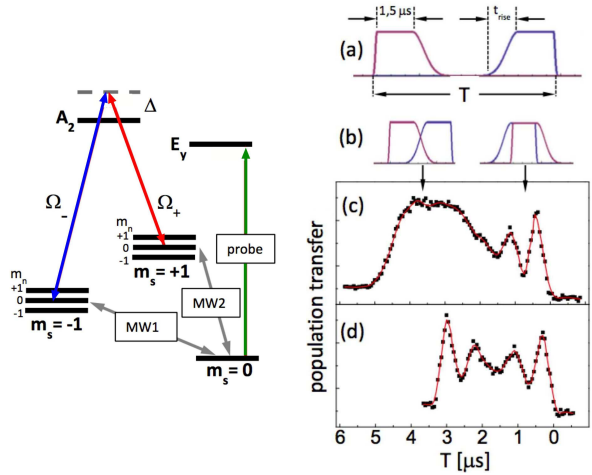}
\caption{\coloronline
STIRAP with NV centers in diamond in the experiment by \mytcite{Gol14}.
Left: relevant level scheme.
The electronic spin sublevels $m_s = \pm1$ are Zeeman split by 150 MHz, and the
 nuclear spin sublevels $m_n$  by 2.2 MHz.
State $m_s = 0 $ is coupled to states $m_s = -1$ or + 1 by microwave pulses MW1 or MW2, respectively.
Right:
(a) pulse shapes and (b) pulse overlap for STIRAP transfer (left side) and two-photon Rabi oscillations (right side);
(c) population of the target state in $m_s = +1$ monitored through fluorescence from state $E_y$
 for  $t_{\rm rise} = 1.2 \mu$s, allowing adiabatic evolution;
(d) same as in (c) but with $t_{\rm rise} = 20$ ns, which is too steep to allow STIRAP transfer, and instead, two-photon Rabi oscillations are observed. %
\locationa{Figs. 1 and 4 of  \mytcite{Gol14}.}
}
\label{Fig:Golter2014}
\end{figure}

Color centers in diamond are a promising candidate for quantum computing and quantum sensing because
 they feature high-fidelity preparation, control and readout, as well as long  coherence times for electron and nuclear spins, even at room temperature.
The most popular color centers are the negatively charged nitrogen vacancy (NV) centers in diamond, in which
 quantum state transfer between electron and nuclear spins \mycite{Chi06,Dut07},
 spin entanglement mediated by dipole coupling, and spin-photon entanglement \mycite{Tog10,Sip12,Ber13c,Neu08,Dol13} have been observed.
 For their control  it is crucial to manipulate electronic spin states through optical transitions, while avoiding the rapid decoherence of these transitions, which includes radiative decay and variation of transition frequencies with time due to the fluctuating influence of the environment (spectral diffusion).
Moreover, the optically-driven spin dynamics should be nuclear-spin selective.
To this end,  \mytcite{Gol14}  found that STIRAP satisfies all these requirements.
They conducted a thorough experimental study of coherent optical control of electronic spin states in single NV centers by observing Rabi oscillations and STIRAP,
and found that Rabi oscillations are still prone to spectral diffusion, while STIRAP is immune to it.
Moreover, they found that the STIRAP efficiency depends on the orientation of the adjacent ${}^{14}$N nuclear spin, i.e. it is nuclear-spin selective.
Thus this experiment proved that NV centers are suitable to mediate coherent spin-phonon coupling, enabling the optical control of spin and mechanical degrees of freedom.

Figure  \ref{Fig:Golter2014} shows the relevant level scheme and a sample of the results.
The experiment starts 
 from the $m_s = 0$ ground state of the system.
A microwave $\pi$-pulse (MW1) transfers the population into the hyperfine manifold of the $m_s = -1$ state from where the STIRAP transfer to the $m_s = +1$ target state starts.
STIRAP is induced by the $\sigma_+$ and $\sigma_-$ polarized $\Omega_+$  and $\Omega_-$ pulses (at 637 nm), respectively, detuned by $\Delta \approx  1$ GHz from the $A_2$ level.
After completion of the transfer, the population in the target state is determined by transferring the population of state $m_s = +1$ by another microwave $\pi$-pulse (MW2) back to $m_s = 0$, from where it is excited to state $E_y$, and the ensuing fluorescence is measured.
Frames (c) and (d) of Fig.~\ref{Fig:Golter2014} show the variation of the population with the pulse delay $T$ for gradual and steep rise of the slope of the pulses, $t_{\rm rise}  = 1.2 \mu$s and 20 ns, respectively.
In frame (c), the slowly rising slopes make adiabatic evolution possible: a broad plateau is seen for $2.5 \mu$s $< T < 4 \mu$s, and Rabi oscillations for $T <1.5 \mu$s  where the pulses largely overlap.
The steep rising slopes in Fig.~\ref{Fig:Golter2014}(d) prevent adiabatic evolution and STIRAP but instead, two-photon Rabi oscillations occur.

\subsection{Superconductors\label{sec-superconductors}}

\begin{figure}[tb]
\includegraphics[width=0.98\columnwidth]{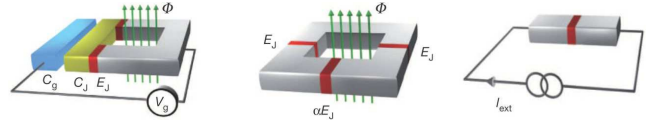}
\caption{\coloronline
Voltage-driven superconducting  Cooper pair box (charge qubit, left), flux-driven loop (flux qubit, middle), current-driven junction (phase qubit, right).
\location{Fig.~1 of   \mytcite{You11}.}
}
\label{Fig:You}
\end{figure}

Superconducting qubits based on the Josephson tunnel junction have emerged as a promising physical platform for quantum computation \mycite{Mak01,Dev04,You11}.
A significant advantage of superconducting qubits is that such  solid-state electrical circuits   can easily be fabricated with techniques used for conventional integrated circuits.
 In contrast to atoms and photons, the superconducting qubits can conveniently be coupled to other circuits thereby facilitating qubit control, gate implementation and readout.

Superconducting qubits are classified into three types, based on their degrees of freedom:
 charge \mycite{Bou98,Nak99}, flux \mycite{Fri00,van00}, and phase \mycite{Mar02}, see Fig.~\ref{Fig:You}.
These qubits have excellent scalability due to the well-established fabrication techniques but   they suffer from short coherence times   $\tau_c$.
Different strategies have been proposed to enhance  $\tau_c$.
One approach \mycite{Mar05} is to improve the properties of the junctions  in order to  suppress sources of $1/f$ noise \mycite{Pal14}.
A popular approach is the elimination of linear noise by operating the qubits at optimal working points called ``sweet spots''.
Using  this  latter approach   \mytcite{Vio02} have demonstrated an increase in dephasing times by three orders of magnitude.
\mytcite{Koc07} introduced a new type of superconducting qubit -- 
 transmon, a capacitively-shunted Cooper-pair box strongly coupled to an electromagnetic transmission line resonator.
Its design is related to the charge qubit but it operates at a greatly increased ratio of Josephson energy $E_J$ and charging energy $E_C$.
The transmon has drastically reduced sensitivity to charge noise and has increased qubit-photon coupling while keeping sufficient anharmonicity for selective control.

\begin{figure}[tb]
\includegraphics[width=0.85\columnwidth]{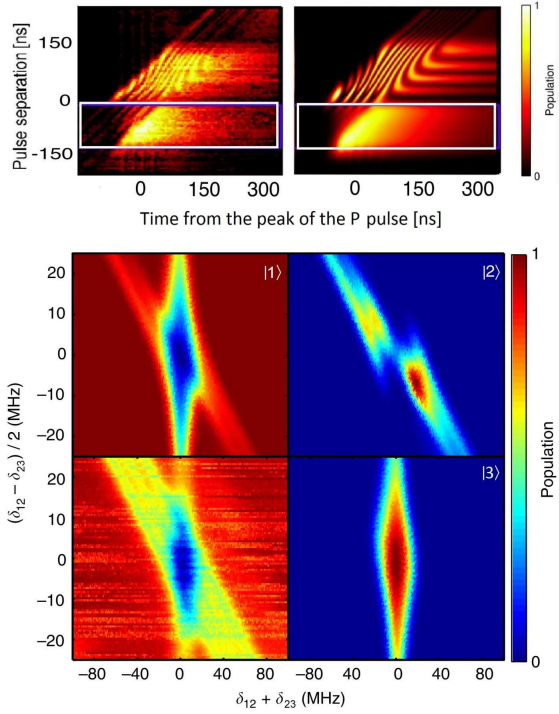}
\caption{\coloronline
Experimental demonstration of STIRAP in a superconducting transmon circuit.
\emph{Top} (left: experiment, right: simulations): Contour plots of populations of state  $\ket{3}$ vs pulse delay and time.
The white boxes bound the STIRAP region.
\emph{Bottom}:
Contour plots of populations vs the single-photon detuning (vertical) and two-photon detuning (horizontal).
The lower left frame shows the experimental data for the population of the ground state $\ket{1}$.
The other three frames show simulations for states $\ket{1}$, $\ket{2}$ and $\ket{3}$.
\locationa{Figs.~2  and 3 of  \mytcite{Kum16}.}
}
\label{Fig:Kumar-2}
\end{figure}

Superconducting circuits are  multilevel systems.
A number of well-known quantum three-level effects have been demonstrated recently, e.g., the Autler-Townes effect with phase qubits \mycite{Sil09,Li12} and transmons \mycite{Bau09}, coherent population trapping with phase qubits \mycite{Kel10}, and EIT with flux qubits \mycite{Abd10}.
 \mytcite{Pas04,Kis04a,Liu05,Sie06,Sie09,Fal13,DiS15,DiS16} have proposed designs of three-state systems with superconducting circuits suitable for implementation of STIRAP.

Very recently, \mytcite{Kum16} demonstrated STIRAP in a comprehensive study.
They used the ladder   linkage formed by   the ground and two excited states of a transmon 
 irradiated by two microwave fields, and they used STIRAP to transfer the population from the ground state $\ket{1}$ to the second excited state $\ket{3}$ with over 80\% efficiency. 
The top portion of Fig. \ref{Fig:Kumar-2} shows  contour   plots that display characteristic STIRAP signatures
in  the populations plotted versus time and pulse delay.
As expected for STIRAP, the maximum population transfer occurs for negative delays (confined between the horizontal lines).
The bottom portion of Fig. \ref{Fig:Kumar-2}  shows  the populations versus the single- and two-photon detunings, with the typical robustness of STIRAP to single-photon detuning and sensitivity to two-photon detuning.

Very recently, \mytcite{Xu16} reported STIRAP with 67\% and 96\% efficiency in superconducting phase and transmon qubits, respectively.

\subsection{Semiconductor quantum dots \label{sec-dots}}

\begin{figure}[tb]
\includegraphics[width=0.95\columnwidth]{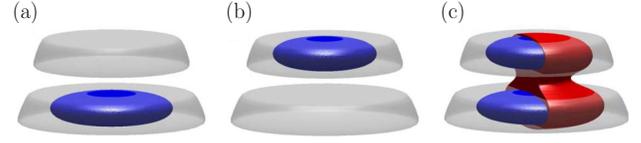}
\caption{\coloronline
Double-dot structure and carrier wavefunctions proposed by  \mytcite{Hoh00,Hoh06}   for implementation of quantum gates by STIRAP.
The confinement potential consists of two truncated-cone-shaped regions of low-band-gap material.
Frames (a) and (b) show, respectively, the squared modulus of the hole ground and first excited wavefunction.
Frame (c) shows the charged exciton state,
 where the electron wavefunction (red, cut off for visibility) extends over the hole wavefunction due to the lighter electron mass, and allows optical coupling to both hole states.
\location{Fig.~2 of  \mytcite{Hoh06}.}
}
\label{Fig:Hohenester}
\end{figure}

Semiconductor quantum dots (QD) are small islands of semiconducting material,
  embedded in a surrounding host material.
Carrier confinement within these islands is achieved by using different semiconductor materials or by applying external gate voltages.
These structures allow the demonstration of fundamental quantum-coherence effects and make semiconductor QDs, also known  as ``artificial atoms'' due to the possibility to engineer their discrete quantum levels, promising candidates for quantum information processing \mycite{Li04,Bra05,Liu10}.
Rabi flopping -- an important test for quantum coherence -- has been observed already in 2001 \mycite{Sti01,Kam01}. 

STIRAP has generated significant theoretical interest in QDs, including proposals for 

\begin{itemize}

\item
  using optical excitations (excitons) in two coupled QDs as qubits, with the Coulomb interactions between the optically excited electrons and holes providing the means to construct conditional quantum gates \mycite{Hoh00};

\item
 a qubit represented by the spin of an excess electron in a vertically coupled double-QD structure (an ``artificial molecule'', see Fig.~\ref{Fig:Hohenester}), using auxiliary states to perform quantum gates by STIRAP, thereby suppressing environment-induced losses \mycite{Tro03,Hoh06,Che04};

\item
measurement schemes for the state of single-QD and double-QD qubits \mycite{Paz01,Paz02};

\item
a scheme  to distill, transport, and detect spin entanglement between two correlated electrons in three coupled QDs   \mycite{Fab05};

\item
 a vacuum-STIRAP scheme with a single QD in a microcavity in the presence of lateral electric field  \mycite{Jar11};

\item
 schemes
 to prepare arbitrary superposition states \mycite{Bra01},
 to reduce the sensitivity to unequal transition dipole moments \mycite{Abe06},
 to reduce the coupling to the phonon degrees of freedom \mycite{Ros05},
 and to optimize fidelity in double-QDs \mycite{Koh13};

\item
 coherent manipulation of an asymmetric double-QD structure 
  \mycite{Vou07}
and
coherent electron transfer between the ground states of two coupled QDs 
\mycite{Fou13}.

\end{itemize}

\begin{figure}[tb]
\includegraphics[width=0.95\columnwidth]{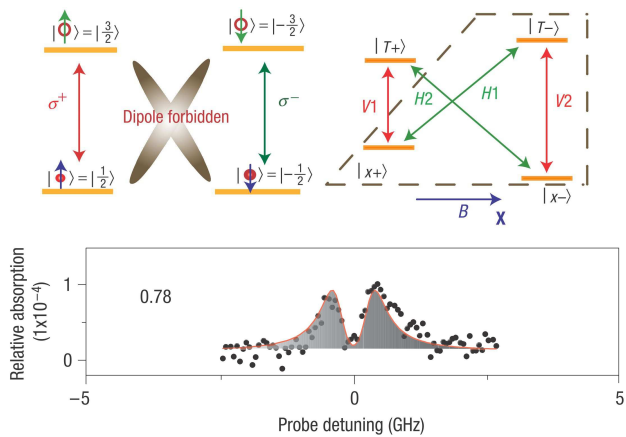}
\caption{\coloronline
Experimental demonstration of  CPT in a single quantum dot by  \mytcite{Xu08}.
\emph{Top}:  Trion energy level diagrams without (upper left) and with (upper right) magnetic field.
 V (H)  means vertical (horizontal) polarization.
At zero magnetic field, the spin-flip Raman transitions are dipole forbidden, while they become allowed for nonzero magnetic field.
The dashed lines   isolate the  three-state   \L~linkage  used in the experiment.
\emph{Bottom}: Experimental evidence of  CPT  in the probe absorption spectrum across transition H1 with the driving field applied on the transition V2. The dip reveals the formation of the dark state.
\locationa{Figs.~1 and 2 of   \mytcite{Xu08}.}
}
\label{Fig:Xu}
\end{figure}

On the experimental side, STIRAP is still to be demonstrated in quantum dots.
Recently, several important steps toward such demonstration were taken.
 \mytcite{Xu08}   demonstrated CPT in the two ground states (spin-up and spin-down) of an electron spin in a single self-assembled InAs quantum dot embedded in a Schottky diode structure, see Fig.~\ref{Fig:Xu}.
 \mytcite{Wei12}  demonstrated CPT reaching zero absorption with the single and triplet ground states of a quantum dot molecule by using a ``sweet spot'' in the bias parameters, which allowed them   to increase the CPT lifetime by two orders of magnitude, to 200 ns.
 \mytcite{Bru09}  demonstrated CPT with a zero-absorption dip of some 100 MHz width on a hole spin {(which has   longer dephasing times than electron spins)} in a single InGaAs quantum dot.
  \mytcite{Hou14}  demonstrated {an  atom-like CPT width of 10 MHz} with a hole spin in a semiconductor heterostructure.

Finally, we point out the successful experimental demonstration by  \mytcite{Sim11}  of rapid adiabatic passage (RAP) {with nearly 90\% efficiency between the ground state and the neutral exciton X$^0$ state of a single InAs quantum dot using a frequency-swept laser pulse.}

These advances present strong evidences for the feasibility of STIRAP in semiconductor quantum dots.

\subsection{Semiconductor quantum wells  \label{Sec:QWells}}

Another semiconductor structure -- a quantum well -- has been proposed as a possible medium for STIRAP.
Quantum wells are easier to fabricate than quantum dots and may be operated at room temperature.
A quantum well consists of a very thin layer of one material, sandwiched by two layers of different material with a larger band gap.
The carriers are trapped inside the middle layer, the thickness of which is comparable to the de Broglie wavelength of the carriers.
Therefore, the energy in the confinement direction is quantized, but the carriers are free to move in the plane of the layer and  hence their energy spectra are continuous.

Nonetheless, many coherent optical effects have been predicted from solutions of multiband nonlinear semiconductor Bloch equations, including Rabi oscillations, photon echo \mycite{Lin92}, self-induced transparency \mycite{Koc92}, adiabatic following \mycite{Bin90}, CPT, EIT, STIRAP \mycite{Lin95,Bin98}, multistate STIRAP \mycite{Jin05}, and chirped-pulse adiabatic passage 
\mycite{Pas10}.
\mytcite{Lin95} predicted that, despite the band energy structure and strong excitonic many-body effects (e.g., electron-electron correlations), signatures of dark/trapping states in semiconductor wells should be observable in experiments with femtosecond pulses.
Although the exact analog of the trapping state does not exist,
\mytcite{Bin98} found that population transfer of STIRAP type in $p$-doped quantum wells only requires an approximate trapping condition to be fulfilled.
\mytcite{Ruf00,Tsu00}
 predicted electron teleportation in a GaAs charge-transfer double quantum well -- STIRAP-type coherent transfer of an electron between one quantum well to its hole-filled neighbor via a common excitonic state -- by mid-infrared femtosecond pulses.

Some of these proposals have been already experimentally demonstrated.
\mytcite{Sch99} observed Rabi oscillations of the heavy-hole exciton density on a subpicosecond time scale.
\mytcite{Ser00} observed EIT in InGaAs  intersubband quantum well system.
\mytcite{Phi03,Phi04}
demonstrated EIT in a GaAs quantum-well experiment, in which the absorption of an exciton resonance was reduced by a factor of 20.
\mytcite{Sla10}
observed EIT in low-doped $n$-type GaAs.
\mytcite{Fu05}
demonstrated CPT in high-purity $n$-type GaAs subjected to a strong magnetic field by using a   \L~system formed of two Zeeman states of neutral-donor bound electrons and the lowest Zeeman state of bound excitons.
\mytcite{Fro06}
achieved light slowing by a factor of 40 and gain without inversion.
\mytcite{Tom12}
 identified and characterized with few-cycle terahertz pulses a three-level system suitable for implementation of STIRAP in a quantum-well microcavity.
When the exciton is nearly resonant with a cavity resonance, the quantum well and the cavity become strongly coupled and give rise to exciton-polariton modes.
The  \L~system is formed of the lower and higher exciton-polariton modes and the optically forbidden $2p$-exciton state.

\section{Classical analogues of STIRAP  \label{sec-classical}}

In this section we discuss a few examples that are implementations of the STIRAP concept beyond quantum physics.
The analogy with STIRAP arises from the similarity of the Schr\"odinger equation and a specific given type of equation of motion.

\subsection{Waveguide optics \label{sec waveguides}}

The similarity of the two-dimensional time-dependent Schr\"{o}dinger equation in quantum mechanics,
\be
\ii \hbar \frac{\partial}{\partial t} \Psi =
  -\left[ \frac{\hbar^2}{2m} \left( \frac{\partial^2}{\partial x^2} + \frac{\partial^2}{\partial y^2} \right) -V \right] \Psi,
\ee
to the  paraxial Helmholtz equation of monochromatic light  propagating along the $z$-axis,
\be\label{eq-propE}
\ii \lambda \frac{\partial}{\partial z} \calE =
 -\left[ \frac{\lambda^2}{2n_0} \left(\frac{\partial^2}{\partial x^2} + \frac{\partial^2}{\partial y^2} \right)+\Delta n \right] \calE,
\ee
suggests identifying the wavefunction $\Psi \left(x,y,t\right) $ with the  electric-field amplitude $\calE(x,y,z)$ \mycite{Lon09}.
The incremental refractive index $\Delta n(x,y,z) $ takes the role of the potential $V(x,y,t) $ and propagation is not in time $t$ but along the $z$-coordinate.
The Dirac constant $\hbar $ becomes the wavelength $\lambda $\ and the particle mass $m$ becomes the background refractive index $n_{0}$.

The transcription from the partial differential equation \eqref{eq-propE} to a set of coupled equations
analogous to those of the TDSE is obtained by introducing discrete field modes analogous to the discrete quantum states.
The result 
  is the coupled-mode formalism \mycite{Pie54, Yar73} descriptive of the 1D propagation of electromagnetic fields along confining paths such as optical waveguides (WGs).
The similarity of the sets of equations arising in two very different contexts makes possible analogies between quantum-state manipulation such as STIRAP and \wg~behavior.

\subsubsection{Light transfer in a set of three waveguides}

\begin{figure}[tb]
\includegraphics[width=0.80\columnwidth]{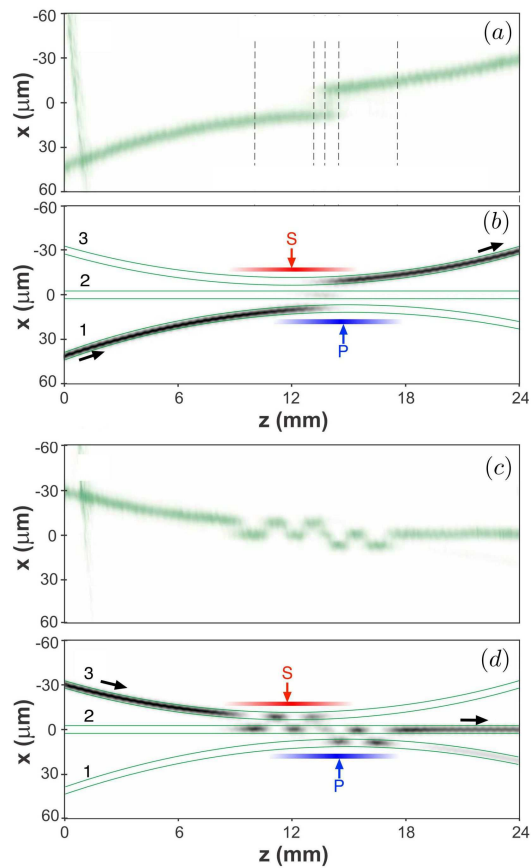}
\caption{\coloronline
Experimental data [frames $(a)$  and $(c)$] and simulation [frames $(b)$ and $(d)$] of light propagating through a set of three \wgs.
Frames $(a)$ and $(b)$ show couplings that correspond to the counterintuitive interaction sequence, producing STIRAP-like transfer of light intensity from \wg~1 to \wg~3,  while frames $(c)$ and $(d)$ show couplings that correspond to the  intuitive interaction sequence leading to Rabi oscillations.
Horizontal red and blue lines added to frames $(b)$ and $(d)$ indicate the extent of $S$ and $P$ interactions.
\locationa{Fig. 8 of \mycite{Lon09}} 
}
\label{fig-Longhi}
\end{figure}

The STIRAP analog  in a  three-\wg~directional coupler was suggested theoretically by \mytcite{Ken01},  further explored by  \mytcite{Pas06,Lon06a,Lon06b}, and experimentally demonstrated  by  \mytcite{Lon07} [see the review by \mytcite{Lon09}].
The  scheme of \mytcite{Lon07}, seen in Fig.~\ref{fig-Longhi}, comprises  three suitably bent \wgs.
Light transfer between adjacent \wgs~occurs due to coupling via the evanescent wave that accompany the transport of light through the central part of the WG structure.
This coupling depends  on the distance between the \wgs~and  the wavelength of the light.
To mimic STIRAP the light is injected into \wg~1,  which approaches the middle \wg~2 only after the initially empty \wg~3 has approached \wg~2,
  in analogy with the counterintuitive sequence of STIRAP. 
The variation of the coupling strength occurs adiabatically and complete transfer between \wgs~1 and 3 results.
Little field energy resides in \wg~2. 
Moreover, because in the adiabatic limit the light transfer is insensitive to the coupling strength, it is largely achromatic.

When the light is injected in \wg~3, it ``sees'' the \wg~couplings in the intuitive ordering, and one observes Rabi oscillations between \wgs~2 and 3, see Fig.~\ref{fig-Longhi}(c) and (d).
In the situation shown the conditions are adjusted such that the light exits through the central \wg~2.

A \wg~analogue of fractional-STIRAP 
was suggested theoretically by \mytcite{Pas06},
 who considered an  equal splitting of optical power between the two outer \wgs.
 This was demonstrated experimentally by    \mytcite{Dre09a}.
The couplings start in counterintuitive order, but terminate simultaneously with equal values of their strengths. 
A tripod-STIRAP \wg~analogue was suggested by \mytcite{Hop15} for use as adiabatic quantum gates that produce a 50:50 and $\frac13:\frac23$ beam splitter, and for a CNOT gate in a planar thin, shallow-ridge \wg~structure.
\mytcite{Men13} used the dependence of the coupling between WGs on the light's wavelength to experimentally demonstrate a STIRAP-inspired optical device that simultaneously behaves as a low- and high-pass spectral filter.

\subsubsection{Multiple waveguides \label{sec-multiple-analogues}}

\begin{figure}[tb]
\includegraphics[width=0.85\columnwidth]{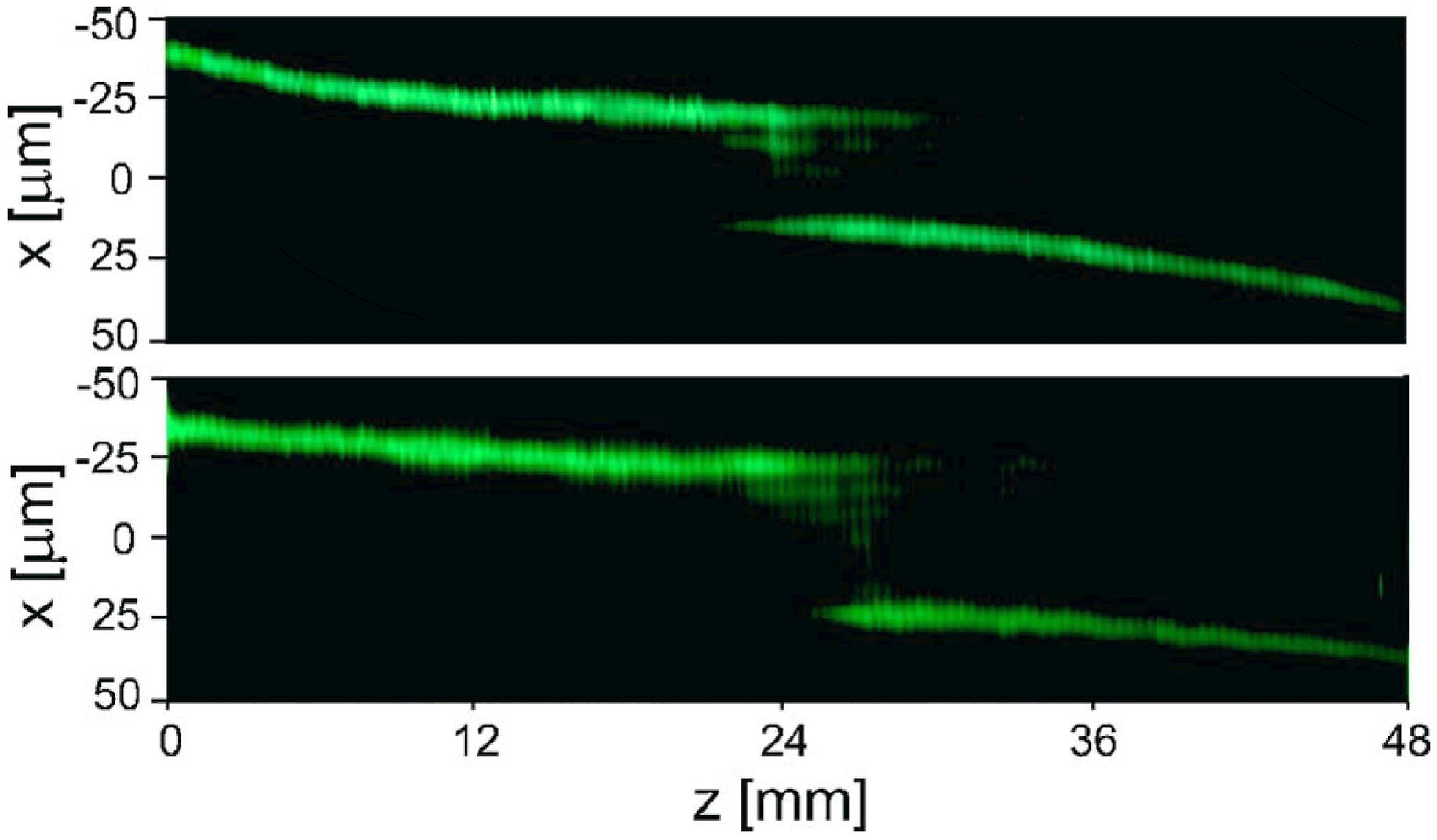}
\caption{\coloronline
Straddle-STIRAP analog in \wg-structure for 3 (top) and 5 (bottom) middle \wgs.
\locationa{Figs. 2 and 3 of \mytcite{Del08}.}
}
\label{fig:DellaValle2008}
\end{figure}

In another development,   \mytcite{Del08} confirmed experimentally the proposal by \mytcite{Lon06b} for a straddle-STIRAP analogue (Sec.~\ref{Sec:straddle}).
 They transferred light between two \wgs~separated by sets of three and five 
  optical \wgs, achieving nearly perfect efficiency, with negligible transient transfer into the intermediate \wgs, see Fig.~\ref{fig:DellaValle2008}.
The achromatic nature of multiple-\wg~STIRAP was demonstrated experimentally by  \mytcite{Cir13} in arrays of up to  nine \wgs.

 An extension of  STIRAP in which the final \wg~is replaced by a set of \wgs, and which produces complete population transfer
to a superposition of this set of \wgs, was proposed by    \mytcite{Ran12}, and
demonstrated experimentally in a reconfigurable light-induced \wg~structure by   \mytcite{Cir12}. 

\begin{figure}[tb]
\includegraphics[width=0.95\columnwidth]{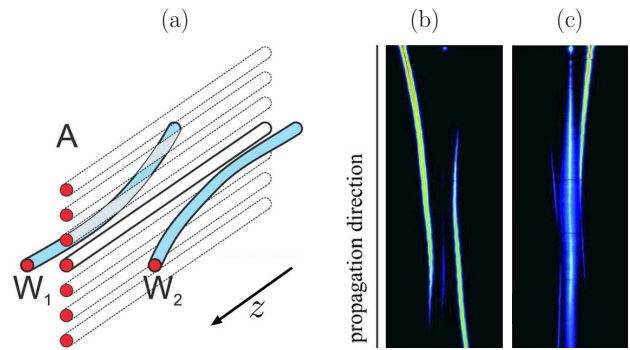}
\caption{\coloronline
STIRAP-like action in \wgs~via a quasi-continuum.
(a) \wg~arrangement: two \wgs~$W_1$ and $W_2$ couple to a common \wg~array $A$.
(b) Experimental results, fluorescence image of light transfer by  counterintuitive ordering of interactions: input enters in $W_1$ and and transfers to $W_2$.
(c) As in frame $(b)$ but with intuitive ordering: input enters in $W_2$ and is lost into the quasi-continuum array $A$.
 \locationa{Figs. 1 and 2 of  \mytcite{Dre09b}.}
 }
\label{fig-Dre09b}
\end{figure}

As it was discussed in Sec.~\ref{Sec:continuum}, replacing the middle state $\s2$ in the $\Lambda$ system of STIRAP by a quasi-continuum of equidistant discrete states allows STIRAP-like population transfer $\s1\to\s3$.
The \wg~analogue of such behavior  was investigated theoretically by   \mytcite{Lon08} and experimentally demonstrated by   \mytcite{Dre09b}.
In the experiment, two weakly-curved single-mode \wgs,
  $W_1$ and $W_2$, approach a \wg~array $A$ from different sides, as depicted in Fig.~\ref{fig-Dre09b}$(a)$.
Input through $W_2$ encounters the array before $W_1$ is present -- this is the ``intuitive'' ordering of pulses.
It  introduces radiation into the intermediate array,   the analogue of exciting an atomic electron into the ionization continuum, Fig.~\ref{fig-Dre09b}(c).
When the coupling strengths are in the ``counterintuitive'' ordering [Fig.~\ref{fig-Dre09b}(b)], then transfer of light from $W_1$ to $W_2$ occurs without engaging the intermediate array.
The fact that a large set of \wgs~in array A 
 does indeed participate in  the transfer process has been confirmed by observing the light exciting this array.

\subsection{Polarization optics \label{Sec:polarization}}

\begin{figure}[tb]
\includegraphics[width=0.98\columnwidth]{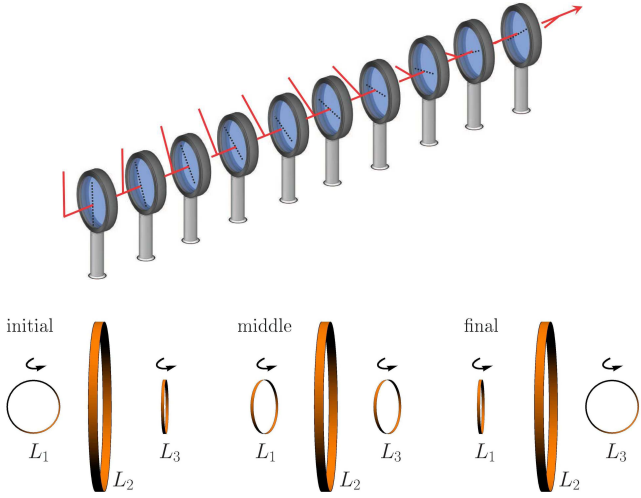}
\caption{\coloronline
\emph{Top}: Light polarization rotator based on piecewise adiabatic passage.
The linear polarization vector (solid lines) adiabatically follows the rotating fast polarization axis of the half-wave plates (dotted lines).
\locationa{Figs. 1 of  \mytcite{Dim15}.}
\emph{Bottom}: STIRAP-inspired wireless energy transfer between three loops $L_1, L_2, L_3$ carrying AC current.
The energy is transferred between loops $L_1$ and $L_3$, each of which undergoes a rotation of  90$^{\circ}$.
The middle loop is much larger (in order to ensure stronger couplings) and hence much lossier but it acquires only negligible energy.
The relative orientations proceed from $ L_1 \perp L_2 \, || \,  L_3 $ to  $ L_1 \, ||  \,  L_2 \perp  L_3 $.
\locationa{Fig. 1 of  \mytcite{Ham09}.}
\black
}
\label{fig:classical}
\end{figure}

As we discussed in Sec.~\ref{sec-two state STIRAP}, on exact resonance STIRAP can be described by a torque-like equation of motion, Eq.~\eqref{torque}.
Similar equations appear in various classical systems.
One important example is the equation describing the evolution of the Stokes polarization vector $\mathbf{S}$ \mycite{Bic85} of light propagating along the z-axis through an optically anisotropic medium with no polarization-dependent losses, 
\be\label{Stokes vector evolution}
\frac{d\mathbf{S}}{dz} = \mathbf{\Omega} \times \mathbf{S},
\ee
 where $\mathbf{\Omega }$ is the birefringence vector of the medium.
The analogue process of STIRAP allows efficient rotation or/and conversion of light polarization,
 which is achromatic and robust to variations in the propagation length and the rotary power \mycite{Ran10}.

Particularly suitable for light polarization manipulation is the discretized version of STIRAP -- piecewise adiabatic passage [Sec.~\ref{Sec:PAP} and \mytcite{Sho16}].
\mytcite{Dim15} experimentally demonstrated an analogue of this technique with the setup in Fig.~\ref{fig:classical}(top).
The Stokes vector $\mathbf{S}(z) $ is initially parallel to the birefringence vector $\mathbf{\Omega }(z)$ (the fast optical axis is parallel to the polarization vector), such that $\mathbf{\Omega } \left(z_{i}\right) \times \mathbf{S}\left(z_{i}\right) =0$.
If the orientation of the fast optical axis is changed slowly from one birefringent element to the next one, the Stokes vector will also follow up.

The polarization optics analogy with STIRAP was used to design an achromatic fibre-optical isolator (an optical diode) \mycite{Ber13a,Ber13b}.
Here the STIRAP-analog mechanism produces broadband reciprocal and non-reciprocal quarter-wave plates, which in combination work as an optical isolator.

\subsection{Further applications of the STIRAP concept in classical systems
\label{sec-further}}

Other notable applications of the STIRAP concept to classical systems include

\begin{itemize}

\item light transfer between \wgs~in the presence of an intensity-dependent index of refraction due to the optical Kerr effect \mycite{Lah08,Kaz10};

\item using $\chi ^{(2)}$ nonlinearity to induce spontaneous parametric down-conversion in a double-STIRAP process in a configuration of six planar \wgs~
\mycite{Wu14};

\item wireless energy transfer \mycite{Ham09,Ham13}, see Fig.~\ref{fig:classical}(bottom);

\item third-harmonic generation via the cascaded processes of second harmonic generation ($\omega +\omega \rightarrow 2\omega $) and sum-frequency generation ($\omega +2\omega \rightarrow 3\omega $) in $\chi^{(2)}$  nonlinear media  without transient generation of the second harmonic $2\omega$ \mycite{Lon07};

\item sum- and difference-frequency generation \mycite{Por12};

\item four-wave mixing demonstrated experimentally in ${}^{87}$Rb atoms with 70\% efficiency \mycite{Vew07};

\item classical data processing \mycite{Rem06,Bei11};

\item a new look at the rotation of magnetization \mycite{Ran09} based on a reinterpretation of the Bloch equations \mycite{Blo46};

\item a new look at the manipulation of the direction of a charged particle under the influence of a Lorentz force \mycite{Ran09};

\item a new look at the action of the Coriolis force on a moving particle \mycite{Ran09}.

\end{itemize}


\section{Perspectives for new applications of STIRAP  \label{sec-perspectives}}

This final section focusses on the prospects for promising future applications of STIRAP.
Some of them are on the way to be implemented. 

\subsection{Optomechanics \label{sec-optomechanics}}

\begin{figure}[tb]
\includegraphics[width=0.80\columnwidth]{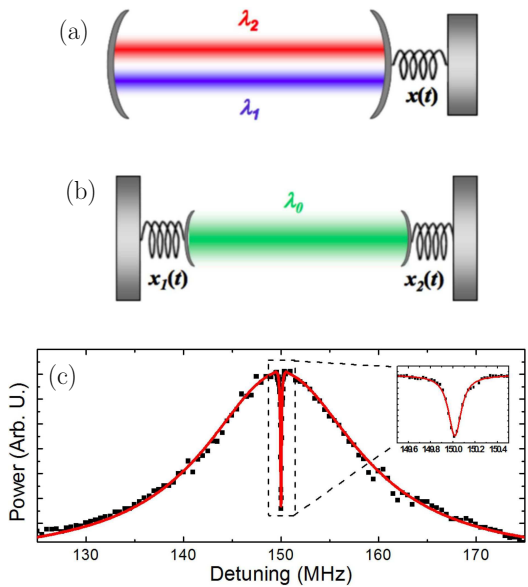} \\
 \caption{\coloronline
Schematics of three-mode optomechanical systems.
(a) A mechanical mode coupled to two optical modes.
(b) An optical mode coupled to two mechanical modes.
 Energy is transferred between optical cavity modes in (a) and the mechanical oscillators in (b).
The optomechanical coupling 
 is controlled by external laser fields (not shown) injected into the cavity with their frequency detuned from the cavity mode by the mechanical oscillators frequency.
\locationa{Fig. 1 of \mytcite{Don12} and Fig. 1 of \mytcite{Don14}}.
(c) The transmission of the weak probe field shows a minimum when the frequency of the control
field is detuned from the cavity resonance (and thus the frequency of the probe field) by one
mechanical resonance frequency. (courtesy of Hailin Wang)
}
\label{fig-Hailin}
\end{figure}

In an optomechanical resonator, circulating optical fields couple to the motion of a mechanical oscillator via radiation pressure force
\mycite{Asp14}, see Fig.~\ref{fig-Hailin}.
A unique property of this type of systems is that an optically-active mechanical mode can couple to any of the optical resonances supported by the resonator. For a three-mode optomechanical system, in which two optical modes
 (e.g., whispering gallery modes supported by the silica microsphere)
 couple to a common mechanical oscillator
 (e.g., a breathing vibration of the silica microsphere), the optomechanical coupling can mediate the transfer of quantum states between the two optical modes, see Fig. \ref{fig-Hailin}(a).
This mechanically-mediated optical state transfer can play an important role in a hybrid quantum network, enabling quantum communication between disparate quantum systems \mycite{Sta10,Tia10,Saf11,Reg11}.
Indeed, theoretical work \mycite{Tia12,Wan12,Wan12b} suggests that STIRAP transfer in such systems is feasible.

 In most transfer schemes overcoming 
the inherent thermal noise of the mechanical system is a challenge \mycite{Wan12}.
STIRAP transfer, however, will proceed via the mechanically-dark state and is thus immune against thermal mechanical noise.
The mechanically-dark optical  mode  has been demonstrated in a recent experiment \mycite{Don12}.
The relatively short photon lifetime, however, has hitherto prevented a direct demonstration of the STIRAP-based optical state transfer in these systems
\mycite{Wan16}.

The phenomenon of  optomechanically induced transparency (OMIT) \mycite{Wei10}  documents the coherent coupling of optical and mechanical degrees of freedom.
A weak probe field is injected into the cavity (Fig.~\ref{fig-Hailin}) at the cavity resonance and its transmission is observed.
Then a strong control field one mechanical frequency away from the cavity resonance is injected.
The beat frequency between these two radiation fields drives the mechanical mode, and thus also the cavity mirror, at its resonance frequency.
The motion of the mirror in turn induces sidebands of the control field, one of which interferes destructively with the weak probe field, leading to the OMIT structure shown in Fig.~\ref{fig-Hailin}.
The signature of the  formation of mechanical dark state, which is established when the radiation modes decouple from the mechanical mode, is the variation of the depth of the OMIT structure with the strength of the probe field {that couples the other optical modes.}

An optical field can also induce the coherent coupling between two mechanical modes in a three-mode system, in which two mechanical oscillators couple to a common optical mode via radiation pressure, Fig.~\ref{fig-Hailin}(b).
Optically-induced mechanical normal modes and so-called mechanical Bogoliubov modes have been realized experimentally \mycite{Shk14,Mas12,Don14}.
Since the mechanical damping rate is typically much smaller than the optical cavity decay rate, this type of three-mode system may provide  an excellent model system for the realization of STIRAP of a mechanical excitation. Specifically, the three-mode system can either  be prepared in a superposition of the two mechanical oscillators, which is decoupled from the optical cavity mode,
or be used to transfer energy between the mechanical oscillators via the cavity mode. 
Such experiments are in progress \mycite{Wan16}.

\subsection{Precision experiments \label{sec-precision}}

STIRAP is likely to play a prominent role in future experimental efforts  {in the search} for the electric dipole moment of the electron $d_e$  (eEDM).
Results from such experiments, which measure extremely small shifts of the energies of quantum states, are of fundamental importance for testing extensions of the standard model of particle physics \mycite{DeM15}.
A nonzero value of $d_e$   would be a source of violation of parity or time-reversal symmetry \mycite{Sak67,Sak91}.
The standard model of particle physics  \mycite{Oer06}  {predicts} a very small value,  $d_e^{SM}  < 10^{-38}$  e cm.
Because the standard model is known to be incomplete, many alternative theories have been proposed,
 nearly all of which predict an eEDM of $d_e  > 10^{-30}$  e cm.
Therefore it is of fundamental interest to measure the eEDM or to determine an upper limit to its value \mycite{Pos05}.

Such efforts have been going  on for several decades [see \mytcite{Ber91}, a compilation of data and references since 1950  can be found in \mytcite{Hes14}].
Progress until 2012 had reduced the upper limit of $d_e$  to about $10^{-27}$  e cm \mycite{Reg02,Hud11,Kar12}.
An order of magnitude smaller upper limit ($d_e  < 10^{-28}$  e cm) was recently determined by the ACME collaboration \mycite{Bar14}.
The next generation of the latter experiment aims at yet another order-of-magnitude  {improved sensitivity to} $d_e$,  which would, if established,  {refute} the predictions of most models in particle physics regarding the value of $d_e$.
The potential of STIRAP for the planned experiments has already been demonstrated \mycite{Pan16}, as will be briefly discussed here.

\begin{figure}[tb]
 \includegraphics[width=0.9\columnwidth]{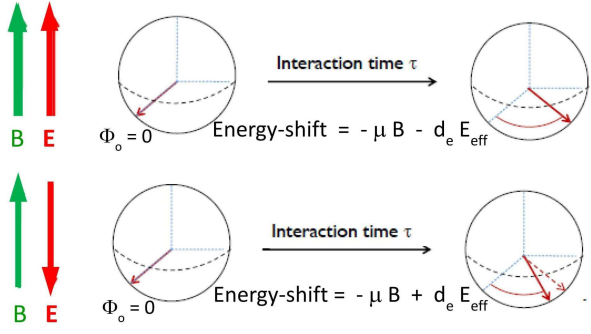}
\caption{\coloronline
A coherent superposition of $M = 1$ and $ -1$ states (leading to an alignment of the electron spin) in the $J = 1$ level of the H  $^3\Delta_1$ metastable electronic state of ThO  is optically prepared by either optical pumping \mycite{Bar14} or, for the next-generation experiments, by STIRAP \mycite{Pan16}.
The spin precession angles $\Phi_+$  and $\Phi_-$  are measured for two opposite directions of the electric field.
The difference $\Delta \Phi  = \Phi_+ - \Phi_-$   yields,  for known effective electric field $E_{\rm{eff}}$, the value of the electric dipole moment $d_e$  of the electron.
 }
\label{Fig:EDM1}
\end{figure}

The central idea is to measure the very small splitting of molecular energy levels (due to different orientations of the eEDM in an electric field).
{This is done}  in a suitable state of the ThO molecule in a  cold molecular beam \mycite{Hut12}.
First, a coherent superposition of  $M = \pm 1$ states in the $J = 1$ level of the H  ${}^3 \Delta_1$  metastable electronic state of ThO is prepared by either optical pumping \mycite{Bar14} or, for the next-generation experiments, by STIRAP \mycite{Pan16}.
The second step is the determination of the precession angle $\Phi$ of the  {electron} spin during the lifetime of a metastable level in a  {magnetic field} in combination with an electric field with the direction of the latter being either parallel (yielding the angle $\Phi_+$) or anti parallel (yielding the angle $\Phi_-$) to the B-field,  see Fig. \ref{Fig:EDM1}.
The precession angle is given by $\Phi_\pm = \Delta E_\pm \tau /\hbar$, were $\Delta E_\pm  = - \mu B  \pm  d_e  E_{\textrm{eff}}$.
Here $\mu$ is the magnetic dipole moment of the given state \mycite{Vut11,Fle14}, $E_{\textrm{eff}}$  is the effective electric field felt by the electron and {$\tau=1.1$ ms} is the time between preparation and detection, limited by the lifetime of the H  ${}^3 \Delta_1$  state or by the flight time of the molecules between the location of preparation and detection.
During a set of experimental runs many experimental parameters are changed to discriminate against systematic errors \mycite{Spa14}.
These procedures are not discussed here.

\begin{figure}[tb]
 \includegraphics[width=0.95\columnwidth]{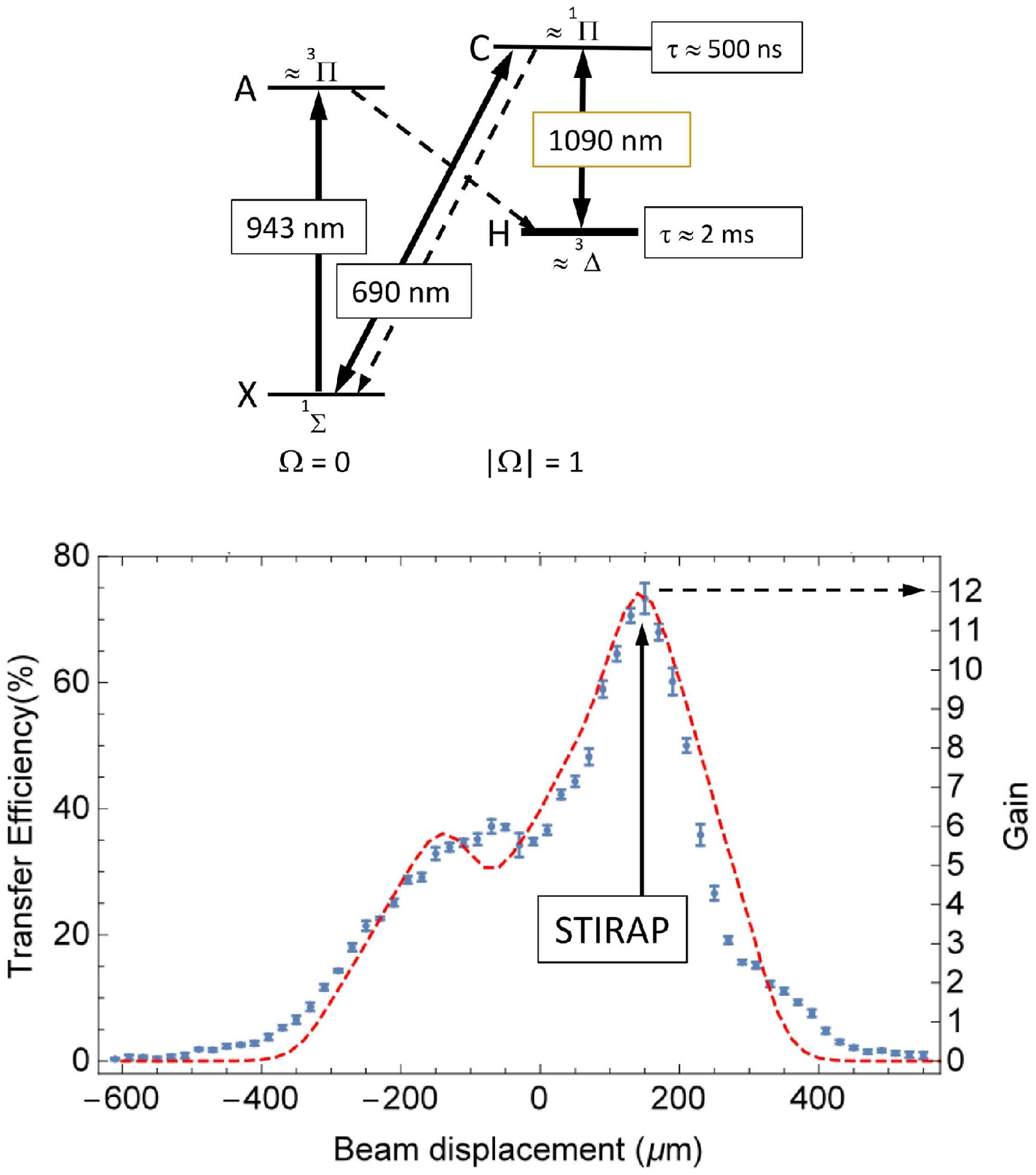}
\caption{\coloronline
\emph{Top}: The relevant level  scheme of ThO for the eEDM  measurement.
 The appropriate level in the H state is either populated by optical pumping via the A state or by STIRAP via the C state.
\locationa{Fig.~2.2 of \mytcite{Spa14}.}
\emph{Bottom}:
STIRAP transfer efficiency (left axis) from the ground state $X^1\Sigma$ of ThO to the state
H $^3\Delta_1$  relevant for the eEDM experiment.
Positive values  of the beam displacement correspond to the STIRAP arrangement.
The right axis shows the enhancement of the population with respect to the optical pumping approach. 
 \locationa{Fig. 3 of \mytcite{Pan16}.}
 }
\label{Fig:EDM2}
\end{figure}

State H  ${}^3 \Delta_1$ of ThO is chosen because its properties are favorable for such measurements: the magnetic moment $\mu$ is very small ($< 10^{-2}$  $\mu_B$, {where $\mu_B$ is the Bohr magneton}) and therefore the dynamics  is not overwhelmed by  the Larmor precession, the polarizability is very large because of very small $\Omega$-splitting (the energy difference between the states $\Omega = +1$  and $ -1$, where $\Omega$ is the projection of the angular moment $J = 1$ on the molecular axis)  and the relativistic enhancement of the externally applied electric field at the location of the electron \mycite{San65,San66} is very large.

Figure \ref{Fig:EDM2}(top) shows the relevant level scheme.
Preparation by optical pumping, as done by \mytcite{Bar14},  proceeds via excitation of state A followed by spontaneous emission to state H.
Preparation of the dark state (by optical pumping) and detection occurs via state C with the fluorescence back to state X being observed.
STIRAP preparation occurs via state C and prepares the needed coherent superposition directly.
A particular noteworthy feature \mycite{Pan16} is that the \Stokes laser, driving the C--H transition, has a power of 10 W  (the power of the laser driving the X--C transition is 50 mW).
The high power is needed to broaden the STIRAP two-photon linewidth to values larger than the Doppler width (in order to address all molecules in the beam).

The gain in signal due to STIRAP over the previous optical-pumping approach is a factor of 12 [Fig.~\ref{Fig:EDM2}(bottom)] with a corresponding gain in  sensitivity of 3.5.
Further modifications  in the experiment are expected to lead to an improvement of the  sensitivity by an order of magnitude over the most recent results \mycite{Bar14} with STIRAP making the largest single contribution \mycite{Gab16}.

An alternative approach for the measurement of eEDM is
followed up by the groups of J. Ye and E. Cornell at JILA/Boulder.
They use a  $^3\Delta_1$ state, however based on a trapped molecular ion, either HfF$^+$  \mycite{Loh13} or potentially ThF$^+$ \mycite{Gre16}.
STIRAP will also play a role in these experiments \mycite{Ye16}.

\subsection{Detection of parity violation in molecules \label{sec-parity}}

\begin{figure}[tb]
\includegraphics[width=0.50\columnwidth]{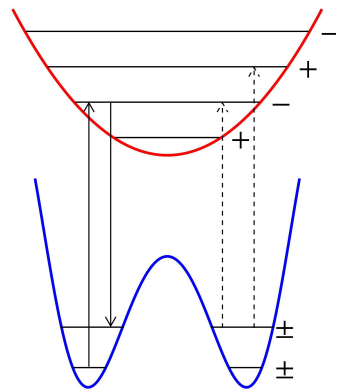}
\caption{ \coloronline
Schematic potential energy curves as a function of a normal coordinate $q$.
A state of well-defined parity is populated through STIRAP from the ground state.
During subsequent evolution and because of parity-violating interaction, the state acquires a contribution from the other parity component, which is detected spectroscopically.
 \locationa{\mytcite{Qua86}.}
}
\label{Fig:Quack}
\end{figure}

A highly significant application of STIRAP has recently been discussed by \mytcite{Die15}.
A fundamental new aspect of the 
 stereochemistry of chiral molecules is the small difference $\Delta E_{\textrm{pv}}$ predicted for the ground states of the enantiomer´s mirror image isomer arising from the parity violating electroweak interaction.
Recent theoretical progress, as summarized by \mytcite{Qua11,Qua14}, predicted that this difference is up to two orders of magnitude larger than predicted by older theories, but still very small, typically in the sub-femto eV range.
So far this effect has not been observed experimentally.
Following a scheme proposed by \mytcite{Qua86} (Fig.~\ref{Fig:Quack}),
 $\Delta E_{\textrm{pv}}$  can be measured in a two-step population transfer scheme to prepare a state of well-defined parity in a molecular beam.
This state subsequently evolves in time and acquires, due to parity violation, a component of the state with opposite parity.
The latter state  is detected very sensitively on the millisecond time scale.

It has been demonstrated in test experiments with ammonia using rapid adiabatic passage \mycite{Lie89} 
 that the sensitivity achievable would be sufficient to detect parity violating energy differences  $\Delta E_{\textrm{pv}}$   as small as 100 aeV.
Model calculations \mycite{Die15} suggest that the high efficiency and robustness of STIRAP will be essential for such an experiment.
Resulting experimental data along with theoretical analysis would provide important information on fundamental parameters of the Standard Model of Particle Physics \mycite{Qua11} and might have, in the longer run, also implications for our understanding of the long standing open question of
homochirality, i.e. the question why the evolution of life has led to the overwhelming dominance of one form of enantiomer over the other in biomolecular systems on Earth \mycite{Qua14}.

\subsection{Chiral molecules  \label{Sec:chiral}}

\begin{figure}[tb]
\includegraphics[width=0.80\columnwidth]{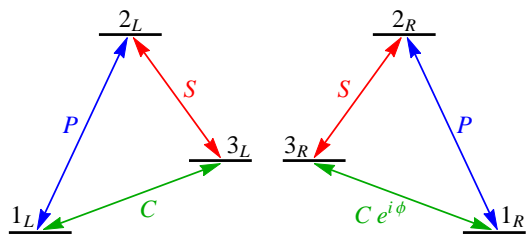}
\caption{\coloronline
Left- and right-handed chirality.
All couplings are the same except for a phase of $\phi$ in the coupling between states \s1 and \s3.
 }
\label{Fig:chiral}
\end{figure}

Chirality is a geometric property of some heteroatomic molecules that do not possess an inversion center.
A chiral molecule is non-superposable on its mirror image.
Chiral molecules of opposite (left and right) handedness are known as ``enantiomers'', and their separation is of significant interest in chemistry.
To this end, \mytcite{Kra01a,Tha03} proposed to use STIRAP for enantiomer separation, and conversion from one  to the other \mycite{Kra03}.
Because of the broken symmetry in chiral molecules, the molecular states do not have a definite parity and all single-photon transitions between the three molecular states are allowed (Fig.~\ref{Fig:chiral}).
The enantiomer separation is possible due to the different phases of the transition dipoles, and hence the couplings. 
Because STIRAP alone is insensitive to the phases of the fields, it is supplemented by another single-photon field on the $\s1\to\s3$ transition, thereby forming a closed loop, which is phase-sensitive.
This allows one to direct the population toward different final states in the two enantiomers, and hence separate them with subsequent state-selective manipulation.
For example, if both enantiomers are initially in state \s1, then, depending on the phase $\phi$, one of them can be transferred to state \s2 and the other to state \s3, by the same driving fields.
\mytcite{Ger04} used this method to simulate purification of a (so-called racemic) mixture of dimethylallene with 95\% efficiency.
Finally, \mytcite{Kra05} proposed to create entanglement between enantiomers using non-classical light.

\subsection{Spectroscopy of core-nonpenetrating Rydberg states \label{sec-core}}

\begin{figure}[tb]
\includegraphics[width=0.90\columnwidth]{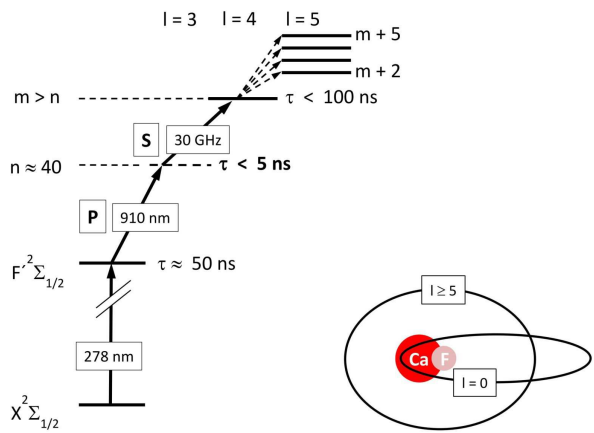}
\caption{\coloronline
\emph{Left}: Level scheme for CaF. Microwave spectroscopy of  $l = 5$ Rydberg states requires efficient population of a state $m > 40, l = 4$.
Traditional stepwise excitation is not possible because states such as the one with $n \approx 40, l = 3$ decay  rapidly.
Pulsed excitation from the ground state to the $F'$ state followed by optical-microwave STIRAP transfer allows bypassing the rapidly decaying state and depositing population into a sufficiently high-lying long-lived Rydberg state with $l = 4$.
\emph{Right}: Schematics of a $l = 0$ (core penetrating) and high $l$ (core non-penetrating) orbital, with the CaF molecule chosen as an example.
}
\label{Fig:core}
\end{figure}

Spectroscopy of molecular Rydberg states is a powerful tool in molecular physics \mycite{Eyl86,Lun05},  until now mostly neglected.
In particular, Rydberg states with angular momentum $l \geq 5$ have negligible overlap with the ionic core and their level structure is hydrogen-like [Fig.~\ref{Fig:core}(right)].
However, the finite extension of the ionic core and its deviation from spherical symmetry lead to small deviations from the hydrogenic level structure \mycite{Kay11}.
Hence these core-nonpenetrating (CNP) Rydberg states provide a platform for precision measurements of the mechanical and electric properties of molecular ions \mycite{Spr14,Jan15,Haa15}.
Moreover, CNP Rydberg states are wanted  for highly efficient Stark slowing and trapping of molecules \mycite{Hog09,Hog12}.

Core-nonpenetrating Rydberg states are stable against nonradiative decay and their electronic lifetimes approach that of long-lived atomic Rydberg states, making them valuable stepping stones and targets for molecular Rydberg experiments.
Lacking overlap with the ground-state wavefunction, CNP states cannot be accessed by one-photon transitions.
Stepwise excitation, however, is usually  blocked by fast nonradiative decay (predissociation, autoionization, and inter-system crossing) in the intermediate electronic states, see Fig.~\ref{Fig:core}(left).

STIRAP is well suited for efficient transfer of population into CNP-Rydberg states while avoiding the rapid nonradiative decay of the intermediate states.
Optical-microwave STIRAP, taking advantage of the very large Rydberg-Rydberg electric dipole transition moments of the order of up to 1000 Debye, is being developed for this purpose \mycite{Kay08,Pro11,Col13,Zho15,Fie16}.

\subsection{Bright beams  of ultracold atoms \label{sec-bright beams}}

Another very promising application of STIRAP has been proposed by \mytcite{Rai14}.
Based on recent success in the preparation of slow atomic beams by magnetic deceleration techniques \mycite{Hog08,Rai09} they developed a scheme for increasing  the phase-space density of cold atomic ensemble,  potentially leading  to very bright low-velocity atomic beams for chemical dynamics studies \mycite{Nar12}.
That approach starts from a pulsed supersonic beam of rare gas atoms (usually Ar or Ne) with atoms entrained by, e.g.,  laser ablation or metastable atoms produced by an electrical discharge, yielding -- in the frame moving at the mean speed of the particles in the  beam -- a cold atomic ensemble.
The atoms are slowed or stopped by magnetic forces using the "coilgun" concept \mycite{Nar08}, which is  the magnetic analogue of the deceleration of neutral dipolar molecules by electric forces \mycite{Bet99,Mee12}.

The process that shall lead to an increase of the phase space density by more than two orders of magnitude is based on a combination of magnetic forces and STIRAP. 
\mytcite{Rai14} explained the approach using as an example  metastable  ${}^4$He with angular momentum  $J = 1$ both in the lower and upper state.
The method can be adapted for other angular momentum situations.
It leads to a compression of both the velocity and the spatial distribution and  exploits not only the robustness of STIRAP with regard to small variations of experimental parameters but also its sensitivity to the two-photon resonance.

Reduction of the width of the velocity distribution, e.g., in the $x$-direction, starts with optically pumping the atoms into the $m = 0$ state followed by STIRAP transfer to the $m = 1$ state by two orthogonally linearly polarized counter-propagating laser beams.
In the counter-propagating configuration the transfer process is velocity selective.
The frequencies of the lasers are tuned such that the two-photon resonance is met for those atoms with a velocity $v_x^*$ of, e.g.,  $v_x > 0$ in the wings of the profile.
The range of velocities, which participate in the transfer process, depends on the width of the two-photon resonance, i.e., on the power of the lasers.
The atoms in $m = 1$ are then exposed to an inhomogeneous magnetic field that decelerates the atoms before they are transferred back to $m = 0$ by optical pumping.
This process can be repeated with slightly detuned laser frequencies in order to address atoms with $v_x  < v_x^*$.
Therefore, by stepwise changing the frequencies of the lasers, atoms, which initially populate the wings of the velocity distribution, are pushed towards  $v_x  =  0$.
The process can be repeated for $v_x  <  0$ and the $y$ and $z$ directions.
We note that this cooling scheme does not rely on the transfer of photon momentum to the atoms by STIRAP, as suggested as a cooling mechanism by \mytcite{Kor96,Iva12}.

Reduction of the width of the spatial distribution starts by limiting the overlap of the STIRAP laser with the atomic ensemble to, e.g., one-half  of the distribution followed by  magnetic forces to spatially shift the atoms in state $m = 1$ until they overlap with the untreated parts.
Although the sequence of processes appears rather complex numerical  simulations suggest that the method is feasible resulting in a significantly enhanced phase-space density, ultimately providing  a new source for beams of ultracold atoms with unprecedented  brightness.

\subsection{Preparation of polarized  diatomic molecules
\label{sec-polarized molecules}}

In most STIRAP experiments the coupling between the initial and final state is via a level in an electronically excited state.
Reaching such levels may require UV or VUV radiation.
\mytcite{Muk10a} suggested the preparation of polarized vibrationally excited states for stereodynamic studies of chemical processes by two time-delayed infrared laser pulses.
With a proper choice of the initial rotational level and suitable polarization of the \pump and \Stokes laser beams an ensemble of diatomic molecules in
a rovibronic level $(j'', v'')$ can be prepared, which is either polarized, aligned or unpolarized [see also \mytcite{Vew10}].
Since the transition dipole moments are strongest for $\Delta v = 1$, it will be easiest to prepare such ensembles in the vibrationally excited state $v'' = 2$.
The authors show through numerical studies, using the properties of the HCl molecule as an example, that complete population transfer can be achieved using the radiation of quantum cascade lasers \mycite{Fai13} with a linewidth of the order of 10 kHz and intensities of the order of 30 mW/mm$^2$.
Although quantum cascade lasers do provide suitable radiation, experimental demonstration of such a STIRAP process with infrared lasers has not yet been reported.

\subsection{Polarization of high angular-momentum states \label{sec-high angular momentum}}

Polarized high angular-momentum states, i.e., states  with all the population in $m = J$  or $ - J$,  are of interest to experiments in quantum optics, atomic physics and metrology \mycite{Auz10}.
Such states can be prepared by optical pumping depletion of all $m$-states except the end states $|m| = J$  \mycite{Hef86} or only one of them.
However, in such schemes, most of the initial population of level $J$ is lost.
Searching for a scheme that minimizes the number of steps of optical interactions (excitation, stimulated emission, and spontaneous decay), \mytcite{Roc16} proposed a scheme, involving STIRAP, in which most, if not all, the population in a given  $J$-state (degenerate ensemble 1 of states) is accumulated at one of  the end states.
However, because STIRAP, like any coherent process, cannot transfer thermal population of different states into a single one, that scheme needs to involve also optical pumping  (to a degenerate ensemble 2 of states) followed by spontaneous emission.
The latter process may lead to some loss of the initial population.
\mytcite{Roc16} suggested using an additional degenerate ensemble 3 of states, serving as shelf states.
Using circularly and linearly polarized radiation, the population is driven by STIRAP back and forth between states of ensemble 1 and 3 via ensemble 2, driving the population towards one of the end states.
After a few transfer  cycles the system needs to relax to the ground state by spontaneous emission before the process is repeated.
At the end, most, if not all, of the initial population is accumulated at  one of  the end states,  $m = J$ or $ - J$.
A scheme using two shelf states was also proposed, with fewer spontaneous emission processes.

\subsection{Nanoscale resolution for fluorescence microscopy \label{sec-microscopy}}

Stefan Hell received the 2014 Nobel prize in chemistry \mycite{Hel15} for the development of the stimulated-emission-depletion method (STED) which allows nanoscale resolution in optical microscopy \mycite{Hel94}.
\mytcite{Mom09}, \mytcite{Vis12} and \mytcite{Rub13} presented simulations, which suggest further improvement of the resolution of fluorescence microscopy when the depletion is done by population transfer via STIRAP rather than STED.
The challenge in the experimental implementation of that approach lies in the application of STIRAP to large molecules, which remains to be demonstrated.

\subsection{Atmospheric chemistry \label{sec-atmospheric}}

As discussed in detail by \mytcite{Ber15},  the development of STIRAP was initially driven by the hope to solve problems related to chemical processes  in the atmosphere.
A major challenge in atmospheric chemistry remains the study of reaction processes of vibrationally excited molecules such as O$_2$, N$_2$  or OH.
Although much progress has been made in recent years, it is still true that many reactions and energy transfer processes involving highly vibrationally-excited species are poorly understood, although they are important for the chemistry of planetary atmospheres \mycite{Vai14}.
For instance, the collision processes of OH($v''\gg 1$) molecules, formed in high vibrational levels through,
e.g., the reaction of ozone and hydrogen, are of interest \mycite{Kal11}, as is the vibrational dependence of the dissociative  combination of O$_2^+$  and CO$_2^+$  \mycite{Pet05b}.

Unfortunately, such experiments have not yet been done, because many molecules have their first electronic state at energies that require radiation fields in the ultraviolet or vacuum-ultraviolet (VUV) region.
Although such radiation sources have been available for many years, their poor coherence properties make them unsuitable for the implementation of STIRAP.

However,  many new radiation sources are currently under development, including large-scale machines such as free-electron lasers, and some of those sources are expected to yield radiation with good coherence properties \mycite{Har13}.
Therefore, it may well be possible in the near future to use STIRAP for efficient and selective vibrational excitation of molecules of interest to atmospheric chemistry, such as N$_2$, O$_2$, H$_2$  and others, to a level with vibrational quantum number $v''\gg 1$.
The appendix of \mytcite{Ber15} shows that for a pulse length of the order of 10 ns and wavelength of about 150 nm a fluence of no more than a few $\mu$J/mm$^2$  suffices to successfully apply STIRAP for the efficient and selective preparation of high vibrational levels in the electronic ground state of H$_2$, O$_2$, and NO.

\subsection{X-rays: Innershell excitation and nuclear physics\label{sec-x-rays}}

\bws
At first glance,  the statements earlier in this article and in particular, in Sec.~\ref{sec-adiabatic}, seem to rule out any possibility of implementing STIRAP with radiation in the x-ray regime.
\bws
However, looking  ahead a few years we can be encouraged by
new radiation sources
\mycite{Hem14}
and the use of
temporally-coherent light and pulse lengths as long as a few ps
\mycite{Ama12,Har13,All12,All13}.
Therefore meeting the adiabaticity criterion for efficient population transfer may soon be possible.

\bws
Based on this perspective, \mytcite{Pic15}
proposed the implementation of STIRAP with two-color high-intensity highly-coherent few-femtosecond x-ray pulses from free-electron lasers.
\bws
Such x-ray STIRAP would allow  one to use inner-shell resonances as the middle state, without populating them and thus avoiding radiation damage. The results of their simulations suggest that robust population transfer in neon atoms and carbon monoxide molecules is feasible. X-rays allow large penetration depths and could be of interest in experiments with liquids or buried interfaces of materials.

An even further-reaching proposal has been made by
\mytcite{Lia11,Lia13},
\bws
who suggested to use STIRAP for population transfer between states of  nuclei with transition energies of a few 100 keV.
The short wavelengths needed in the frame of the nuclei would be achieved by accelerating them to the relativistic regime and thereby Doppler-shift the frequency of the x-ray radiation to match the target resonance. 
 \mytcite{Lia11} proposed two scenarios.
In one of them a single-wavelength source would be used with the \Stokes and  \pump  radiation crossing the trajectory of the nuclei at different angles to realize the needed Doppler shift.
The challenge of this approach is the required high precision of the timing of the radiation pulses. 
Alternatively, it is more realistic to use pulses of two different x-ray wavelengths, propagating collinearly and crossing the trajectory of the nuclei at an angle of (nearly) 180$^{\circ}$.
Indeed, calculations suggest 
 that the adiabaticity criterion for STIRAP 
 can be met with the upcoming coherent x-ray sources.

STIRAP transfer between states of nuclei would be particularly interesting when metastable, isomeric nuclei states are involved.
Such states may have energies of several MeV above the ground state and thus can store a large amount of energy over a long period of time
\mycite{Wal99}.
 STIRAP could transfer nuclear-state population from isomeric to fast-decaying states of the nucleus leading to release of the energy stored in the isomer.
 Such controlled depletion of isomeric state population was suggested by  \mytcite{Lia13} as a potential nuclear battery, offering  clean storage of nuclear energy.
It  would constitute an important step in the newly developing field of nuclear quantum optics.

\subsection{Concluding remarks \label{sec-concluding}}

The outlook toward promising and fascinating upcoming applications of STIRAP presented in this Sec.~\ref{sec-perspectives} shows the same rich variety of systems and problems  as the many experiments  discussed in the earlier parts of this article.
This unequivocally demonstrates that STIRAP has become a powerful enabling tool for quantum technology and,  in particular, for  quantum-state control in many areas of science.


 \acknowledgments

 We thank many colleagues for very helpful discussions of details of their work, in particular
 Dima Budker,
 Michael Drewsen,
 Giuseppe Falci,
 Robert W. Field,
 Gerald Gabrielse,
 Andrew D. Golter,
 David Grimes,
 Christoph Keitel,
 Hanns-Christoph N\"agerl,
 Adriana P{\'a}lffy,
 Cris Panda,
 Sorin Paraoanu,
{Manos Paspalakis,}
 Martin Quack,
  Mark Raizen,
  Alexander Szameit,
 Hailin Wang,
 Jun Ye, and
 Martin Zwierlein.
NVV acknowledges support from the Alexander-von-Humboldt Foundation.

\appendix
\section*{Acronyms and variations of STIRAP}

The following acronyms appear in this article. A longer list of related acronyms appears in \mycite{Sho13}.

\begin{itemize}

\item APLIP: adiabatic passage by light-induced potentials (Sec.~\ref{Sec:APLIP}); 


\item bright-STIRAP (Sec.~\ref{sec-ordering}); 

\item cavity-STIRAP (or vacuum-STIRAP) 
(Sec.~\ref{Sec:vacuum-STIRAP});

\item composite-STIRAP (Sec.~\ref{Sec: composite Stirap});

\item continuum-STIRAP 
(Sec.~\ref{Sec:continuum});

\item CTAP: coherent tunneling by adiabatic passage (Sec.~\ref{sec-CTAP}); 

\item DAP: digital adiabatic passage (Sec.~\ref{Sec:PAP}); 



\item electron-STIRAP and hole-STIRAP (Sec.~\ref{Sec:hole-STIRAP});

\item f-STIRAP, fractional STIRAP, and half-STIRAP (Sec.~\ref{sec-fractional}); 

\item Feshbach-STIRAP (Sec.~\ref{sec-ultracold molecules}); 


\item LICS-STIRAP (Sec.~\ref{sec-into-continuum});

\item multistate-STIRAP 
(Sec.~\ref{Sec:multi});

\item PAP: piecewise adiabatic passage (Sec.~\ref{Sec:PAP}); 


\item parallel-STIRAP (Sec.~\ref{Sec: parallel Stirap}); 

\item SAP: spatial adiabatic passage (Sec.~\ref{sec-CTAP}); 

\item STIHRAP: stimulated hyper-Raman adiabatic passage (Sec.~\ref{sec-STIHRAP}); 

\item straddle-STIRAP (Sec.~\ref{Sec:straddle}); 

\item tripod-STIRAP (Sec.~\ref{Sec:tripod}); 

\item two-state STIRAP 
(Sec.~\ref{sec-two state STIRAP});

\item waveguide-STIRAP (Sec.~\ref{sec waveguides}); 

\item x-ray STIRAP (Sec.~\ref{sec-x-rays}). 

\end{itemize}



\end{document}